# Techniques and Challenges in Speech Synthesis

## Final Report for ELEC4840B

**David Ferris - 3109837**

04/11/2016

# Abstract

The aim of this project was to develop and implement an English language Text-to-Speech synthesis system. This first involved an extensive study of the mechanisms of human speech production, a review of modern techniques in speech synthesis, and analysis of tests used to evaluate the effectiveness of synthesized speech. It was determined that a diphone synthesis system was the most effective choice for the scope of this project. A diphone synthesis system operates by concatenating sections of recorded human speech, with each section containing exactly one phonetic transition. By using a database that contains recordings of all possible phonetic transitions within a language, or diphones, a diphone synthesis system can produce any word by concatenating the correct diphone sequence.

A method of automatically identifying and extracting diphones from prompted speech was designed, allowing for the creation of a diphone database by a speaker in less than 40 minutes. The Carnegie Mellon University Pronouncing Dictionary, or CMUdict, was used to determine the pronunciation of known words. A system for smoothing the transitions between diphone recordings was designed and implemented.

CMUdict was then used to train a maximum-likelihood prediction system to determine the correct pronunciation of unknown English language alphabetic words. Using this, the system was able to find an identical or reasonably similar pronunciation for over 76% of the training set. Then, a Part Of Speech tagger was designed to find the lexical class of words within a sentence (lexical classes being categories such as nouns, verbs, and adjectives). This lets the system automatically identify the correct pronunciation of some Heterophonic Homographs: words which are spelled the same way, but pronounced differently. For example, the word "dove" is pronounced two different ways in the phrase "I dove towards the dove", depending on its use as a verb or a noun. On a test data set, this implementation found the correct lexical class of a word within context 76.8% of the time.

A method of altering the pitch, duration, and volume of the produced voice over time was designed, being a combination of the time-domain Pitch Synchronous Overlap Add (PSOLA) algorithm and a novel approach referred to as Unvoiced Speech Duration Shifting (USDS). This approach was designed from an understanding of mechanisms of natural speech production. This combination of two approaches minimises distortion of the voice when shifting the pitch or duration, while maximising computational efficiency by operating in the time domain. This was used to add correct lexical stress to vowels within words.

A text tokenisation system was developed to handle arbitrary text input, allowing pronunciation of numerical input tokens and use of appropriate pauses for punctuation. Methods for further improving sentence level speech naturalness were discussed. Finally, the system was tested with listeners for its intelligibility and naturalness.

# Acknowledgements

I want to thank Kaushik Mahata for supervising this project, providing advice and encouragement as to the direction my research should take, and discussing this project with me as it developed over time.

I also want to thank my friends Alice Carden, Josh Morrison-Cleary, and Megan McKenzie for sitting down and making weird sounds into a microphone for hours to donate their voices to this project. I promise never to use your voices for evil.

In addition, many thanks to those who listened to the synthesis system and provided feedback, from the barely intelligible beginnings to the more intelligible final product. Particular thanks to KC, You Ben, That Ben, Dex, and Kat, who helped with formal intelligibility and naturalness tests.

More broadly, I want to thank the staff of the Engineering and Mathematics faculties of the University of Newcastle for teaching me over the course of my degree. Without the knowledge imparted through their classes, this project would have been impossible.

Finally, I want to thank my parents and friends for putting up with me excitedly talking about various aspects of linguistics and signal processing for an entire year.

# List of Contributions

The key contributions of this project are as follows:
- Completed an extensive background review of acoustics and linguistics,
- Reviewed and compared the various tests used to evaluate the intelligibility and naturalness of speech synthesis systems,
- Reviewed and compared different techniques currently used for speech synthesis,
- Designed and implemented a system to automatically separate a speech waveform of a prompted monophone into sections of initial excitation, persistence, and return to silence,
- Designed and implemented a system to automatically extract a prompted diphone from a given speech waveform,
- Used these to automate the construction of several English language diphone databases,
- Designed and implemented a computationally efficient method of smoothly concatenating recorded diphone waveforms,
- Used the above in conjunction with a machine-readable pronunciation dictionary to produce synthesized English speech,
- Designed and implemented a data-driven method of converting arbitrary alphabetic words into corresponding English-language pronunciations,
- Designed and implemented a trigram-based Part of Speech tagging system to identify the lexical class of words within an input sentence,
- Used this Part of Speech data to determine the correct pronunciation of Heterophonic Homographs,
- Designed and implemented a novel method for arbitrarily altering the volume, fundamental frequency, and duration of synthesized speech on the phone level,
- Designed and implemented a text pre-processing system to convert arbitrarily punctuated input text into a format which the system can synthesise,
- Designed software allowing a user to define the target volume, pitch, and duration of produced speech based on sentence punctuation, lexical class of input words, and word frequency to improve system prosody,
- Combined all of the above into a comprehensive English language Text To Speech diphone synthesis system, and
- Experimentally evaluated the intelligibility and naturalness of said system.

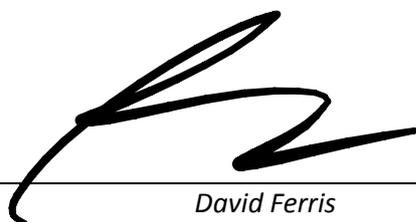
*David Ferris*

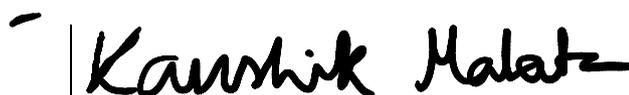
*Kaushik Mahata*

# Table of Contents

# List of Figures

# List of Tables

# 1. Introduction

Computer speakers are one of the most universal computer output devices. In personal computer usage, audio cues are used to inform the user of program behaviours and to supplement visual data. When an action completes, we hear a beep of confirmation. When our computer encounters an error, we are informed with a harsh buzz. This subtle second layer of user feedback complements the computer experience on an intuitive level.  Anyone familiar with personal computing understands this language of beeps and buzzes. Yet, perhaps ironically given that the apparatus of electronic audio output is referred to as a "speaker", personal computers rarely "speak" to us in the tongue that we use to communicate to each other.

This makes sense from a design perspective. The human reading speed is substantially faster than the maximum speed of human speech. If a system can output visual data to a user, then that is almost always a faster and more robust method of communication. If we misread a word in a displayed sentence, no further interaction with the computer is required, as we can simply re-read the word; if we mishear an acoustic cue, we need to request that the computer speak it again. Considering this, we should rarely want our devices to talk to us when they can simply pass on information visually. Interface designers have known this for years - show, don't tell.

So why are consumer electronics talking to us more than ever before?

The resurgence of natural language interaction with electronics can be traced to the advent of several technologies in recent years. The use of Global Positioning System navigation is now ubiquitous, with spoken instructions being given to the driver. Speech recognition technology has advanced astonishingly: the effective deep neural network approach has been widely adopted, giving us two-way communication between user and electronics through sound. The continuing miniaturisation of technology lets us keep smartphones in our pocket, each one containing greater processing power than cutting-edge desktops from a decade ago. The podcast is the new incarnation of the radio show, which we can play through our earphones no matter where we are. New improvements to wireless networks let us stream huge amounts of data.

The question then becomes, with all of these technologies already talking to us, why consider speech synthesis at all? It would be an understandable mistake to believe that the field has already been mastered – that speech synthesis is "solved". But this is not the case. Consider, for example, how almost all audiobooks are still manually dictated into a microphone by a human being, rather than being automatically spoken by a computer. Synthesized speech is rarely used in music - at least, not outside of niches where the robotic speech qualities match the song's aesthetic.

Despite its newfound importance as a communication mechanism with our devices, we have not achieved mastery of speech synthesis. While with current techniques we can understand what a computer is saying to us, we might not enjoy how it's being said. Over a long period of time, synthesized voices can seem repetitive, robotic, and monotonic. Small systemic errors which might remain unnoticed in a short sentence may become more obvious and grating over time.

Our greater exposure to synthesized speech has resulted in higher expectations of synthesis systems. We want these voices to sound like a real human being, emulating the natural human speech actuations. What can we do to make these systems sound more human? What methods do we currently use to synthesize speech, and why? How do we compare the effectiveness of two different speech synthesis systems? Which signal processing techniques can be used to produce better sounding speech?

These are some of the questions which this project aims to answer.

## 1.1. Report Outline
This report consists primarily of eight chapters:
- Chapter 1, this chapter, contextualises the motivations for studying speech synthesis, and outlines the content of the report.
- Chapter 2 contains an extensive background review of concepts in acoustics and linguistics, which are necessary to address and analyse the problem. It also discusses how acoustic information is captured and electronically encoded, and considers problems and challenges specific to recording human speech.
- Chapter 3 formalises the concepts of intelligibility and naturalness as they relate to speech synthesis systems, reviews various tests to evaluate intelligibility and naturalness, and determines which testing methodologies are most applicable for evaluating the effectiveness of this project.
- Chapter 4 is an overview of the techniques currently in use for speech synthesis, primarily separated into the paradigms of sample-based and generative approaches. It also discusses the advantages and disadvantages of each technique, and explains why a diphone synthesis technique was selected for this project.
- Chapter 5 explains the initial implementation of the diphone synthesis system. A method of automatically identifying and extracting prompted diphones from recorded speech is developed and implemented. A simple lookup-table approach is used to determine the pronunciation of known space-separated English words, and then a method of smoothly concatenating diphone waveforms is used to produce intelligible speech.
- Chapter 6 expands the capabilities of the synthesis system. It details and implements methodologies for determining a pronunciation for unknown alphabetic words. Then, a method of identifying the lexical class of words within a sentence is developed. This information is used to determine correct pronunciations for words which are spelled the same but which can be pronounced differently. A method for scaling the volume, pitch, and duration of synthesized speech on the level of individual phones is then developed.
- Chapter 7 discusses the final revisions to the speech synthesis system. It implements techniques for text tokenisation, so that an input string containing punctuation or numbers can be correctly synthesized. We also consider some techniques which could further improve the naturalness of our synthesized speech. Finally, a graphical user interface is designed to allow users to customise the system's sentence-level prosodic variation.
- Chapter 8 experimentally analyses the effectiveness of our system, evaluates our results, and concludes with potential avenues of future research.

## 1.2. Other Remarks on this Report
Speech synthesis, being a subset of computational linguistics, is an expansive cross-disciplinary study involving signal processing, mathematics, linguistics, acoustics, psychoacoustics, psychology, and many other tangential areas of interest. This report is written with an engineer in mind as the target audience. To address the challenges in speech synthesis, aspects of other academic disciplines must be considered in this paper. These topics may seem to be beyond what is strictly in the realm of engineering, but we can only make effective design decisions with a detailed comprehension of the systems we are working with – or, in speech synthesis, the systems we are trying to imitate.

As engineers, we must understand the problems which we are designing solutions for, even if parts of those problems go beyond what traditionally is associated with our field. The study, research, understanding, and analysis of such multidisciplinary challenges is a fundamental component of modern engineering practice. As such, the substantial depth with which we will examine these topics in this report should not be considered extraneous detail, but a vital and inherent part of the engineering design process.

# 2. Background on Linguistics and Technologies

Speech synthesis, unsurprisingly, is a field which requires some knowledge of linguistics and phonetics to consider and analyse. As this paper is targeted towards an Engineering audience, an initial examination of terms and concepts in linguistics is necessary to discuss the problem. Similarly, it is important to understand some general concepts in acoustics, as well as how computers are able to record, encode, and represent audio data. This section will therefore focus on reviewing these fields with sufficient depth to discuss speech synthesis in detail.

In this section, we will introduce various linguistic terms, with definitions derived from The Concise Oxford Dictionary of Linguistics [1]. Whenever a new definition from this source is introduced in this chapter, it will be ***in bold italics*** and indented as a bullet point for emphasis.

## 2.1. Acoustics

Acoustics is the study of mechanical waves, of which sound waves are a subset. Most sound waves are longitudinal waves, being periodic fluctuations of pressure within a medium. The fluctuation of pressure in most audible sound waves is very small relative to the ambient pressure level. Like all mechanical waves, sound waves propagate at a finite speed. This speed is determined by the properties of the medium in which it is propagating.

In our consideration of human speech generation, we will only consider propagation within atmospheric air. We must make some definitions about this medium, as actual, real-world air can vary in composition and pressure. The International Organization for Standardization (ISO) defines the atmospheric pressure at sea level to be 101325 Pascals and the temperature to be 15 °C, providing a resulting air density of 1.225 kilograms per cubic meter [2]. This gives a speed of sound of approximately 343.2 metres per second. For any simulation or model of sound propagation, it is important that these variables are as accurate as possible to ensure that the model meaningfully reflects real world behaviours.

The magnitudes of sound waves are typically measured in decibel units of sound pressure level, abbreviated as dB SPL, or often simply dB. The use of a logarithmic scale is effective due to the wide range of possible sound wave magnitudes which the human ear can detect. Like all logarithmic scales, the decibel scale is based off a reference value set at 0 dB. For sound propagating in air, this is chosen to be the quietest sound an average human can hear: a sound wave with root mean square value of 20 micropascals. Decibel scales are typically logarithmic with a base of 10, such that an increase of 10 dB is an increase in power by a factor of 10. However, as sound waves are a field quantity, propagating along a two-dimensional plane or surface rather than an axis, the conversion from a sound wave's RMS value in SPL, $p_{RMS}$, to magnitude in dB SPL, $L_p$, is given by:

$$L_p = 20 \, log_{10} \left( \frac{p_{RMS}}{20 \mu Pa} \right) \, dB \, SPL$$

In examining acoustics to study speech, it is important to be aware of the property of acoustic resonance, as it is integral to the production of voiced speech. Resonance occurs when interactions in an acoustic system (such as reflection and constructive interference) cause certain frequencies of sound to be greatly magnified, making those frequencies more audible than others. While we must know of its existence and what causes it in order to discuss speech generation, resonance is a complex and nuanced topic; detailed comprehension of it is not necessary in this report.

We also wish to consider the distinction between turbulent flow, when the flow of air in a region is highly chaotic, and laminar flow, where the flow of air is mostly smooth and linear. The distinction between the two is best described by a dimensionless quantity known as the Reynolds number, given by the ratio of inertial forces to viscous forces within the air [3].

The Reynolds number is a good descriptor of which type of force dominates: with a low Reynolds number, viscous forces dominate, and flow of air is smooth and laminar; with a high Reynolds number, inertial forces are dominant, and flow of air is unstable and turbulent. The Reynolds number is formulaically given by:

$$Re = \frac{QD_H}{vA}$$

Where  Re is the Reynolds number;
       Q is the volumetric flow rate in $m^3$/s;
       $D_H$ is the hydraulic diameter of the pipe in metres;
       v is the kinematic viscosity of the fluid in $m^2$/s;
and    A is the cross-sectional area through which the fluid is flowing in $m^2$.

From this formula, we know that the Reynolds number is inversely proportional to the cross sectional area of the material it is moving through. As such, if we have air flowing through a pipe and the pipe constricts at a certain point, that constriction will reduce the Reynolds number and can potentially cause turbulent airflow. This effect is shown in Figure 1.

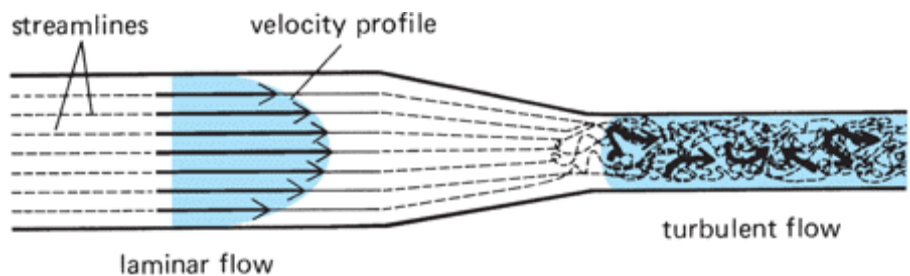

*Figure 1: Laminar and Turbulent Flow in a Constricting Pipe [4]*

Acoustically, laminar flow lends itself better to the production of sound at a certain frequency, as sound waves (being propagations within the medium of air) will maintain their coherence. Turbulent air will behave chaotically, giving rise to broadband signal noise. Both types of flow play a major role in speech production.

## 2.1.1. The Human Ear

While most of this section examines human speech generation, it is important to make a few remarks on the way that human ears receive information about sound and communicate it to the brain. The branch of acoustics concerned with sounds produced and received by living beings is called bioacoustics.

Humans, like most mammals, have two ears. This allows people to determine the relative position of a sound source, which the brain deduces from information such as the different relative loudness of the same sound source as heard by each ear. Determining the direction of a sound source is made easier by the auricle, that part of the ear which is outside of the head, which helps to directionally gather sound energy and focus it through the external auditory canal.

This sound then reaches the tympanic membrane (commonly referred to as the eardrum) as shown in Figure 2. The eardrum separates the outer ear from the middle ear behind it. This sound is then transmitted through the auditory ossicles, interlocking bones which help magnify sound and transmit it further into the ear; these are the malleus, incus, and stapes. The stapes is attached to the vestibular or oval window, which connects it to the cochlea of the inner ear. The cochlea is filled with fluid, which (due to the mechanical sound amplification from the ossicles) vibrates strongly. Finally, this fluid vibrates against stereocilia, hair cells which convert the vibrations into neural impulses and transmit them through the cochlear nerve. [5]

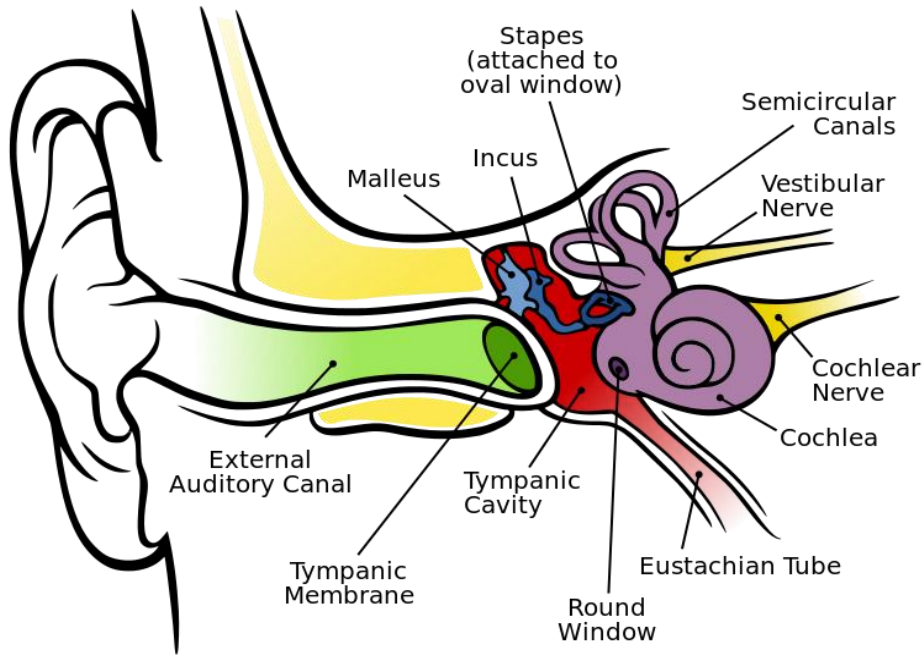

*Figure 2: Diagram of the Human Ear [6]*

Due to the physical properties of the ear, the frequency response of human hearing is nonlinear; that is, sound waves of the same magnitude but different frequencies will subjectively appear to have different volumes. As such, subjective human assessment of loudness of a sound is not always an effective assessor of its actual dB SPL. The subjective loudness of a sound is therefore a separate value to its SPL. Loudness is measured using the phon, with loudness in phons being equivalent to the dB SPL of a sound wave with a frequency of 1kHz with the same subjective loudness.

The frequency response of the human ear is most commonly represented using a series of equal-loudness contours, as shown in Figure 3. Each curve represents the actual sound pressure levels of sounds at different frequencies that a listener perceives to be of the same loudness. [7]

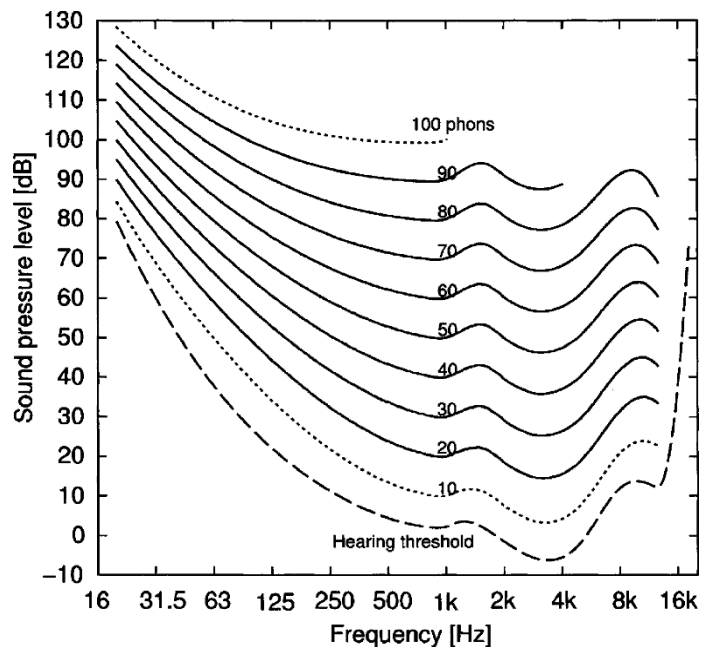

*Figure 3: Average Human Equal Loudness Contours [7]*

This perceptual difference between different frequencies can factor into several design decisions for speech synthesis. It is especially important when considering suitable sampling frequencies for digital audio signals. The maximum sound frequency within the human hearing range is typically around 20 kHz. As such, if we want to be able to represent any human-audible sound, then by the Nyquist sampling theorem we know that our sampling frequency should be at least double this, at a minimum of 40 kHz. Sampling frequencies for sound waves are thus chosen above 40 kHz in order to prevent signal aliasing. Historically, a sampling rate of 44.1 kHz was most commonly used for recordings of human speech, as it is divisible by both 50 and 60, making it cross-compatible with both 50 and 60Hz power frequencies. In the modern day, 48 kHz frequencies are also common.

## 2.2. Human Speech Production

The field of phonetics is the subset of linguistics concerned with the sounds of human speech. There are two main areas of phonetics which we are interested in:

- **Articulatory Phonetics**: *The study of the production of speech sounds.*
- **Acoustic Phonetics**: *The study of the physical properties of the sounds produced in speech.*

This section will focus primarily on articulatory phonetics, but will also touch on topics in acoustic phonetics where appropriate.

The human body has various ways of producing sound that we would class as speech. The most universal source of sound in human speech is excitation of the vocal folds.

- **Vocal Folds/Cords:** *Parallel folds of mucous membrane in the larynx, running from front to back and capable of being closed or opened, from the back, to varying degrees. In normal breathing they are open; likewise in producing speech sounds that are voiceless. In the production of voice they are close together and vibrate as air passes periodically between them.*
- **Glottis:** *The space between the vocal folds; thus also "glottal", relating to the glottis.*
- **Phonation:** *The specific action of the vocal cords in the production of speech. Phonation types include voice, in which the vocal cords vibrate, vs. voicelessness, in which they are open.*

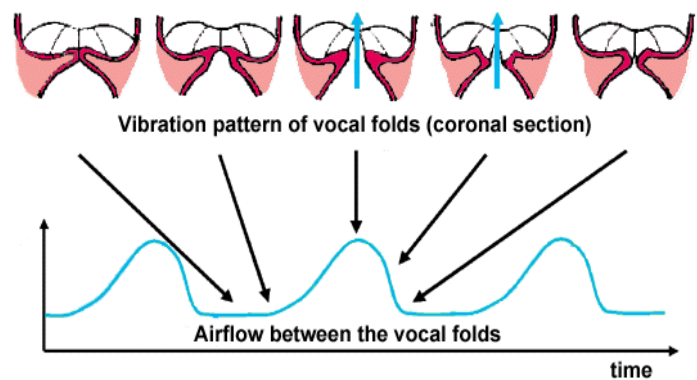

*Figure 4: Illustration of Glottal Airflow [8]*

In producing voiced speech, the vocal folds open and close at a certain frequency, transmitting and blocking air from the lungs with a regular period. As shown in Figure 4, this creates a pulse sequence in the sound waveform: an increase in pressure as air passes through the vocal folds, and then a decrease as the folds close. The frequency at which this occurs is referred to as the fundamental frequency of voiced speech. Due to physiological sex differences, the fundamental frequencies of male and female voices typically occupy different ranges. The average range for male speakers is between 85 and 180 Hz, while the average adult female speaker's range is between 165 and 255 Hz.

- ***Vocal Tract:*** *The passages above the larynx through which air passes in the production of speech: thus, more particularly, the throat, the mouth (or **oral tract**), and the passages in the nose (or **nasal tract**).*

While the vocal cords contribute to the fundamental frequency of human voice, the sound that they generate resonates within the vocal tract. This results in additional frequencies resonating, depending on the configuration of the vocal tract. A labelled diagram of the different sections of the vocal tract is shown in Figure 5. This also indicates the different sections of the tongue, as will later be referred to in our consideration of phonetics.

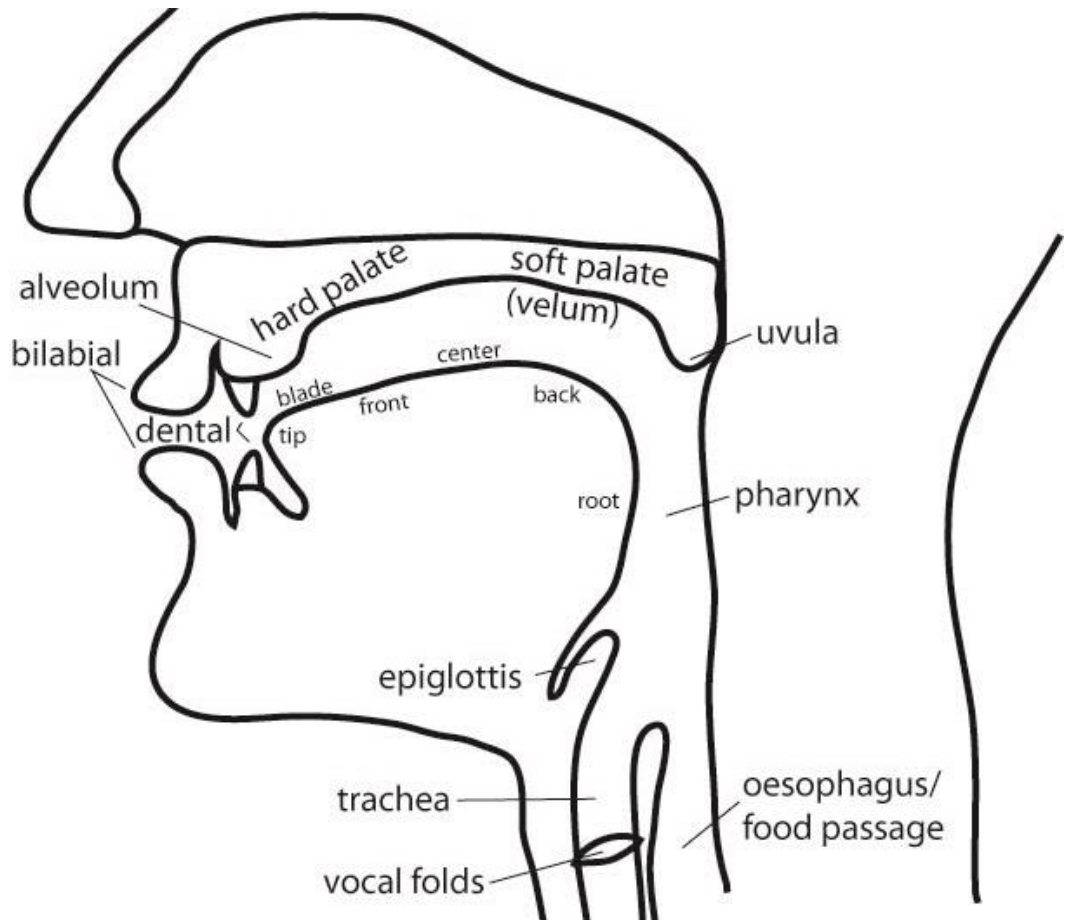

*Figure 5: Labelled Human Vocal Tract [9]*

- ***Articulator***: *Any vocal organ used to form specific speech sounds. Places of articulation are defined by the raising of a movable or active articulator, e.g. a part of the tongue, towards a fixed or passive articulator, e.g. a part of the roof of the mouth.*
- ***Stricture****: Any constriction of part of the vocal tract in the articulation of a speech sound, varying from Open Approximation, to Close Approximation, and then Closure.*
    - ***Open Approximation:*** *A narrowing of the space between articulators which is not enough to cause turbulence in the flow of air through the mouth.*
    - ***Close Approximation:*** *Narrowing of the space between articulators sufficient to cause turbulence in a flow of air.*
    - ***Closure****: Contact between articulators by which a flow of air is completely blocked.*

Varying approximation in the vocal tract produces different resonant frequencies in the speech sound produced, causing peaks to occur in the frequency domain of the speech waveform. For instance, even if the fundamental frequency of voiced speech stays the same, the tongue being raised at the back will produce an audibly distinct sound from it being raised at the front.

Later in this report, we will discuss the use of signal processing techniques on speech waveforms to modify their pitch and duration, as well as methods for programmatically synthesizing waveforms to imitate real world speech production. As there are a wide range of possible speech sounds, we must often consider the acoustic characteristics of the produced sound waves. It is therefore important to understand the differences between the waveforms of distinct speech sounds. To illustrate this, we will provide some waveforms and spectrograms of different kinds of speech sounds on the same time scale. As has been discussed, speech sounds can be voiced or unvoiced, and be produced with open approximation, close approximation, or closure. Unvoiced speech with open articulation has no source of sound production, so we do not need to consider it.

If produced speech is voiced, with open approximation, then the produced speech sound's waveform will be periodic according to the fundamental frequency of phonation. There will then be some additional frequency contributions due to resonance within the vocal tract. Figure 6 shows three waveforms of voiced speech produced with open articulation. They are all produced with similar fundamental frequency, but different articulation, giving rise to distinct waveform shapes.

Speech sounds which are unvoiced, with close approximation, are mostly produced by turbulence within the mouth. As such, their waveforms appear as noise with a high frequency relative to the frequency of phonation. Distinct sounds can be produced depending on the particular articulator and the degree of closeness. While it is difficult to distinguish between such waveforms in the time domain, their spectral characteristics more clearly show their differences. Figure 7 shows three waveforms of unvoiced speech with close approximation, as well as their spectrograms.

Voiced speech with close approximation has both of these characteristics: from Figure 8, we can see large-scale/low frequency periodicity due to phonation, but also small-scale/high frequency noise due to the close approximation. Again, these sounds are best distinguished by their spectrograms.

Speech sounds contributed to by closure cannot be persisted in the same way other speech sounds can: as closure completely blocks airflow, the speech sound cannot be produced over a longer period of time. Sounds from closure are very brief relative to the duration of other speech sounds, and are often produced by percussive effects in the mouth or short-term air expulsion. Figure 9 shows speech waveforms produced from a starting point of closure, with one being voiced and the other not.

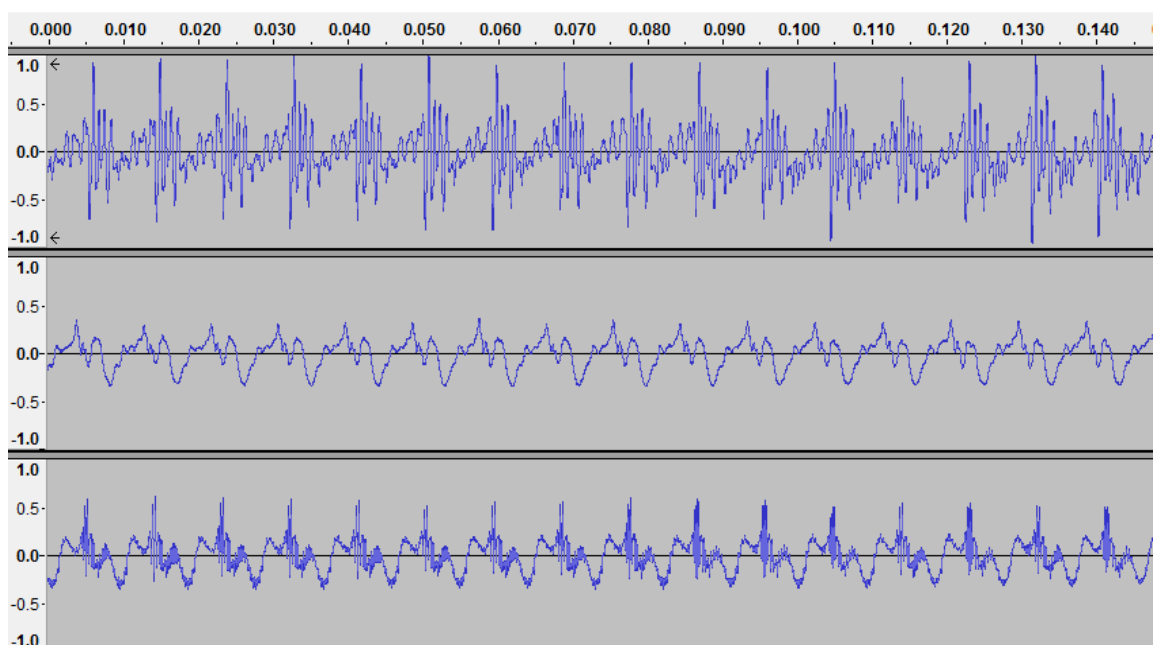

*Figure 6: Voiced Speech Sounds with Open Approximation*

*2. Background on Linguistics and Technologies*

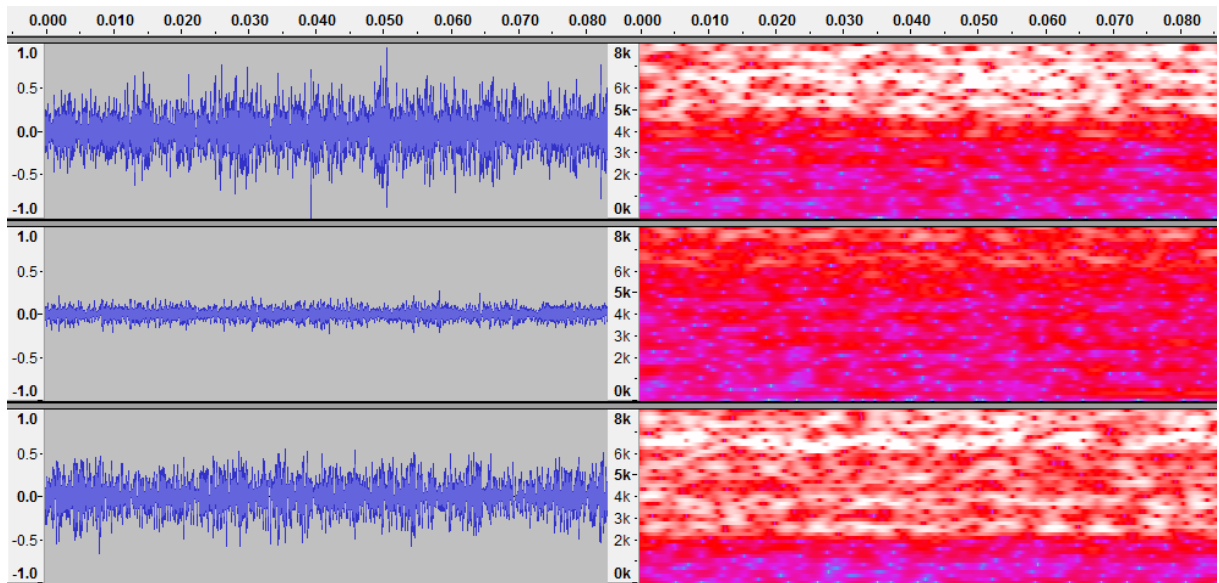

*Figure 7: Unvoiced Speech Sounds with Close Approximation*

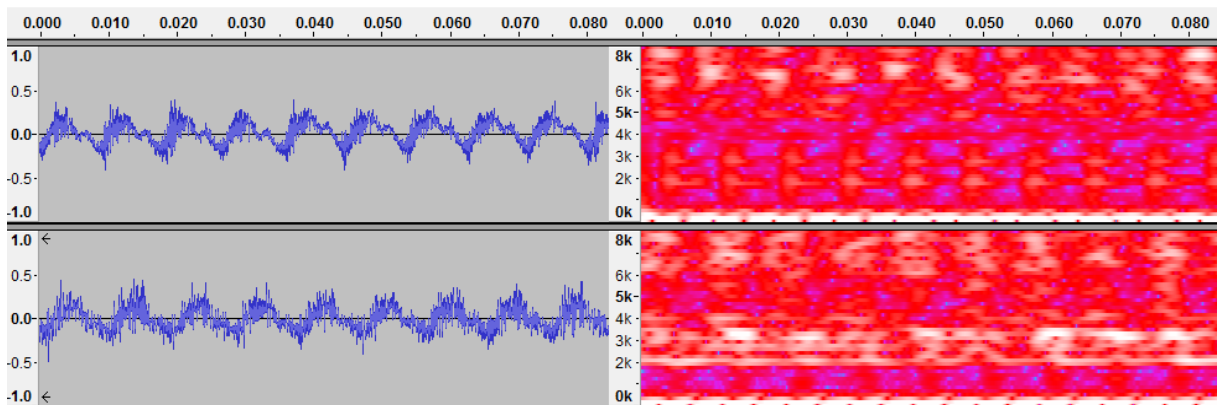

*Figure 8: Voiced Speech Sounds with Close Approximation*

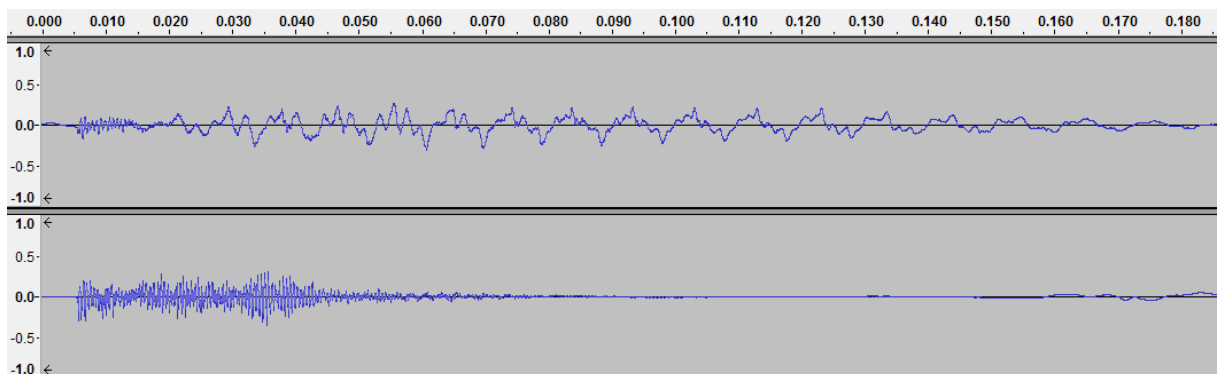

*Figure 9: Voiced (above) and Unvoiced (below) Speech Sounds from Closure*

The unique characteristics of these different speech sounds will require particular consideration in many parts of our design process: speech waveforms are composed of periodic and non-periodic sound sources, and the relative duration of different kinds of speech sounds varies greatly. Another factor which complicates the problem is that natural speech moves between these different points of articulation and levels of phonation over time.

We will postpone discussions on this until later in the report. For now, we will continue to review the background knowledge necessary to discuss the problem in greater detail.

## 2.2.1. Classification of Speech Units

While the previous section explained how the physical articulation of different sounds in speech occurs, it is also necessary to identify between different speech sounds within a hierarchy.

- ***Phonology:*** *The study of the sound systems of individual languages and of the nature of such systems generally. Distinguished as such from phonetics.*

Where phonetics is purely interested in the speech sounds themselves, their properties, and their methods of articulation, phonology seeks to describe and categorise speech sounds in a more specific fashion, and broadly to examine properties of languages.

- ***Phoneme****: The smallest distinct sound unit in a given language.*
- ***Phone****: A speech sound which is identified as the realization of a single phoneme.*
- ***Syllable****: A phonological unit consisting of a unit that can be produced in isolation.*
- ***Nucleus****: The central element in a syllable.*

It is important to understand the distinction between these terms. Where a phoneme is a sound unit, the term phone refers to the sound itself. A phoneme is a smaller unit than a syllable; a syllable is composed of one or more phonemes which together can be produced in isolation.

Here we also introduce the term **diphone**, referring to the sound of two phones as one flows into another. Diphones capture the transition between two phones, and are of particular interest in speech synthesis; this will be discussed in greater detail later in this report.

- ***Vowel****: A minimal unit of speech produced with open approximation which characteristically forms the nucleus of a syllable.*
    - ***Quality****: The auditory character of a vowel as determined by the posture of the vocal organs above the larynx.*
    - ***Formant****: A peak of acoustic energy centred on one point in the range of frequencies covered by the spectrum of a vowel. Vowels have several formants, but the distinctions as perceived between them lie, in particular, in the three lowest.*

The quality is that aspect of a vowel defining it within articulatory phonetics, whereas the formants of a vowel define it within acoustic phonetics. Any particular vowel will be composed of multiple formants, determined by the oscillations of the vocal folds and the configuration of the articulators in the vocal tract. Vowels of different quality are often grouped according to their phonetic closeness or openness and whether they are articulated in the front, centre, or back. The International Phonetic Alphabet (IPA) vowel quadrilateral is shown in Figure 10. The IPA will be discussed in further detail in Section 2.3.1 on Page 14.

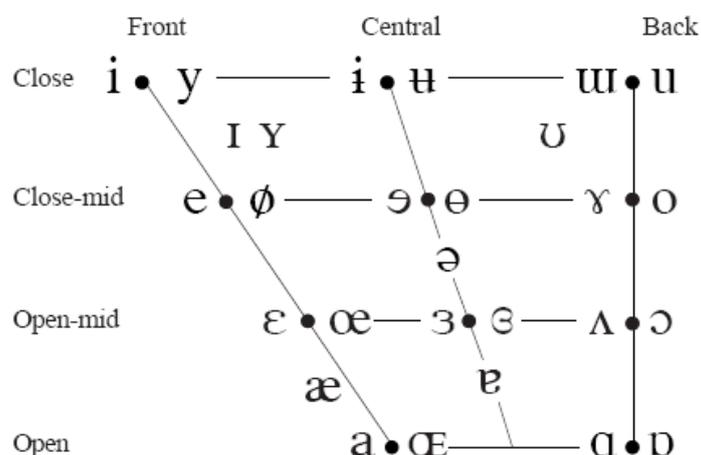

*Figure 10: IPA Vowel Chart [10]*

- **Monophthong**: *A vowel whose quality does not change over the course of a single syllable.*
- **Diphthong**: *A vowel whose quality changes perceptibly in one direction within a single syllable.*
- **Triphthong**: *A vowel whose quality changes in two successive directions within a single syllable.*
- **Hiatus:** *A division between vowels belonging to different words or syllables.*

The different types of vowel transitions play a vital part in language comprehension; in English, the distinction between a Diphthong and a Hiatus can be especially important. For example, in the word "iodine", the initial "i" and the "o" belong to distinct syllables rather than being a transition within the same syllable; it is therefore a hiatus transition. In the word "noun", the "o" and the "u" contribute to different sounds but belong to the same syllable, meaning that they form a diphthong.

- **Consonant**: *A phonological unit which forms parts of a syllable other than its nucleus.*
- **Semivowel**: *A unit of sound which is phonetically like a vowel but whose place in syllable structure is characteristically that of a consonant.*

Semivowels, being phonetically similar to vowels, are often grouped as such in articulatory classification systems, but can also be grouped with consonants in a more phonological categorisation. The broader category of consonants can be articulated in a wide variety of ways. In English, all consonants fall within a two-dimensional grid, with one dimension being the place of articulation, and the other being the manner of articulation. Consonants within that grid can also either be voiced or unvoiced. The IPA consonant table for American English is shown in Table 1.

The following are manners of articulation which occur in English:
- **Fricative**: *Consonant in which the space between articulators is constricted to the point at which an air flow passes through with audible turbulence.*
  - **Sibilant:** *Fricative characterized by turbulence that produces noise at a high pitch, e.g. [s] and [z] in sin and zip.*
- **Stop**: *Consonant in whose articulation a flow of air is temporarily blocked: e.g. [p] and [t] in pit.*
  - **Affricate**: *A stop consonant released with a fricative at the same place of articulation.*
  - **Plosive:** *A stop produced with air flowing outwards from the lungs: e.g. [t] in tea.*
- **Nasal**: *Consonant produced with lowering of the soft palate, so that air may pass through the nose; opposed to oral. Thus [m] and [n] in man [man] are nasal. The nasal cavity is the additional resonating chamber formed by the passages through the nose when the soft palate is lowered.*
- **Liquid**: *Cover term for 'r's and 'l's, especially in languages where their roles in phonology are similar.*
- **Glide**: *An audible transition from one sound to another, typically of semivowels.*

The following are places of articulation which occur in English:
- **Bilabial:** *Articulated with the lower lip against or approximated to the upper lip. E.g. p in pit [p] is a bilabial stop.*
- **Labiodental:** *Articulated with the lower lip against or approximated to the upper teeth: e.g. [f] in fin or [v] in veal.*
- **Dental/Interdental:** *Articulated with the tip or blade of the tongue against or approximated to the upper teeth: e.g. t [t̪] in Italian, where the diacritic distinguishes dentals from alveolars.*
- **Alveolar:** *Articulated with the tip or blade of the tongue against or approximated to the ridge behind the upper teeth: e.g. [t] and [d] are normally alveolar in English.*
- **Palatal:** *Articulated with the front of the tongue against or approximated to the hard palate.*
- **Velar:** *Articulated with the back of the tongue against or approximated to the soft palate (or velum). E.g. [k] in [kat] (cat) is a velar stop.*
- **Glottal:** *Articulated with the vocal cords.*

*Table 1: Classification of North American English Consonant Phonemes in IPA [11]*

| Manner of Articulation | Place of Articulation | | | | | | |
|---|---|---|---|---|---|---|---|
| | Bilabial | Labiodental | Dental | Alveolar | Palatal | Velar | Glottal |
| Stop Voiceless Voiced | p  b | | | t  d | | k  g | |
| Fricative Voiceless Voiced | | f  v | θ  ð | s  z | ʃ  ʒ | | h |
| Affricate Voiceless Voiced | | | | | tʃ  dʒ | | |
| Nasal Voiced | m | | | n | | ŋ | |
| Liquid Voiced | | | | l | r | | |
| Glide Voiced | w | | | | y | | |

## 2.2.2. Classification of Words

Now that we have examined how speech sounds are articulated and structured within a language, we also wish to define properties of language on a larger scale, such as within words and sentences, which allow meaning to be imparted.

- **Semantics:** *The linguistic study of meaning.*
- **Grammar:** *Any systematic account of the structure of a language and the patterns that it describes; typically restricted to the study of units that can be assigned a meaning.*
- **Morphology:** *The study of the grammatical structure of words and the categories realized by them. Thus a morphological analysis will divide girls into girl and -s, which realizes 'plural'; singer into sing and -er, which marks it as a noun referring to an agent.*
- **Syntax:** *The study of relations established in a grammar between words and other units that make up a sentence.*
- **Parts of Speech/Lexical Categories:** *A set of word classes, primarily as distinguished by the syntactic constructions its members enter into.*

The following definitions are non-exhaustive examples of different lexical categories. These and further categories will be further discussed later in this paper as they become relevant.

- **Noun:** *A word characterized by members denoting concrete entities, e.g. tree, sun, moon.*
- **Pronoun:** *A unit with syntactic functions similar to a noun phrase whose meaning is restricted to those distinguished by specific grammatical categories, e.g. him, her, it.*
- **Verb:** *One of a class of lexical units characteristically of words denoting actions or processes.*
- **Conjunction:** *A word etc. which joins two syntactic units. e.g. "but" is a conjunction which joins the units in [[came] but [didn't stay]], imparting particular meaning.*

## 2.2.3. Prosodic Features of Speech

Previously we described phonemes, letting us analyse the acoustic features of small, individual segments of speech. In this section, we define features of speech which become important when considering larger phonetic groupings, such as of different syllables within a word or different words within a sentence. These are referred to as the suprasegmental or prosodic elements of speech.

- **Stress**: *Phonological feature by which a syllable is heard as more prominent than others. Also of sentence stress and prominence within larger units generally. The phonetic correlates vary: in auditory terms, stress can mean a difference in length, in perceived loudness, in vowel quality, in pitch, or in a combination of any of these.*
- **Pitch:** *The property of sounds as perceived by a hearer that corresponds to the physical property of frequency. Thus a vowel which from the viewpoint of articulatory phonetics is produced with more rapid vibration of the vocal cords will, from an acoustic viewpoint, have a higher fundamental frequency and, from an auditory viewpoint, be perceived as having higher pitch.*
- **Loudness:** *The auditory property of sounds which corresponds in part to their acoustic amplitude or intensity, as measured in decibels.*
- **Timbre:** *The auditory properties of sounds other than those of pitch and loudness: hence sometimes, in phonetics, in the sense of vowel quality.*
- **Length:** *Phonetic or phonological feature, especially of vowels. A phonological distinction described as one of length may well be realized, in part or entirely, by differences other than physical duration.*
- **Pause:** *Any interval in speaking between words or parts of words.*

Prosody is vital for imparting speech with information beyond the literal meaning of the sentence being said. The prosody of a sentence can communicate the emotion of the speaker, irony, sarcasm, or emphasis on important aspects of a sentence. In normal speech, humans will vary these prosodic aspects of speech intuitively to impart additional meaning to the spoken word.

## 2.3. Transcribing Phonetic Information

In discussing various aspects of linguistics, it is necessary to transcribe information about a language. Unfortunately, in many languages, the native writing system does not effectively communicate the phonetic character of words.
- **Grapheme:** *A character in writing, considered as an abstract or invariant unit which has varying realizations. E.g. the grapheme <A> subsumes the variants or 'allographs' 'A' (Roman capital), 'a' (Roman minuscule), and so on.*

Describing the problem in set theoretic terms, the Latin alphabet used to write the English language does not have a one-to-one or bijective mapping to the phonetic content of the words they correspond to. Identical graphemes do not always correspond to the same phoneme, and transcriptions of identical phonemes do not always correspond to the same grapheme. For example, the phrase "I shed a tear as I saw him tear up the paper" uses the word "tear" twice, but the word is pronounced differently each time. These are called heterophonic homographs, as the phonemes that constitute them are different while the graphemes that constitute them are the same. [12]

The converse problem also exists; in the words "right" and "rite", both words are spelled differently but are pronounced identically. Therefore it is possible for different graphemes in English to map to the same phonemes. Such examples are called homophonic heterographs.

Due to the existence of sociolinguistic accents (local variations on how words are pronounced) even the same word in the same context might be pronounced differently. In English, most consonants are pronounced identically regardless of locale, but vowel pronunciations tend to vary from place to place. For example, the vowel "a" in the word "tomato" makes a different sound depending on whether it is pronounced with an Australian or a General American accent. Within the umbrella of English in general, this makes the word tomato a heterophonic homograph. Further, local spelling variants as aluminum/aluminium and realise/realize are usually pronounced the same, acting as homophonic heterographs. Thus, sociolinguistic accents can complicate the problem even more.

These problems are partly due to the influence of many other languages on English as it evolved, but also arise from the lack of any central standardisation of languages in general. There have been many unsuccessful efforts throughout history for English-language spelling reform to establish a one-to-one correspondence from graphemes to phonemes. Such a transcription system would be entirely homographic, such that every transcription of a word in that language communicates how it is pronounced. Perhaps unfortunately for our purposes, none have been widely adopted for general usage.

English therefore has both heterographic and heterophonic elements. The existence of heterophonic homographs in the English language makes speech synthesis exclusively from text input more difficult. The synthesizer needs to find the correct phonetic information corresponding to each word in the sentence, and cannot do this from such a word in isolation. We therefore need to distinguish which pronunciation of the word should be used based on context, which is a rather difficult problem. This will be discussed further in Section 6.2. on Page 66.

For now, it is important to establish a transcription system which has a one-to-one mapping between graphemes and phonemes: we need every written symbol to correspond to exactly one phoneme, and every phoneme within the language we are transcribing to correspond to exactly one grapheme. By using such a system, we can describe how words are pronounced or pronounce a transcribed word with no ambiguity in either direction.

## 2.3.1. International Phonetic Alphabet (IPA)

The International Phonetic Alphabet, or IPA, is the most widely used phonetic notation system. It is composed of various symbols, each of which is either a letter, transcribing the general character of the phoneme, or a diacritic, which clarifies further detail. There are 107 letters and 52 diacritics in the current IPA scheme, though most of these are not necessary to transcribe English language speech. [10]

To distinguish phrases which are transcribed phonetically rather than regular text, IPA is typically enclosed in either square brackets, as in [a], or slashes, as in /a/. This is a notation which will be used throughout this paper whenever IPA is used. Most of the IPA graphemes which are used in transcribing the English language have already been shown in Figure 10 and Table 1. Transcriptions in IPA can either be narrow, where diacritics are used extensively to detail the exact properties of the phonemes transcribed, or broad, only including approximate detail and few diacritics [13]. Some sample broad transcriptions of English sentences in IPA are shown in Table 2.

*Table 2: Example Broad IPA Transcriptions in North American English*

| English | IPA |
| --- | --- |
| The cloud moved in a stately way and was gone. | /ðə klaʊd muvd ɪn ə ˈsteɪtli weɪ ænd wʌz gɔn/ |
| Light maple makes for a swell room. | /laɪt ˈmeɪpəl meɪks fɔr ə swɛl rum/ |
| Set the piece here and say nothing. | /sɛt ðə pis hir ænd seɪ ˈnʌθɪŋ/ |
| Dull stories make her laugh. | /dʌl ˈstɔriz meɪk hɜr læf/ |
| A stiff cord will do to fasten your shoe. | /ə stɪf kɔrd wɪl du tu ˈfæsən jʊər ʃu/ |

To again use set theoretic terms, IPA not only attempts to establish an injective mapping from graphemes to phonemes, but a bijective one, filling all of the phoneme space used in human spoken language. IPA can even encode some elements of prosody, such as breaks (through the major and minor prosodic units) and intonation (through the various tone letters). There is even a set of letters and diacritics composing Extensions to the IPA, which can be used to transcribe elements of speech impediments.

IPA is a powerful, unambiguous notation system which is useful for a wide range of linguistic applications. However, it uses characters which are difficult to encode in software, which becomes relevant when we want a machine-readable transcription for software to interpret. For this reason, the Speech Assessment Methods Phonetic Alphabet, or SAMPA, was developed [14]. SAMPA encodes various IPA characters with ASCII characters, such as replacing /ə/ with the /@/ symbol, and replacing /œ/ with the /9/ symbol. This allows for broad phonetic transcriptions, however, SAMPA does not cover the entire range of IPA characters; the more advanced X-SAMPA rectifies this and allows for easier computer input of the entire IPA system.

## 2.3.2. Arpabet

Arpabet is a phonetic notation system which was designed by the U.S. Department of Defense's Advanced Research Projects Agency (ARPA, hence Arpabet). Rather than attempting to describe all possible elements of human speech, it is exclusively concerned with encoding all phonemes which are used in the General American English sociolinguistic accent.

It is composed of 48 distinct symbols (all of which are composed of one or two capital letters) and 3 classes of stress marker (which are denoted at the end of a vowel using the digits 0, 1, and 2). It also uses punctuation marks in an identical way to written English; an Arpabet transcription of an English phrase will retain any characters such commas, semicolons, colons or periods. These can effectively act as prosodic markers within the sentence, equating to short and long pauses. Some example Arpabet transcriptions are shown in Table 3.

*Table 3: Example Arpabet Transcriptions (with slashes replacing inter-word spaces for clarity)*

| English | Arpabet |
|---|---|
| The cloud moved in a stately way and was gone. | DH AH0/K L AW1 D/M UW1 V D/IH0 N/AH0/S T EY1 T L IY0/ W EY1/AH0 N D/W AA1 Z/G AO1 N/. |
| Light maple makes for a swell room. | L AY1 T/M EY1 P AH0 L/M EY1 K S/F AO1 R/AH0/S W EH1 L/ R UW1 M/. |
| Set the piece here and say nothing. | S EH1 T/DH AH0/P IY1 S/HH IY1 R/AH0 N D/S EY1/ N AH1 TH IH0 NG/. |
| Dull stories make her laugh. | D AH1 L/S T AO1 R IY0 Z/M EY1 K/HH ER1/L AE1 F/. |
| A stiff cord will do to fasten your shoe. | AH0/S T IH1 F/K AO1 R D/W IH1 L/D UW1/T UW1/F AE1 S AH0 N/ Y AO1 R/SH UW1/. |

When used to transcribe English language phonetic data, Arpabet has some advantages and disadvantages compared to the IPA. One advantage is that it uses only ASCII characters, making Arpabet transcriptions easy to store in text on a computer. For this reason, it is more immediately readable to someone who already speaks English, as it only uses Latin symbols which they would already be familiar with. It is also encoded using only characters which exist on an English keyboard, which means that it is trivial for a human to input phonetic transcriptions. In using Arpabet instead of IPA, there is no need for an intermediary system like SAMPA.

By only aiming to describe English language speech sounds, Arpabet uses fewer distinct symbols than the IPA, making it easier to fully learn and understand in a short time. Additionally, as Arpabet only describes the General American English accent, we do not have to be concerned with handling different pronunciation variants between sociolinguistic accents. There may be multiple IPA transcriptions of the same word due to sociolinguistic accent, but using Arpabet, any particular word is always transcribed in an identical fashion.

Arpabet encoding also specifically denotes some English diphthongs with distinct characters from their two constituent monophthongs, where a broad IPA transcription would not contain this information. This can be seen in Table 4 in the rows for the Arpabet graphemes AW, AY, EY, OW, and OY, where two separate characters would be used in IPA. As some English to IPA dictionaries only contain broad transcriptions (due to using SAMPA for transcription), this means that Arpabet transcriptions can often contain more detail on the behaviour of vowel transitions within the word.

*Table 4: Arpabet and IPA Correspondence with Example Transcriptions in General American English*

| Arpabet | IPA | Example | Arpabet Transcription |
|---|---|---|---|
| AA | /ɑ/ | odd | AA D |
| AE | /æ/ | at | AE T |
| AH | /ə/ | hut | HH AH T |
| AO | /ɔ/ | ought | AO T |
| AW | /aʊ/ | cow | K AW |
| AY | /aɪ/ | hide | HH AY D |
| B | /b/ | be | B IY |
| CH | /tʃ/ | cheese | CH IY Z |
| D | /d/ | dee | D IY |
| DH | /ð/ | thee | DH IY |
| EH | /ɛ/ | Ed | EH D |
| ER | /ɝ/ | hurt | HH ER T |
| EY | /eɪ/ | ate | EY T |
| F | /f/ | fee | F IY |
| G | /g/ | green | G R IY N |
| HH | /h/ | he | HH IY |
| IH | /ɪ/ | it | IH T |
| IY | /i/ | eat | IY T |
| JH | /dʒ/ | gee | JH IY |
| K | /k/ | key | K IY |
| L | /l/ | lee | L IY |
| M | /m/ | me | M IY |
| N | /n/ | knee | N IY |
| NG | /ŋ/ | ping | P IH NG |
| OW | /oʊ/ | oat | OW T |
| OY | /ɔɪ/ | toy | T OY |
| P | /p/ | pee | P IY |
| R | /r/ | read | R IY D |
| S | /s/ | sea | S IY |
| SH | /ʃ/ | she | SH IY |
| T | /t/ | tea | T IY |
| TH | /θ/ | theta | TH EY T AH |
| UH | /ʊ/ | hood | HH UH D |
| UW | /u/ | two | T UW |
| V | /v/ | vee | V IY |
| W | /w/ | we | W IY |
| Y | /j/ | yield | Y IY L D |
| Z | /z/ | zee | Z IY |
| ZH | /ʒ/ | seizure | S IY ZH ER |

## 2.4. Encoding and Visualising Audio Information

In the real world, the pressure within a sound wave varies continuously over time. However, if we wish to store a recorded sound in a computer, we need to sample it and store it as a discrete-time signal. Similarly, if we wish to play back that audio information using a speaker, we must be able to read the information recorded. It is also often useful to represent the data visually so that the information encoded in the sound can be more easily understood. This can provide greater intuitive understanding of the processes involved in generating that sound. In this project, we will be extensively considering audio signals, so it is important to understand the processes involved in digitising sound. We also wish to analyse and address problems specific to digitally recording speech.

### 2.4.1. Recording Digital Audio Signals

Sound is stored on a computer by way of a microphone. There are various specific designs of microphone, which mostly differ in the mechanism of transforming the mechanical sound wave into an electrical signal. They can operate based on electromagnetic induction, capturing a sound wave with the vibrations of a coil. Other designs operate by measuring the signal produced by a changing distance between two plates of a capacitor, with one plate free to vibrate due to the incident sound wave. Despite these differences, most microphones have similar mechanical designs: one section of the microphone remains stationary while another is free to move along an axis, kept in place by a membrane which captures sound. In that fashion, their operation is analogous to that of the human ear, with the membrane acting as an electronic eardrum. Speakers operate in the reverse fashion, where the membrane is stimulated by an electrical signal to produce a mechanical sound wave.

Once the sound wave has been captured as an electrical signal by a microphone, it must be transformed from an analog to a digital signal if we wish to store it on a computer. As previously discussed, due to the frequency response of human hearing, 44.1 and 48 kHz sampling frequencies are widespread in use and sufficient for most applications. Lower sampling frequencies cannot represent the higher end of the range of human hearing when the signal is reproduced as a sound wave, while higher sampling frequencies will capture no additional data within the hearing range of the average listener, needlessly increasing the storage space necessary to keep it on a system. As such, for this project, all audio signals will be recorded at 48 kHz.

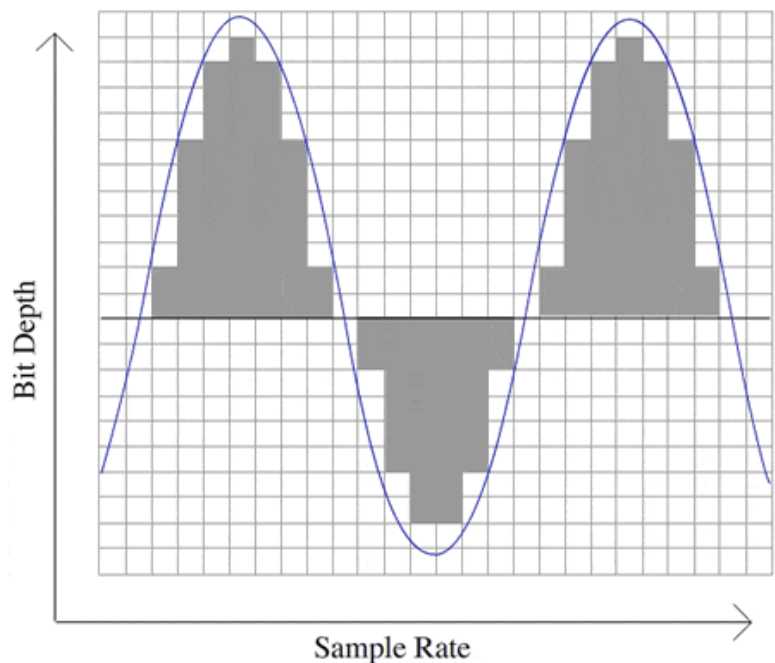

*Figure 11: Sample Rate and Bit Depth in Digital Audio Signals [15]*

*2. Background on Linguistics and Technologies*

An audio signal is also captured using a certain number of bits per sample, or bit depth. This brings up two main problems. The first, signal clipping, is when an input sound displaces the membrane of the microphone so much that even the largest magnitude expressible in a digital format cannot capture it. This is easily resolved by appropriately adjusting the gain of the microphone. The second problem is of having a reasonable signal-to-quantisation-noise ratio; by sampling the signal at some quantisation, we may introduce noticeable signal noise. As with frequency, we should make our decision based on the properties of human hearing. Human hearing has a dynamic range of approximately 120dB, so we wish to keep this noise ratio below that [16]. The relationship between bit depth, sampling rate, and the digital signal which we record is shown in Figure 11.

Modern computers usually represent digital audio signals with pulse-code modulation using an integer number of bytes per sample. The most common modern bit depths are 16, 24, and 32 bit sound, which give 96 dB, 144 dB, and 192 dB signal-to-quantisation-noise ratios respectively. The tradeoff between storage space and sound quality means that 16 bit sound is often used, but 24 bit sound gives us a signal-to-noise ratio beyond human hearing, making it suitable for almost all audio playback applications. Using 32 bit sound can provide an even more detailed wave, which may be more useful when we wish to perform signal processing tasks. However, it also makes the signal far more sensitive to electronic and acoustic noise, and in audio quality gives no great improvement over 24 bit sound waves. As such, all signals used in this project will be recorded at 24 bit depth.

In most audio formats, different streams of audio data can be stored in different channels; most commonly, it is stored in either one channel (mono sound) or two channels (stereo sound). In conjunction with the use of multiple speakers, such as in stereo headphones, a stereo recording can emulate elements of natural human hearing's directionality by playing sounds from multiple sources at different volumes in each ear. While this can add richness and realism to a sound, this project is interested solely in emulating the sound from a single source, which we will assume to be directly in front of the virtual listener, with no acoustic effects from the environment. This means that, even if we were to record in stereo, the audio data of both the left and right channels should be almost identical. Therefore, all recordings in this project are single-channel audio signals.

Most audio formats compress the sound data in either a lossless or lossy fashion. Lossless formats can perfectly reconstruct the original audio signal encoded, while lossy formats only reconstruct an approximation of the original signal. When playing back recorded speech, the reconstruction from a lossy format can have various abnormalities which might annoy the listener. If we have a lot of audio recordings and computer storage space is an issue, there may be an acceptable tradeoff between low storage space and high fidelity of our recording. Even lossless compression may cause an issue; if we are performing operations on multiple recordings, the speed at which the recording is decompressed may become a factor that interferes with the time efficiency of the process. For simplicity, this project will exclusively store recordings in an uncompressed format.

Ideally, audio recordings of speech could be taken in total isolation, only capturing the desired sound signal. In practice, it is impossible to remove all elements of extraneous sound; thus, there exists some audio-signal-to-audio-noise ratio. The most effective method of mitigating this is through the use of soundproofing to construct a recording chamber, thus minimising external sounds. The walls of such chambers are most commonly lined with an absorptive layer of rubber foam or sponge, which serves as an effective sound barrier. However, due to space or cost restrictions, it is often not possible to construct such an environment. If that is the case, there are also various methods of software noise reduction; these methods can also mitigate the effect of electronic noise on the signal.

## 2.4.2. Recording Human Speech

Now that we have considered broadly some of the challenges in digital audio recording, we will consider some issues which specifically apply to the recording of human speech. When a person is speaking directly into a microphone from a short distance away, we might expect to be able to capture the most effective recording. In such a setup, the microphone should be able to detect every element of speech with minimal effects from sound propagation. Unfortunately, recording in this fashion introduces some problems associated with air flow resulting from speech production.

As previously discussed, most human speech is dependent on the exhalation of air from the lungs through the vocal tract. In listening to human speech from a distance, we hear the mechanical sound wave which propagates through the air. However, when a person speaks directly into the membrane of a microphone, the laminar airflow exiting their mouth or nose pushes against the membrane. This effect is especially noticeable with plosives, as rapid, short-term exhalation is the mechanism of articulation. This exhalation is not a part of the sound wave being generated and is inaudible over longer distances, but as discussed above, a microphone captures sound exclusively through the movement of its membrane. As such, the digital signal received is not representative of the actual nature of the sound wave produced, often clipping the microphone. This effect can be mitigated by placing what is known as a pop filter between the speaker and the microphone.

In pop filter design, we want a material which permits the sound wave to propagate through, but prevents direct airflow into the microphone membrane. This is done by using one or more layers of a tight mesh material (typically either nylon or metal) over a frame to keep the filter in place. This mesh absorbs the energy within laminar airflow; it resists the motion of air through it, drastically reducing the effect of exhalation on our recording. However, as there is still a propagating medium between speaker and microphone, the speech sounds which we want to capture still reach the microphone, while the undesirable effects of aspiration are greatly reduced [17]. This effect can be seen in Figure 12: the sudden exhalation of air causes undesirable microphone clipping or vibration, which does not accurately represent the speech sounds we wish to capture.

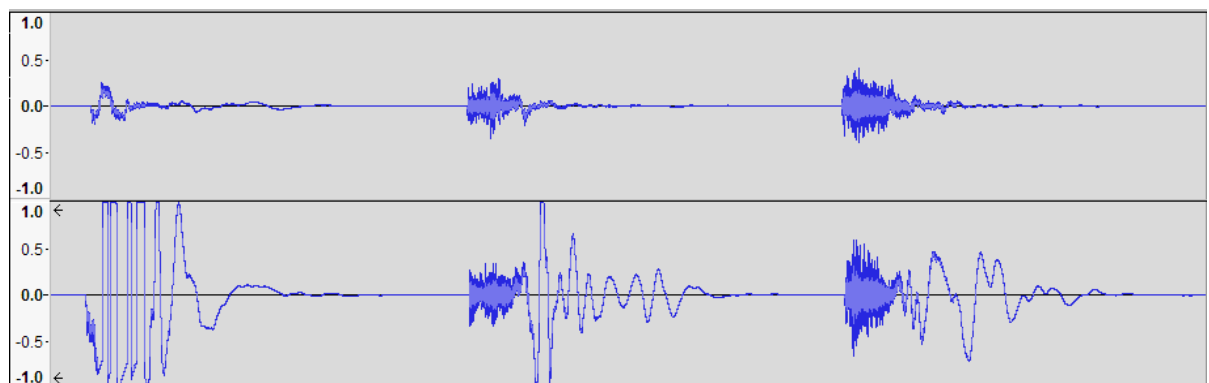

*Figure 12: Wave forms of /p/, /k/, and /t/ captured with a pop filter (above) and without (below)*

The other main problem with digital recordings of speech is the effect of sibilance. As previously defined, sibilants are fricatives which generate turbulence with a high pitch; sibilance is in turn the acoustic characteristic of the sound produced by the sibilant. When listening to a recording of speech, the acoustic bias on many microphones is such that sibilance is undesirably amplified. This can also make sibilants less distinguishable from one another, as they all occupy similar frequency bands at high energy relative to vowels. To counteract this, we can perform any of a range of techniques, collectively referred to as "de-essing" due to "s" being a common sibilant we wish to fix.

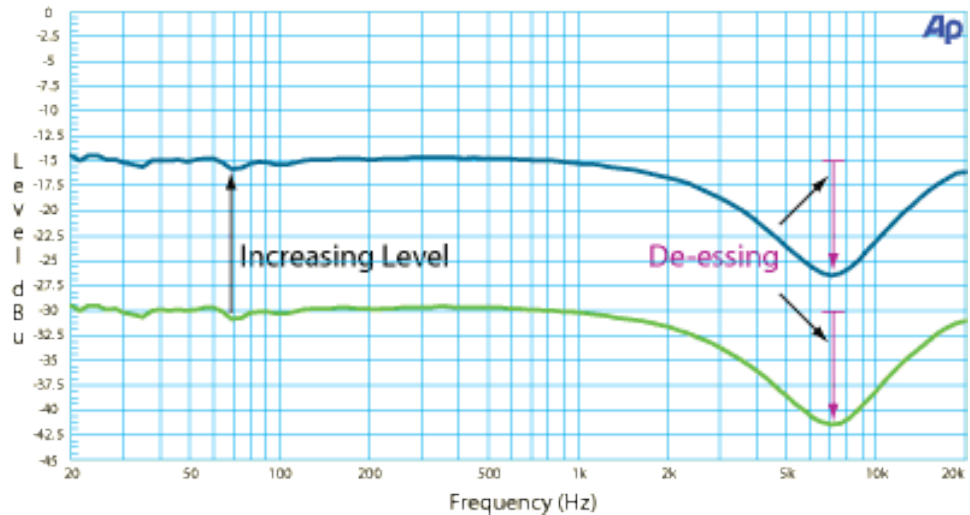
*Figure 13: De-essing Filter [18]*

Most de-essing techniques involve a limited bandstop filter in the frequency region that sibilants occupy, designed only to reduce the volume within those frequencies rather than remove the signal completely [18]. These filters are typically quite advanced, comparing the relative amplitudes of the sibilant and non-sibilant frequencies, and reducing the sibilant frequencies by the appropriate ratio at that time. An example of this style of filter design can be seen in Figure 13.

## 2.4.3. Time Domain Representation

Often, we wish to represent our digital audio data visually, allowing us to observe the details of recorded sound in a static fashion. The obvious way of representing digital signals in this fashion is simply plotting the digitally captured waveform over time. By representing digital signals in such a way, we can make observations about how the volume changes over time by observing the amplitude of the wave over the duration of an utterance, as shown in Figure 14.

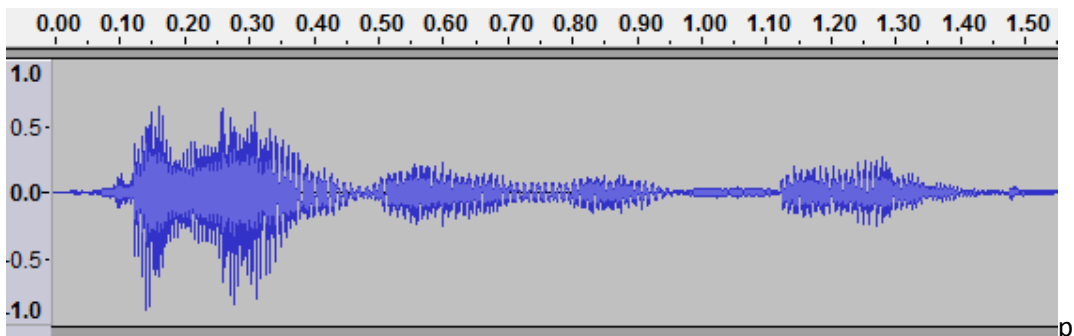
*Figure 14: Time Domain Representation of the phrase "Hello there, my friend".*

If we look at a signal in the time domain on a smaller timescale, we can make certain observations and deductions about how speech was produced, based on our knowledge of acoustic phonetics. For example, based on the regular periodicity and lack of high-frequency noise in the waveform shown in Figure 15, we can conclude that the sound is being pronounced with open aspiration, and from the period of the signal we could determine its fundamental frequency. However, visually inspecting this sound wave in the time domain does not let us easily determine which particular phone is being produced.

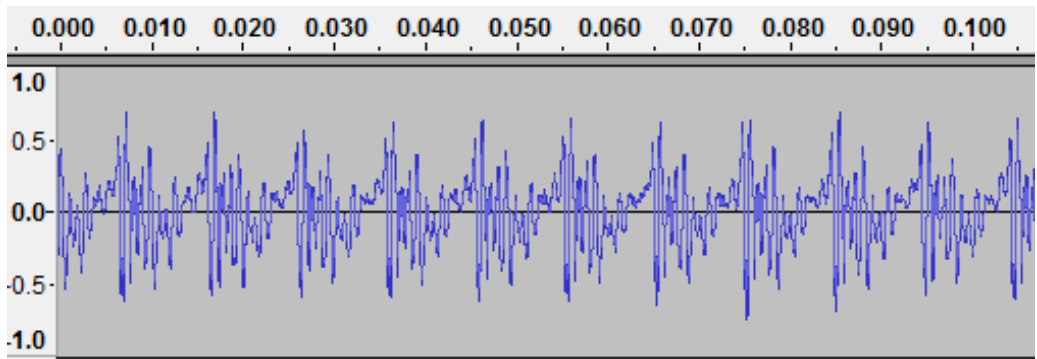
*Figure 15: Time Domain Representation of a sustained /ɑ/ vowel.*

## 2.4.4. Frequency Domain Representation

To determine the frequency characteristics of a particular periodic digital signal, electrical engineers typically use a discrete Fourier transform (DFT), which gives us the spectrum of the signal we transform as a whole. However, in representing a recorded sound signal in the frequency domain, we wish to see how the frequency spectrum of the signal changes over time. When this information is represented visually, it is referred to as a spectrogram.

A spectrogram can be obtained from a digital signal by performing a Short-Time Fourier Transform (STFT) on it. To do this, the time-domain signal is broken up into many overlapping time frames, which are then shaped by a window function. A DFT is then performed on each of these time frames, obtaining the spectral magnitudes and phases which are represented within that particular frame. In representing this visually, we typically discard the phase information, and represent the amplitude at each frequency over time by using a colour plot.

As seen in Figure 16, this helps us distinguish between different phones on a larger timescale far easier than examining the time domain representation. At 0.7 seconds, we can easily see the formant frequencies transition, changing from the /ɑ/ into a /i/ phone. Similarly, at 2.3 seconds, we can see the /s/ become voiced by noting the introduced spectral power in the lower frequency bands, becoming a /z/ phone. Spectrograms are an effective tool for analysing the content of a speech waveform, and can help us to determine specific content of a speech signal, or to determine when phonetic changes occur.

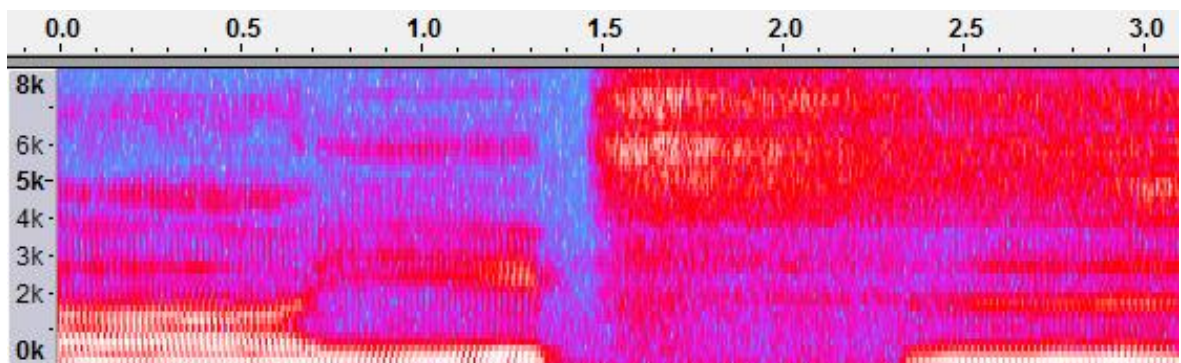
*Figure 16: Spectrogram of the diphone /ai/ followed by /sz/.*

# 3. Formalising Objectives in Speech Synthesis
In any engineering project, it is important to clearly define the main objectives of our research, unambiguously defining an end goal to work towards. This section will discuss how success can be evaluated in speech synthesis, objectives for completion, potential avenues for investigation, and the overall expectations for the project.

## 3.1. Evaluating the Effectiveness of Speech Synthesis Systems
There are two primary qualitative properties which are used to evaluate the effectiveness of synthesized speech, referred to as intelligibility and naturalness. As both are subjective methods of evaluation, their respective meanings are best communicated by example.

A phrase which has **high intelligibility** is one where its words can be easily and accurately identified by a listener. There should be no ambiguity as to which words were being said, and a listener should not need to pay particular attention or focus in order to understand the phrase. It should be clear when one word ends and another word begins.

A phrase with **low intelligibility** is one where words within it can easily be mistaken for other words, or not be recognised as words by a listener at all. The break point between sequential words in a sentence may be ambiguous. If it is possible at all to comprehend what is being said, the listener must pay very close attention to the sound.

Intelligibility is in almost all applications the most important property of a speech synthesis system, as the ultimate objective of the system is to communicate meaningful information quickly and without effort on behalf of the listener. If a human cannot easily and immediately understand the meaning of synthesized speech from a system, then that system has ultimately failed. Thus, the first objective of any speech synthesis system should be to reach a base level of intelligibility.

A phrase which has **high naturalness** is one which is not overly disruptive to the listener, with the system varying the intonation, volume, and speed of its speech to convey emphasis or meaning where suitable [19]. A phrase with high naturalness should flow between different words and sounds in a smooth fashion, and pause where appropriate within a sentence. It should sound as close as possible to the way that a human would pronounce that phrase.

A phrase with **low naturalness** may be monotonic, robotic, or grating to listen to over longer periods of time. It may contain slight auditory abnormalities which, though not altering the intelligibility, disrupt the listener's attention or otherwise annoy them.

Relating these to definite system properties, intelligibility is dependent primarily on segmental attributes of speech, while naturalness is dependent primarily on suprasegmental, prosodic attributes. Depending on the application, we may desire highly intelligible speech but with no particular regard for naturalness. For example, speech which is synthesized at a high speed normally has low naturalness, but can still have a high intelligibility, resulting in faster communication. There is often a necessary tradeoff in design methodologies between maximising the two properties.

In summary, properties of highly intelligible speech include:
- Correct pronunciation of words
- Clear acoustic distinction between similar sounds
- Single words easily identifiable from each other
- Words clearly distinguishable from one another within a sentence
- Speech content highly comprehensible to a human listener

Whereas properties of highly natural speech include [20]:
- Intonation and emphasis suitable to sentence structure
- Volume, pitch, and talking speed varying over time in a human-like fashion
- Appropriate pauses within or between sentences
- Aesthetically pleasing to human listeners

It is also important here to note the distinction between general speech synthesis systems and Text To Speech (TTS) Systems. In general, a speech synthesizer only needs to generate sounds similar to those of real-world speech, with no particular concern for the input method. A speech synthesizer may require as input the phonetic sequence to be pronounced, or to manually input the prosodic information of a sentence. A TTS system requires only written text as input. From the input text, the correct pronunciation of each word and the intonation and prosodic aspects of speech are determined.

The aim of this project is to implement an intelligible TTS system, and then investigate techniques to maximise its naturalness. The structure of this report reflects the progression of the project over time: the first objective is constructing an intelligible speech synthesis system, then improving the robustness of that system, and finally to focus on techniques for maximising speech naturalness.

## 3.2. Testing Methodologies for Intelligibility and Naturalness

In approaching this problem as an engineer, we want to establish objective methods of evaluating the intelligibility and naturalness of a speech synthesis system. As any individual's evaluation of these qualities of speech is inherently subjective, we must use a sample group of listeners and analyse the feedback provided by that group. There have been many attempts to standardise the assessment of intelligibility, each investigating different components of intelligible speech.

However, we should recognise that no matter how standard the test methodology, assessments of the intelligibility of speech can vary wildly from person to person. For example, some people may have trouble understanding speakers with particular speech impediments or different sociolinguistic accents. In assessing the intelligibility of a TTS system, we are reliant not only on the ability of the system to produce speech, but on the familiarity of the sample group with the language being spoken. We want to choose our sample groups such that their own biases are approximately representative of the greater population, so that their evaluations of intelligibility and naturalness are similar to those an average listener would make. The field of assigning definite, quantifiable values to subjective evaluations (such as of intelligibility and naturalness) is referred to as psychometrics, and is a complex field within psychology.

As engineers, we are used to being able to test and refine our approaches in computer simulations or by building models. Speech synthesis offers a unique challenge in that we endeavour to emulate the general quality and aesthetics of human speech, rather than aiming to achieve an easily evaluated and numerically quantifiable goal. We can only test the effectiveness of a TTS system by asking human listeners to evaluate it. Thus, it is important to ensure that the tests we intend to use are effective in determining intelligibility and naturalness.

### 3.2.1. Evaluating Intelligibility

There are many methods which can be used to evaluate the intelligibility of synthesized speech. Some of these tests are not specifically designed for use with speech synthesis systems, but have been proven to be of general applicability and usefulness. For example, intelligibility tests are often used to evaluate the listening ability of the hearing impaired [21], or the effectiveness of communications over different transmission media, such as Voice over Internet Protocol [22].

### 3.2.1.1. Diagnostic Rhyme Test (DRT)

The Diagnostic Rhyme Test, or DRT, is a test which consists of 96 monosyllabic (single-syllable) word pairs which are distinct from each other only by one acoustic feature in the initial consonant [23]. These fall into one of the six categories of Voicing, Nasality, Sustenation (whether the consonant is sustained or interrupted), Sibilation (turbulence in high frequencies), Graveness (concentration of formant energy in low frequencies), and Compactness (concentration of formant energy in a narrow frequency range) [24]. Some illustrative word pairs from the DRT are shown in Table 5; note the similarities and differences between the initial consonants of each word pair. A complete table of words in the DRT is available in the Appendix on Page A-2.

*Table 5: Selected word pairs from the Diagnostic Rhyme Test*

| Voicing | | Nasality | | Sustenation | | Sibilation | | Graveness | | Compactness | |
|---|---|---|---|---|---|---|---|---|---|---|---|
| veal | feel | meat | beat | vee | bee | zee | thee | weed | reed | yield | wield |
| bean | peen | need | deed | sheet | cheat | cheep | keep | peak | teak | key | tea |
| gin | chin | mitt | bit | vill | bill | jilt | gilt | bid | did | hit | fit |
| dint | tint | nip | dip | thick | tick | sing | thing | fin | thin | gill | dill |
| zoo | sue | moot | boot | foo | pooh | juice | goose | moon | noon | coop | poop |

For example, the only phonetic difference between the word "veal" and the word "feel" is that the consonant /v/ is a voiced version of the consonant /f/. To perform the test, one of the two words of each pair is synthesized by a TTS system. These isolated words are played back to a listener, who records which of the two possible words they believed it to be. The test is then assessed according to how many pairs the listener accurately identified; these are then averaged over our sample group to determine an overall score. In analysing results from a larger sample group, we can also make more detailed observations, such as noting which word pairs were less frequently identified correctly. This can help determine particular problems within the system, which can be helpful in targeting areas of the system which should be improved. As each word pair varies by only a single linguistic feature, it can be difficult for a listener to distinguish between these words in a poorly designed synthesis system. This makes the DRT an effective evaluator of system intelligibility.

There are two notable, similar tests derived from the DRT called the Diagnostic Medial Consonant Test (DMCT) and the Diagnostic Alliteration Test (DAT), both of which are composed of 96 word pairs falling into the same 6 categories as the DRT and performed in an identical fashion [25]. Rather than word pairs varying in the initial consonant of a monosyllabic word, DMCT word pairs vary in the central consonant of a disyllabic, or two-syllable word; for example, "stopper" and "stocker" only vary in their central consonant. The DAT word pairs are monosyllabic words that vary only in the terminating consonant, such as between "pack" and "pat". Depending on how the underlying synthesis system is operating, administering the DMCT and DAT can provide different results for the same system than the DRT, which is again useful for targeting areas for improvement.

### 3.2.1.2. Modified Rhyme Test (MRT)

The Modified Rhyme Test, or MRT, is a slightly different test to the DRT. In the MRT, the listener is given more than two choices, which can help us to more quickly identify potential issues in our TTS system [26]. The MRT contains 50 distinct sets of 6 monosyllabic words. Of them, 25 sets vary only in the first consonant, while the other 25 sets vary only in their final consonant. Example word sets from the full MRT list are shown in Table 6; note that the first three sets contain words which vary only in the initial consonant, while the final three sets vary only in the terminating consonant. A complete table of all word sets in the MRT is available in the Appendix on Page A-3.

*Table 6: Selected word sets from the Modified Rhyme Test*

| shop | mop  | cop  | top  | hop  | pop  |
|------|------|------|------|------|------|
| coil | oil  | soil | toil | boil | foil |
| same | name | game | tame | came | fame |
| fit  | fib  | fizz | fill | fig  | fin  |
| pale | pace | page | pane | pay  | pave |
| cane | case | cape | cake | came | cave |

To perform the test, a random word from each word set is produced by the synthesis system, and a listener must select which word from the set they believe was spoken. Unlike the DRT, the MRT can be performed with a carrier sentence, where the word is inserted into a prompting sentence such as "Please circle the word _____ now". Carrier sentences can help to prompt listeners before the specific word is spoken, letting them shift their focus to listening to the word. By providing more than two choices, the feedback from the MRT can identify if a consonant is similar to any other in the same set. For each question that is answered correctly, we are ensuring that the pronunciation was distinct from five other pronunciations rather than just one. If a question is answered incorrectly, then we can easily identify the two problem consonants that sound unacceptably similar.

The MRT therefore lets us test over a similar range of consonant characteristics as the previously discussed tests, but with fewer total questions (50 word sets compared to 96 word pairs), and with the benefit that it examines consonants at both the start and end of a syllable. This lets us test our system for a broad variety of problems over a shorter time, at the cost of some specific detail. While an MRT is an effective diagnostic for assessing general intelligibility, the DRT family of tests can provide more detailed and useful information on problems within the system, at the cost that they take longer per listener to assess. Therefore, in development of a TTS speech synthesis system, the MRT can be more useful for rapid iterative evaluation, while the DRT is more useful in identifying specific weaknesses of the system in greater detail.

### 3.2.1.3. Phonetically Balanced Monosyllabic Word Lists (PAL PB-50)

While the DRT and MRT allow us to examine how well a listener can distinguish between multiple similar sounding words, they are fundamentally limited in being multiple choice questions rather than open-ended. These tests assume that the listener is always capable of identifying the synthesized word as being one of the options available, with only slight ambiguity. This has the advantage that our results can be easily analysed due to binary results of the word being identified correctly or not. However, we are biasing the responses that listeners provide by prompting them with a set of responses. Another test must be used in order to evaluate unprompted effectiveness of the system. In such a test, rather than having to distinguish between different words, the listener should attempt to correctly identify a synthesized word with no prompted information biasing their answer. This is best performed with a standardised word list which lets us compare results.

The Phonetically Balanced Monosyllabic Word lists, abbreviated as the PB-50, are a collection of 20 distinct word lists, each of which is composed of 50 monosyllabic words [27]. Each word list is phonetically balanced to the English language, meaning that each group contains different phonemes occurring at a similar ratio to their occurrence in normal English speech. In performing a PB-50 test, one of the word lists is chosen, and each word within that group is spoken in a random order as a part of a carrier sentence, typically of the form "Please write down the word _____ now". Accurate transcriptions are marked as correct, giving a final mark out of 50. Detailed examination of PB-50 transcriptions can indicate which particular phonemes are being systemically misheard or mispronounced, which can later be analysed in more detail using the DRT. Table 7 shows a typical list within the PB-50; the full word list is available in the Appendix on Page A-4.

*Table 7: List 2 of the PB-50*

| ache  | crime | hurl  | please | take   |
|-------|-------|-------|--------|--------|
| air   | deck  | jam   | pulse  | thrash |
| bald  | dig   | law   | rate   | toil   |
| barb  | dill  | leave | rouse  | trip   |
| bead  | drop  | lush  | shout  | turf   |
| cape  | fame  | muck  | sit    | vow    |
| cast  | far   | neck  | size   | wedge  |
| check | fig   | nest  | sob    | wharf  |
| class | flush | oak   | sped   | who    |
| crave | gnaw  | path  | stag   | why    |

In areas of speech analysis other than synthesis, similar phonetically balanced word lists are used for various purposes, though they are typically designed for more specialised purposes. For example, the Central Institute for the Deaf (CID) W-22 word list is often used to evaluate the degree of hearing deterioration in a listener. Studies indicate that most such lists are comparably useful for evaluating different aspects of speech comprehension, even with slightly different phonetic balances [28]. This is similarly true for applications in speech synthesis, where our aim with this type of list is typically to rapidly evaluate general effectiveness of the synthesis method.

The American National Standards Institute (ANSI) specifies three word lists for evaluating the intelligibility of various aspects of speech communication [29]. These lists are the PB-50 list and the previously discussed DRT and MRT. As such, these three tests are often used in conjunction with one another for intelligibility evaluations.

### 3.2.1.4. SAM Standard Segmental Test and Cluster Identification Test (CLID)

The SAM standard segmental test, like the DRT and MRT, attempts to examine intelligibility of a synthesizer on a level smaller than words, but like the PB-50, the listener's range of responses is open ended. The SAM test is performed by synthesizing a set of mostly semantically meaningless phonetic phrases, of either the form CV, VC, or VCV, where C is a consonant phone and V is a vowel phone [30]. This set should contain all valid vowel/consonant adjacencies in the phonetic range of the language for each consonant tested. A transition from one phone to another, as previously discussed, is called a diphone. A SAM test therefore examines the comprehensibility of all possible vowel/consonant diphones within a language.

The synthesized phrases are played back to a listener, who is asked to transcribe the phrase in any fashion they feel comfortable. The typical prompt is of the form: "Write down what you hear in such a way that, if someone reads it back to you again, they would convert your notations into spoken language in such a way that you would identify the same sequence of sounds." A percentage mark is then given depending on how many consonants were correctly identified in whatever system the listener chose to use, as consonants are typically more difficult to identify than vowels.

The SAM test also has the advantage that, unlike the previous tests, it is independent of language; it can be performed on synthesizers for a subset of phoneme transitions which intersects between the two different languages that they use. This lets us compare the effectiveness of synthesizers of different languages, and evaluate their relative strengths and weaknesses. Another advantage of the test is that, by using nonsensical phone segments, listeners do not anticipate the pronunciation of an existing word within their vocabulary. Because of this, correct identifications are based purely on the sound produced by the synthesizer, removing the influence of listener's biases.

The major downside of the SAM is that it is only concerned with assessing the intelligibility of vowel/consonant diphones. It is therefore not especially useful for assessing vowel/vowel or consonant/consonant phone transitions. We could easily extend the testing methodology to examine all instances of both of these cases, but in practice most vowel/vowel diphones are trivially comprehensible, making exhaustive testing of them a waste of testing time. As such, we wish to examine varying consonant/consonant transitions.

The Cluster Identification test (CLID) resolves this problem. All synthesized sections are of the form CVC, where V is a vowel, but here, C can be any cluster of sequential consonants that occurs within the language. The CLID test is otherwise performed identically to the SAM test. [31]

While these tests are the most broadly useful of the tests so far for evaluating total a TTS system's low-level phonetic accuracy, it takes substantially longer to perform them: a streamlined SAM test takes approximately 15 minutes, or 30 minutes for a more exhaustive one; for a complete CLID trial, times are typically in the region of 2 hours.

This level of time investment is wasteful if the results simply indicate that the system is still mostly unintelligible, since we could establish this such using a less intensive test. Thus, the SAM and CLID tests are most useful when a reasonable level of intelligibility has already been achieved, in order to examine the full gamut of possible low-level utterances within a language. Of course, if a single synthesizer is intended to produce speech in a wide variety of languages, then phonetically balanced word lists will be insufficient to determine general low-level system intelligibility. These are the scenarios where the SAM and CLID test are the most useful – but as we aim to only produce an English-language speech synthesizer, they are less useful for our objectives.

### 3.2.1.5. Harvard Psychoacoustic Sentences

The evaluation techniques which have been discussed are effective at determining how well a human can identify words or low-level utterances produced by a speech synthesis system. From an engineer's perspective as a designer of such systems, we are considering the inverse: that these tests are effective evaluators of the system's ability to produce intelligible words. These tests are therefore useful evaluators of the segmental intelligibility of speech.

However, these tests only target on the level of word-by-word comprehensibility, or even smaller speech sound combinations. While suprasegmental aspects of speech are mostly correlated with the naturalness of a system, intelligibility is also a function of larger-scale properties to some degree. For example, it is important that listeners are able to distinguish between the start and end of each word within a sentence, which we cannot verify with the previously discussed testing methods.

To evaluate intelligibility on this larger scale, tests need to include synthesis of entire sentences rather than individual words. The most widely used test sentence banks are the Harvard Psychoacoustic Sentences, the Haskins Sentences, and the Semantically Unpredictable Sentences [32]. While all of these sentences are different sets of data, the way that these tests are administered is almost identical, with only minor changes to the instructions to listeners.

First, a subset of the larger sentence dataset is chosen through some random process, and the sentences within that subset are synthesized by the system. These sentences are played back to a listener, who transcribes the words in the sentence as best as they can. There are specific keywords within each sentence; these do not include connective words such as "a" or "the", which can easily be identified. A percentage of correctly transcribed keywords is then calculated as the score. Clearly, some level of segmental intelligibility is prerequisite for these tests, since if singular words cannot be correctly transcribed, it is unlikely entire sentences will.

The Harvard Psychoacoustic Sentences are a set of 72 numbered lists of 10 sentences. These sentences are designed to be both syntactically and semantically normal [33]. As previously stated, syntax refers to the structure of sentences; a syntactically normal sentence obeys the rules of English language sentence composition. Semantic normality refers to if meaning is being communicated by the sentence; as Harvard sentences communicate meaningful information to a listener, they are considered to be semantically normal. List 30 is shown in Table 8; the full set of Harvard Psychoacoustic Sentences is available in the Appendix on Page A-6.

*Table 8: List 30 of 72 in the Harvard Psychoacoustics Sentences*

**List 30**
1. The mute muffled the high tones of the horn.
2. The gold ring fits only a pierced ear.
3. The old pan was covered with hard fudge.
4. Watch the log float in the wide river.
5. The node on the stalk of wheat grew daily.
6. The heap of fallen leaves was set on fire.
7. Write fast, if you want to finish early.
8. His shirt was clean but one button was gone.
9. The barrel of beer was a brew of malt and hops.
10. Tin cans are absent from store shelves.

As the Harvard sentences are syntactically and semantically normal, they accurately represent both the structure and meaning of typical English speech. The sentences are also phonetically balanced, making each sentence list a reasonably robust sample of common English diphones and speech elements. As someone listening to the Harvard sentences is parsing them as entire, meaningful sentences, there are further cues within the sentence as to which word is being said.

This test can provide a closer representation of real-world synthesis intelligibility than low-level segmental tests such as the MRT and DRT. This is because even though those tests compare words which sound similar enough to be mistaken for each other in isolation, those words are often members of different lexical classes (groupings such as nouns, verbs, adjectives, and so on). For example, the words in the DRT pair "sing/thing" are less likely to be mistaken for one another when used in a syntactically correct sentence, as one is a verb while the other is a noun. In the two Harvard sentences "Let's all join as we sing the last chorus" and "The sink is the thing in which we pile dishes", a sing/thing substitution would make the sentences not only syntactically incorrect but semantically nonsensical.

The Harvard sentences are the recommended sentence set for evaluation of speech quality by the IEEE. They are therefore widely used for standardised testing of telephone systems, as well as test inputs for speech recognition systems [33].

### 3.2.1.6. Haskins Syntactic Sentences
The Haskins syntactic sentences are a collection of 4 series of sentences, each series containing 50 sentences in total. While the Harvard sentences are designed to be both syntactically and semantically normal, the Haskins sentences are designed to be syntactically normal but semantically anomalous. In other words, while the syntax obeys the rules of English language sentence composition, the semantic, actual meaning of each sentence as a whole is unclear, unusual, ambiguous, or non-existent [34]. This can be seen in the example sentences shown in Table 9; the complete list is in the Appendix on Page A-14.

*Table 9: The first 10 of the 50 sentences in Series 1 of the Haskins Syntactic Sentences*

**Series 1**
1. The wrong shot led the farm.
2. The black top ran the spring.
3. The great car met the milk.
4. The old corn cost the blood.
5. The short arm sent the cow.
6. The low walk read the hat.
7. The rich paint said the land.
8. The big bank felt the bag.
9. The sick seat grew the chain.
10. The salt dog caused the shoe.

When used for assessing a speech synthesis system, evaluators find that the Haskins sentences are more difficult to correctly transcribe than the Harvard sentences. This is because, as the sentences are semantically unusual and mostly meaningless, information on the content of a word cannot be deduced by other elements of sentence structure. As such, the listener is relying more on the acoustic properties of the sentence than with the Harvard sentences. This leads to greater variation in scores when comparing the performance of multiple synthesizers, but usually retains the relative ranking of each method to each other. This effect can be seen clearly from the results shown in Table 10 [14]. Due to this higher variation in results, the Haskins sentences can help us to better discriminate between the intelligibility of systems with close scores using the Harvard sentences.

*Table 10: Comparison of Harvard and Haskins sentences for human speech and speech synthesizers*

| Speech type | Harvard (meaningful) | Haskins (unpredictable) |
|---|---|---|
| Human | 99 | 98 |
| DEC Paul v1.8 | 95 | 87 |
| MITalk-79 | 93 | 79 |
| DEC Betty v1.8 | 90 | 75 |
| Prose-2000 prototype | 84 | 64 |

One downside of the Haskins sentences is that learning effects of the listener may undesirably alter the results. Each sentence within the list is constructed in the form "The (adjective) (noun) (past tense verb) the (noun)". As such, it is possible for a listener to learn which lexical class a word is likely to belong to, based on its position within the sentence, and use this to more accurately guess the correct word – introducing unwanted listener bias [35].

In practice, the Haskins sentences are often used in conjunction with the Harvard sentences, as they are standard and unchanging datasets (making them easy to implement or record data from in a consistent manner) which assess slightly different properties of synthesized speech. If used together, it is possible to very quickly evaluate the general sentence level intelligibility of a speech synthesis system in a way that lets us compare performance reliably to other synthesizers.

### 3.2.1.7. Semantically Unpredictable Sentences (SUS Test)

Unlike the previous two sentence-level intelligibility sets, the Semantically Unpredictable Sentences are not a specific set of input sentences, but instead constitute a specific methodology to generate different types of sentences from databases of valid words. These databases require a selection of nouns, verbs, adjectives, relative pronouns, prepositions, conjunctions, question-words and determiners within a language, as well as their relative frequencies in regular usage [36].

Sentences generated by the SUS methodology are categorised into one of five different syntactic structures, which are similarly valid in most languages. In the original paper outlining the SUS testing technique, examples are given in English, French, German, Italian, Swedish, and Dutch. The SUS generation methodology therefore remains useful regardless of the language being used; the same particular implementation can be used to assess multiple languages, requiring only new word databases and marginal changes in syntactic structure. An example of one of each of the five sentence structures are shown in Table 11.

*Table 11: Example Sentences of Each Syntactic Structure in the SUS Methodology*

**Sentences of Each Structure Type in the SUS Test**
1. The table walked through the blue truth.
2. The strong day drank the way.
3. Draw the house and the fact.
4. How does the day love the bright word?
5. The plane closed the fish that lived.

To create a set of sentences to use for testing, we generate an equal number of sentences using each grammatical structure, with the probability of any particular word appearing in a sentence being proportional to its frequency within the language. This ensures that on a meta-sentence scale, the dataset being generated is an average sample of words from the chosen language. If our databases are sufficiently broad, this means that each SUS dataset will on average be phonetically balanced. Most sentences generated in this fashion will be semantically meaningless, but it is possible for a randomly generated sentence to have some degree of meaning. If desired for the test, semantically meaningful sentences can be removed manually from the dataset.

Notably, the different sentences generated with an SUS methodology are not all syntactically structured as factual statements, as was the case with both the Harvard and Haskins sentences. Specifically, Structure 3 produces imperative sentences, or ones where an instruction is given, and Structure 4 produces interrogative sentences, where a query is asked of the listener. These sentences, if spoken by a human, would have different prosodic elements associated with them (though, for the most part, they continue to be semantically meaningless). As such, generated SUS datasets can find further potential use in evaluating naturalness.

The primary advantage of the SUS test is that, as sections of the standard structures are easily interchangeable depending on the scale of the lexicon used, there are orders of magnitude more possible sentences to use. However, this breadth is at the cost of both consistencies of tests as administered by different research groups, as well as specialty of focus. Both the Harvard and Haskins sentences are specifically designed and curated datasets, having specific properties, whereas SUS test data is typically randomised. [14]

The SUS can perform an exhaustive test of the intelligibility of certain words within different sentence structures, in a sentence-level parallel to the CLID test used for segmental intelligibility. Indeed, the two tests share most of their advantages and disadvantages: the feedback from these tests can provide both broader and more detailed examinations of the performance of a synthesizer, and both can be used for any language. This exhaustiveness comes at a cost of efficiency; like the CLID test, a SUS test designed for exhaustive examination of a word corpus can take multiple hours to complete. Nevertheless, such tests can be performed by larger companies and research groups with greater monetary and financial resources, and the resulting data is almost guaranteed to help find areas of possible improvement in the synthesis system.

## 3.2.2. Evaluating Naturalness

All of the previously discussed testing methods can help us to evaluate whether synthesized speech is intelligible. To evaluate the net effectiveness of a TTS system, we should use a combination of the above tests to ensure intelligibility of the system both on the scale of words and on the scale of sentences. Of course, the overall effectiveness of a synthesis system is not solely based on intelligibility; as such, we must also use other tests to evaluate a system's naturalness.

Unfortunately, standardised tests for the naturalness of speech are difficult to set. Intelligibility, conveniently, has a mostly binary success or failure condition in whether or not a listener can understand speech produced by the system. There is no similar binary for naturalness. We are aware of elements of speech which correspond strongly with speech naturalness [37], but there exist no absolute assessments of those elements. Instead, assessments of naturalness must always be relative [38].

### 3.2.2.1. Mean Opinion Score (MOS) and derivatives

The MOS is a standard test to determine the quality of electronic speech. This test was designed for evaluating the quality of telephony networks, and is recommended by the International Telecommunications Union (ITU) for evaluating TTS systems [39]. The MOS test is administered by playing back a sample of speech generated by a TTS synthesis system. This speech is then assessed by a listener in various categories on a scale from 1 to 5, typically where a 1 is the least desirable to a 5 as the most desirable [40]. The specific categories and questions can vary, but typically fall into some subset of the categories indicated in Table 12; they encapsulate properties of both intelligibility and naturalness. These are then averaged for an overall score. The text being synthesized should be multiple sentences long, so the listener has an accurate understanding of how the system sounds over longer periods of time. Similarly, the text should be phonetically balanced.

*Table 12: Typical questions, categorisations, and response scales in an MOS test [41]*

| Question (Category) | Scale from 1 to 5 |
|---|---|
| **How do you rate the sound quality of the voice you have heard? (Overall Impression)** | Bad to Excellent |
| **How do you rate the degree of effort you had to make to understand the message? (Listening Effort)** | No Comprehension Even With Effort to Complete Relaxation |
| **Did you find single words hard to understand? (Comprehension)** | All The Time to Never |
| **Did you distinguish the speech sounds clearly? (Speech-sound Articulation)** | No, Not At All to Yes, Very Clearly |
| **Did you notice any anomalies in the naturalness of sentence pronunciation? (Pronunciation)** | Yes, Very Annoying to No |
| **Did you find the speed of the delivery of the message appropriate? (Speaking Rate)** | Faster Than Preferred to Slower Than Preferred (3 being Optimal Speed) |
| **Did you find the voice you heard pleasant? (Voice Pleasantness)** | Very Unpleasant to Very Pleasant |

One of the main concerns with the MOS test is that it can be difficult to compare results from tests carried out at different times and under different conditions, especially as specific categories and phrasing of questions for those categories can vary [42]. There are several approaches to resolve this. One solution is for natural speech to be included in MOS tests, which should receive a score of a 5 in all areas. While this can serve as a reasonable reference point for the scale, the evaluation is still inherently subjective. Ideally, if we wish to use the MOS to compare several different systems, the same sample group should perform MOS tests on all speech to be compared under identical conditions. This lets us consider the average and variance of each individual's grading criteria, which we can analyse to extract a more standardised evaluation.

Another concern with the MOS test is that listeners, to some extent, are not meaningfully evaluating the naturalness of the system, but simply how much they like listening to the voice [41]. As people will inherently have aesthetic preferences for voice, it is possible that individuals may not prefer a voice due to personal bias. While this should only change a categorical rating of Voice Pleasantness, there is often a bleed effect between different categories. For example, people are more likely to pay attention to and understand a pleasant voice, thus somewhat correlating Listening Effort and Voice Pleasantness.

This can be mitigated through the use of a larger sample size, where individual biases of listeners should average out to be representative of opinions of the overall population. If this is the case, then while these biases cause some inaccuracies in the evaluation of naturalness, they are still useful in determining the general opinion of voice character.

The raw feedback from a MOS test performed with a sample group can be analysed using different statistical analysis techniques. As the assessment is subjective, finding the average difference between the scores individuals provided can be a more useful metric than simply averaging the raw scores. We must use our knowledge of statistics to consider these datasets appropriately.

One major problem with the MOS test is that the results can easily exhibit what are known as ceiling and floor effects. These are when due to the grading method of the test the distinction between different scores may lose detail, particularly when scores are close to the maximum or minimum value on the scale. For example, if a single reference evaluation is scored highly, and then most other references score low on the scale, the closeness of those lower results can make meaningful distinction between them difficult.

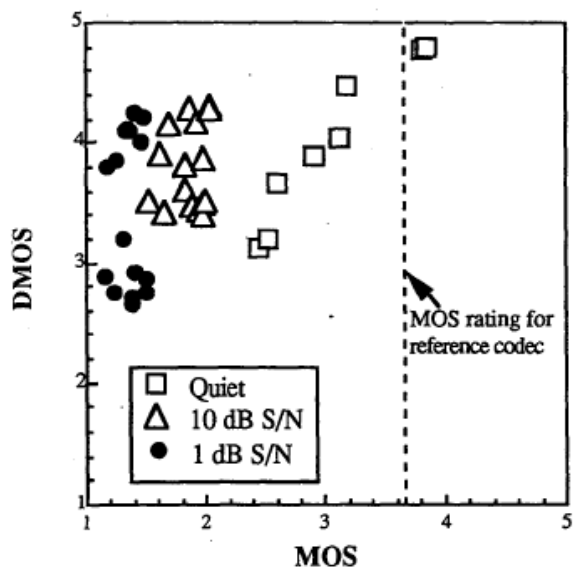

*Figure 17: Comparison of MOS and DMOS on speech for varying signal to noise ratios [43]*

The Degradation Mean Opinion Score (DMOS) attempts to resolve this problem. Rather than evaluators listening to the direct output of the TTS system, the sound signals to be tested are degraded in some fashion. This may be through the introduction of random noise to the signal or by passing it through a filter of some kind. The DMOS system was designed for evaluating the relative effectiveness of different lossy speech codecs (where the degradation of the codec is the main characteristic to be evaluated), but can easily be used for naturalness tests [44].

Relative to the MOS, the DMOS has a far greater variance between different scores, as can be seen in Figure 17. When both the MOS and the DMOS are performed without any ceiling or floor effects, they tend to preserve the ordering of different synthesizers. Thus, if the ceiling effect arises in the MOS evaluation of speech, and it is difficult to meaningfully assess which of two systems has a greater naturalness, we can use a DMOS test to skew the evaluation into a region where the difference between the two can be assessed more clearly.

While the DMOS gives us greater variance in our numerical scores, it gives us no greater depth of insight into the various aspects of speech being assessed. The development of the Revised MOS (MOS-R) and Expanded MOS (MOS-X) had the objective of improving the breadth of data collected, while also improving the reliability, validity, and sensitivity of the pre-existing MOS design [45]. The MOS-R is performed with a 7 point scale rather than the 5 point scale used in the MOS test, giving the marker a greater region over which to evaluate quality. It also introduces additional questions regarding speech comprehension and naturalness.

The MOS-X test is an even more developed and standardised version of the MOS. It is composed of 15 specific questions which are designed to separately assess distinct aspects of intelligibility, naturalness, prosody, and the social impression of the synthesized voice. Each question is evaluated on a scale from 1 to 7. These questions are shown below in Table 13; a sample MOS-X Test form is on Page A-16.

*Table 13: The 15 Questions in the MOS-X Test*

| | |
|---|---|
| 1. | **Listening Effort:** Please rate the degree of effort you had to make to understand the message. |
| 2. | **Comprehension Problems:** Were single words hard to understand? |
| 3. | **Speech Sound Articulation:** Were the speech sounds clearly distinguishable? |
| 4. | **Precision:** Was the articulation of speech sounds precise? |
| 5. | **Voice Pleasantness:** Was the voice you heard pleasant to listen to? |
| 6. | **Voice Naturalness:** Did the voice sound natural? |
| 7. | **Humanlike Voice:** To what extent did this voice sound like a human? |
| 8. | **Voice Quality:** Did the voice sound harsh, raspy, or strained? |
| 9. | **Emphasis:** Did emphasis of important words occur? |
| 10. | **Rhythm:** Did the rhythm of the speech sound natural? |
| 11. | **Intonation:** Did the intonation pattern of sentences sound smooth and natural? |
| 12. | **Trust:** Did the voice appear to be trustworthy? |
| 13. | **Confidence:** Did the voice suggest a confident speaker? |
| 14. | **Enthusiasm:** Did the voice seem to be enthusiastic? |
| 15. | **Persuasiveness:** Was the voice persuasive? |

Within the MOS-X, we can average different groupings of questions to determine different useful metrics. By averaging items 1 to 4, we get a measure of intelligibility; averaging 5 to 8 gives us naturalness; averaging 9 to 11 gives a judgement of prosody; averaging 12 to 15 gives an evaluation of social impression; and of course, averaging all questions gives an overall score. The standardisation, expansion, and improvement of the MOS-X test relative to the more general class of MOS tests makes it more useful for our objectives; where the general MOS is a grading system which is simply borrowed from telephony networks, the MOS-X is a more specialised tool for evaluating TTS systems.

## 3.2.2.2. Preference Tests

Preference testing, also known as discrimination testing, is perhaps the most simplistic approach to evaluating naturalness. Rather than assigning any numerical score to a particular synthesized voice (as with the MOS test), the listener merely has to choose a preferred choice from two options.

There are two primary methodologies for preference testing. Those falling into the first category are referred to as a pairwise tests, also known as AB or XY tests. To perform a pairwise test, the same sample of text is synthesized by two different synthesizers, A and B. The listener is then asked one or more questions: these can be of any form, such as "which sounded more natural", or more vaguely, "which voice did you prefer". They can then choose from either A or B; we can also choose to include a "no preference" option.

In ABX tests, three different recordings are used: like before, the two synthesised speech samples to be given are A and B, but here a reference X is used to evaluate. This reference can be another synthesized voice, but most often it is a sample of real-world speech. Assessors must determine which of A or B (or neither) is closer to X in some characteristic, typically naturalness or intonation.

While these are very simple tests, they can be powerful in determining orderings of multiple synthesizers over a short time span. ABX tests can be used very effectively to determine which synthesizer was more effective at emulating a real human voice. However, unlike MOS-style tests, we do not get a magnitude of difference between A and B, which potentially makes it more difficult to draw useful conclusions from our results.

## 3.3. Planned Testing Method for this Project

Having examined various testing techniques for intelligibility and naturalness, we want to determine which tests should be used to assess the effectiveness of our system in synthesizing speech. In choosing the tests to use for this project, it is important to consider what each methodology requires of the listeners performing these tasks. Since we need to prioritise the development of our system, we will have only a short time span allotted for testing, making the more exhaustive testing methods undesirable. We also wish to minimise the demands on the listener, so that a larger sample size can be gathered from a general group, providing more useful information [46].

In choosing sample groups, it is important to avoid any sampling biases which might alter the outcome. One group of subjects who should be actively avoided are people with knowledge of linguistics or speech sciences. By having a better understanding of speech vocalisation and categorisation, and likely having been previously exposed to anomalous speech patterns and trained to understand them, such people typically score substantially higher on tests for intelligibility [36]. As the objective of some of these tests is to evaluate intelligibility for the average listener, removing such people from our pool of listeners should provide more useful results.

Using the DRT, PB-50, and Harvard sentences, we can assess intelligibility on the segmental level, word level, and sentence level within a reasonable timeframe. As each dataset is fixed and in common use, we can also compare our results to other synthesizers based on how they performed in those tests. For naturalness, the MOS-X would be the most useful metric for conclusive general evaluation, while AB preference testing can determine if a particular change to the system is a measurable improvement relative to an older version, or one with different prosodic attributes. Therefore, we will use these tests to evaluate the effectiveness of our TTS system.

Now that we have determined how we will evaluate our system, we will turn our attention to the different methods and techniques of synthesis which are in common use, and determine which of them is most suitable for this project.

# 4. Review of Speech Synthesis Techniques

There are two primary approaches in speech synthesis, which can be further subdivided into various specific techniques. Systems can either produce a speech waveform which is constructed by using a database of real-world speech recordings, or generate a waveform by using a software model of either the underlying mechanisms of speech production or the natural speech's acoustic properties.

## 4.1. Sample-Based Synthesis

Sample based synthesis is perhaps the most intuitive approach to speech synthesis. If we desire to maximise the naturalness of speech, we might consider actual human speech to be maximally natural, as it is a perfect representation of human speech patterns. As such, using recorded samples from the real world should result in a reasonable approximation of human naturalness.

### 4.1.1. Limited Domain Synthesis

Limited domain synthesis is one of the easiest types of synthesis to implement, but is (as might be expected from the name) only useful for specific scenarios. A small number of specific phrases are recorded for use in the system, and then a combination of the recordings is played back in sequence to communicate information [47]. Limited domain synthesis is commonly used for purposes such as automated attendants or announcement systems, where all messages will conform to the same syntactic format with minimal change. As such, specific recordings can be made for each type of announcement, and the variable sections replaced depending on requirements.

For example, a train announcement system can be made by recording the general phrase, "The train to XXX is arriving on platform YY in ZZ minutes". To replace XXX, we record the name of every location that trains might go to from this station; to replace YY, we record every number for as many stations as we have, and to replace ZZ, we again simply record a grouping of numbers. By concatenating these different parts of the phrase together, we can generate our synthesized waveform.

Limited domain synthesis is very easy to implement and record new samples for, since we only need a few recordings from the speaker. Due to the small database size and negligible processing power required, it is possible to implement limited domain synthesis on almost any platform. This technique can generate speech with exceptionally high naturalness, as the speaker can record a statement with exactly human pronunciation. As we can rapidly create the limited database for new speakers, this approach lends itself well to certain types of research. However, beyond very specific scenarios, this approach is not helpful: we cannot generate a waveform given any arbitrary input speech pattern, making this approach useless for TTS synthesis.

### 4.1.2. Unit Selection Synthesis

Unit selection synthesis is effectively an expansion of limited domain synthesis to a more general domain. A unit selection synthesis database may include many recordings of common words or phrases – the "units" – in their entirety. This can still allow for the capturing of suprasegmental elements of speech, which can retain naturalness of the original speech in the synthesized voice.

The main downside of this approach is that it requires a very large database of recordings, which can take an exceptionally long time to record. A database for unit selection synthesis might contain hours of human recorded speech and, depending on the sampling frequency and encoding of the recordings, multiple gigabytes of data. In addition, analysis to determine the best sequence of samples to use can often be complex. For these reasons, unit selection synthesis is not suitable for embedded systems applications, but the approach can be quite easily implemented on modern computers.

To be able to pronounce speech for arbitrary word input into a system, we also need this system to be able to produce words which do not exist within our recording database. A first consideration of this more general sample-based synthesis might be to simply record every phone within a language. If we used this in conjunction with a general set of rules for finding the phonemes corresponding to the graphemes within a language, then we could simply play recordings of the phones in order to generate speech. This lets us generate a waveform for an arbitrary sequence of words.

While on the surface this seems reasonable, the result typically has very low intelligibility and naturalness. This is because some phones, when generated persistently by a human being, are acoustically identical. For example, a sustained /m/ sound and a sustained /n/ sound are effectively generated in exactly the same way: as both are nasal consonants, when sustained, air only escapes through the nasal cavity. Therefore the position of the lips (closed for /m/, open for /n/) does not alter the sustained sound, since the configuration of the mouth does not alter the nasal resonance. Instead, the distinction between the two is in how they alter the transition between phones. The two sounds /na/ and a /ma/ are distinct if pronounced as a transition, but sound identical if each phone is produced individually in sequence.

A unit selection synthesis system therefore includes many different phonetic transitions; we introduce units smaller than phrases or words, by recording groupings of multiple syllables, individual syllables, diphones, and individual phones. Often these recordings will be highly redundant, with the same unit being recorded multiple times with varying intonation, speed, and pitch. We then concatenate these units together to produce the most natural speech possible: a weighted decision tree determines which unit chain is optimally natural.

The appropriate weighting of this decision tree can be difficult. Clearly, the recordings of larger units will provide greater naturalness, so they should be favoured over the same utterance composed of smaller units. We also wish for the units concatenated to have a continuously changing pitch rather than an erratic one; thus, in moving from one unit to another, we favour units which are close in pitch to the previous unit. Typically the weighting assigned to each aspect is manually tuned, according to the results of naturalness tests.

Unit selection is the most powerful sample-based speech synthesis approach. However, the prohibitive time and effort to create a unit selection database means that there are no freely available databases, and it is impractical for even a small team to generate one in a reasonable timeframe. While unit selection synthesis can produce the most natural speech of any sample-based synthesis technique, it is correspondingly the most time-consuming to implement.

## 4.1.3. Diphone Synthesis

Diphone synthesis is functionally a reduced form of unit selection synthesis. Rather than using a large database including sentences, words, and syllables, diphone synthesis only uses a database containing every diphone within the language, and therefore every phone transition. This database can have some redundancy in its recordings, but often contains only one recording of each diphone. This results in a system with a far smaller database size than unit selection synthesis.

It is vital that these diphone samples are recorded in such a way that concatenating them will not result in abnormalities in the waveform, or introduce unusual variations in pitch over the course of the sample. Thus, a speaker's voice should remain relatively constant in pitch over the diphone. Sample waveforms should also be at the same point (typically zero) at the beginning and end of the waveform, such that when concatenated the waveform remains continuous. There are many techniques for smoothing the waveform after concatenation, which can further improve the quality of the synthesized speech.

Most languages contain between 800 and 2000 possible diphones. For example, the Italian language uses approximately 850 diphones, while English uses around 1800 diphones [48]. Because of this, constructing an effective diphone database can be a reasonably time-intensive process – though still taking far less time to produce than a unit selection database. Often, diphones are automatically extracted from natural speech by using some form of speech recognition system and identifying when phone transitions occur. Automating this process can help us to construct a complete diphone database in the span of only a few hours.

Once we have constructed a suitable database, diphone synthesis can generate speech with a relatively high degree of intelligibility for arbitrary input text. The size of the recording database remains small relative to those in unit selection synthesis, allowing us to implement diphone synthesis on platforms with restricted data storage capacity. As the synthesis step only involves a word to phoneme mapping and then sequentially concatenating waveforms (rather than a more advanced decision tree), diphone synthesis can also be effectively implemented on systems with little processing power, such as embedded systems or mobile devices.

One of the main problems with diphone synthesis is that, while it usually provides a high level of intelligibility, most implementations have a low level of naturalness. This is because samples are often only recorded at one particular pitch and played back verbatim, so there is no variation of pitch over the course of a sentence. Similarly, diphone samples are usually played back at the same speed, whereas natural speech varies talking speed over the course of a sentence. There are several techniques which can be used to resolve this, permitting prosodic overlay onto the system.

A naïve approach would be to simply record more samples: if we record each diphone at different pitches and speeds, we can choose the most appropriate sample from our database. However, this approach is almost never used, as the database grows substantially larger, which counteracts the primary advantage of diphone synthesis. It also takes more time and effort on behalf of the speaker to construct an expanded database.

Therefore, in diphone synthesis implementations where we wish to apply prosody, a signal processing technique must be used to perform separate time scaling and pitch shifting. This keeps the size of our database the same, instead only increasing the processing power required to run the implementation.

One might think that the easiest way to increase the speed of a digital audio recording is simply to play it back at a faster sampling rate. While this does increase the speed of playback, it also increases the pitch of the sound proportional to the change in playback speed. The reason for this is simple; if we play back a digital signal containing a 100 Hz audio wave at twice the original recording's sampling rate, then the wave will oscillate with twice the frequency at 200 Hz.

While doing this is computationally cheap, this approach greatly reduces the naturalness of our sampled speech diphones. Vowels and voiced phones are generated by the opening and closing of the vocal flaps. Each glottal excitation then resonates within the vocal tract, producing acoustic peaks depending on articulator configuration. Simply increasing the speed of our recording's playback will result in an inaccurate representation of how human speech actually sounds. This is because these resonant frequencies should remain relatively constant, while only the voiced fundamental frequency of speech changes. We therefore want to use a technique which takes into consideration how actual acoustic changes change our output waveform when human speech is articulated in a different fashion.

Another problem is that many consonants are not voiced at all; that is, their acoustic frequency ranges should not change with a different harmonic pitch. For example, stop consonants such as /p/ and /t/ are produced solely by actions of the lips or teeth, and affricates and fricatives such as /tʃ/ and /s/ are produced by turbulence. These elements of speech should, as much as possible, have minimal pitch shifting. For example, pitch shifting a diphone sample such as /ta/ might retain naturalness for the /a/ phone section, but the /t/ phone should not be pitch shifted in the same way, as its method of articulation does not permit changes in pitch in the same fashion.

### 4.1.3.1. Phase Vocoders

The term "vocoder" (a contraction of "voice encoder") in a general sense refers to any device or program which analyses and then resynthesizes human vocal input. In most applications, vocoders analyse the frequency spectrum of an input as it changes over time (typically using a Short-Time Fourier Transform), perform some operations on the result, and then synthesize the new waveform back in the time domain. This process is illustrated in Figure 18.

Alternatively, this process can be visualised as getting the spectrogram of the input waveform, changing the spectrogram as desired, and then returning to a time-domain signal. By moving the frequency peaks of the spectrogram representation, it is possible to alter the frequency of the input voice independent of speed. Alternatively, by expanding or contracting the spectrogram in time before transforming back into the time domain, we can effectively keep the frequencies within the speech similar while playing them over a shorter time period. Using this technique, we can perform distinct operations within frequency and time, allowing us to modulate the pitch and duration of the waveform independently as desired.

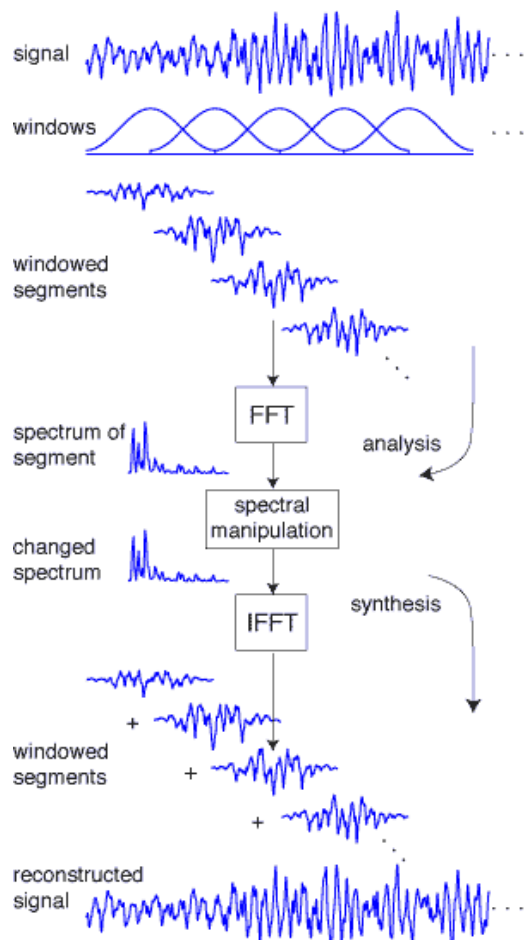

*Figure 18: Procedure in a Phase Vocoder implementation [49]*

A phase vocoder in particular also records the phase of the signals before the transformation into the frequency domain. This is vital for effective resynthesis of the signal. To show this, consider an input audio signal which we have transformed into the frequency domain over time. If we attempt to reconstruct this signal without phase information, each sampling window of our STFT is taken as the start point of the wave – giving us no effective way to meaningfully reconstitute even the original signal. To correctly transform a signal back into the time domain, we therefore need to know the phases of the signal relative to each sampling window of the STFT, so we know how to correctly offset our resynthesized samples.

This can introduce a problem once we attempt to perform operations on the signal. As the STFT windows will intersect with each other, each adjacent window will be very similar to the previous window. To accurately reconstruct the signal after transformation, these frames should represent the same, or a very similar, underlying signal. It is possible for a naïve implementation of a desired transformation to drastically alter the phase correlation between adjacent frequency bins or adjacent time frames. This means that the reconstructed time-domain waveform may have some undesirable discontinuities, which we would have to smooth in some way to reconstitute the waveform. This smoothing could introduce undesirable wave distortion, resulting in an audible abnormality. For example, Figure 19 shows the underlying original waveform in grey, while the different frames' waveforms are shown in red and green. The area circled in blue indicates the incoherence: the two waves are unequal at this connecting point.

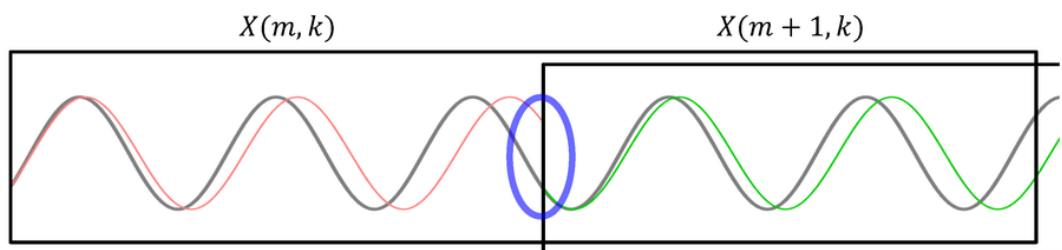

*Figure 19: Horizontal Incoherence in a Phase Vocoder [50]*

The property of continuity between different frequency bins is referred to as vertical phase coherence, while continuity between time frames is referred to as horizontal phase coherence; this is due to their relative dimensions on a typical spectrogram [51]. To reconstruct a signal effectively, we desire high coherence in both of these. However, most algorithms are only able to effectively preserve one of the two properties.

For example, if we wish the pitch over a certain diphone sample to vary over time, we will be performing different frequency transformations on different time windows. As such, the intersecting components of two adjacent windows will be shifted by a different amount, meaning that neither is effectively representing the same underlying waveform. Thus, to minimise this effect, we need to construct an average of the two time-domain signals. This is done by using intersecting sample windows of varying magnitude (most commonly, Hann windows are used, due to the small impact they have in the frequency domain), and then adding the intersecting sections of the transformed signals together. This allows the construction of a more continuous waveform, at the cost of an increase in computational complexity.

While this approach is usually effective, and most implementations are within modern computational limits to perform in real time, we still encounter the problem that non-voiced elements of speech are pitch-shifted even when we do not wish them to be. A scaling or shifting in the frequency domain will still undesirably modify frequency ranges produced by turbulence within the vocal tract. While the effect of this may not be noticeable to a casual listener for small shifts, our objective of natural sounding speech makes this technique less desirable than its alternatives.

## 4.1.3.2. Spectral Modelling

Spectral modelling synthesis is a more advanced shifting approach which exploits an acoustic understanding of how human speech is generated. In the spectrum of the produced speech waveform, this model considers sounds in human speech to be a part of one of two categories.

The first category contains elements which involve the harmonic, or deterministic, content of the waveform. These harmonic sounds in speech are produced by voicing, and have distinct and measurable frequencies. This also includes formant peaks from vocal tract resonance.

The second category contains elements involving noise, or stochastic elements of speech. This includes vocal elements generated by turbulence, stops, and percussive elements. These aspects of speech, if examined in the frequency domain, exist over a certain range of frequencies. In this model of human speech, these are modelled as an acoustic source of white noise which is then shaped over a range of frequencies, with the range also changing over time.

If we can separate out these components, then we can separately perform operations on each contributing factor to the speech waveform without altering the other. This allows us to overcome the shortcoming of the phase vocoder technique: the tonal elements of speech can be adjusted as desired in the frequency domain, while minimally altering non-harmonic elements of speech, such as stops, affricates, and fricatives.

By using this approach, we aim to achieve the same result as if the speaker's vocal folds were vibrating at a different frequency, while retaining the acoustic characteristics of unvoiced speech sounds.

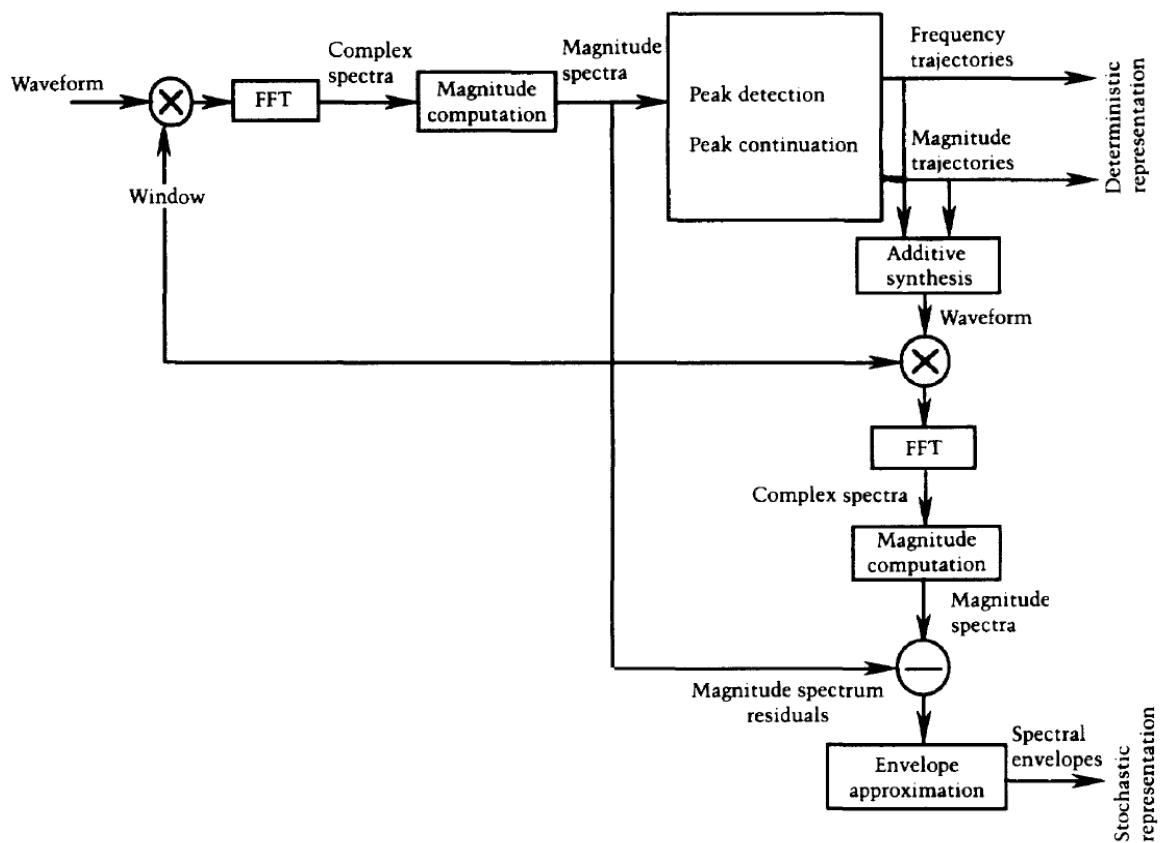

*Figure 20: Spectral Modelling Analysis and Separation Block Diagram [52]*

To implement this, we need a method to detect and separate these distinct categories of speech sounds. As in the phase vocoder algorithm, we perform a STFT on the speech waveform to be shifted. Then, we use a peak detection algorithm on the magnitude spectrum to determine which frequencies are notably higher than those neighbouring them, which correspond to acoustic formants. These frequency peaks are separated and subtracted from the original magnitude spectrum to find a residual spectrum of the original waveform, which represents the stochastic elements of speech. A block diagram of this separation technique is shown in Figure 20.

Once these spectra are separated, the peaks of the voiced elements of speech can be adjusted as desired and resynthesized in the time domain as sinusoidal signals. The residual component can be used as a spectral envelope to shape a generated white noise signal, synthesising the non-harmonic speech elements. While in other applications of the spectral modelling algorithm we may wish to perform frequency operations on these elements, in speech synthesis we do not wish to adjust the frequency spectrum of the noise components. Thus, we can effectively reconstruct the waveform after we have pitch-shifted it.

Similar to a phase vocoder, we can adjust the time scale of the signal separately from pitch by simply expanding or contracting the STFT frequency representation in the time dimension; this is a transformation we perform on both the harmonic and non-harmonic elements of speech in the same way. Therefore, using spectral modelling, we are able to alter both the pitch and time scales of our speech waveform independently, with the added advantage that we can independently adjust the voiced components of speech separately from the unvoiced sections. A block diagram of the resynthesis stage is shown in Figure 21.

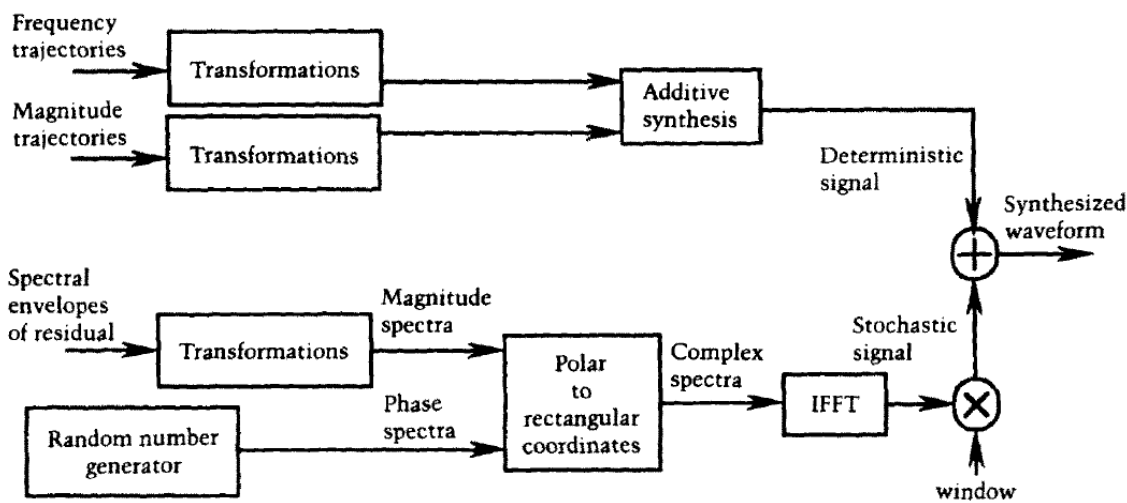

*Figure 21: Spectral Modelling Resynthesis Block Diagram [52]*

One of the disadvantages of spectral modelling is that some of the detail of the speech waveform can be lost in the process of separating it. In considering voiced speech as a sum of sinusoids of maximum magnitude in the frequency domain, some trace, transient elements of speech can be lost. In resynthesizing the signal from sinusoids, characteristics contributing to naturalness that occur in human speech might not be represented in the output waveform. Similarly, while the approximation of non-voiced speech as a frequency shaped white noise source is a useful approximation, we may lose some small amount of detail in our output waveform. Both of these can be counteracted by using a smaller sampling window for our STFT, at the cost of greater required computation, and with a fundamental limit due to the finite sampling frequency of all digital signals. Considering our aim of natural speech and computational efficiency, this is a less than ideal solution.

## 4.1.3.3. Pitch Synchronous Overlap Add (PSOLA)

The main assumption of the PSOLA algorithm is that speech is mostly composed of individually identifiable sections. For each oscillation of the vocal folds, an impulse of pressure is created which resonates through the vocal tract. The fundamental frequency of voiced speech is therefore determined by the spacing in time between these impulses. Thus, if we wish to alter the frequency of speech, we can section it up into these various impulses and vary the pitch by moving each section closer together for a higher frequency, or by moving them apart for a lower frequency. Figure 22 shows how PSOLA can be used to increase or decrease the pitch of a voiced waveform.

To vary the duration of a speech recording, the same section can be removed or repeated multiple times with the same spacing. This keeps the pitch of the recording the same, while the addition or removal of repeated sections is on a sufficiently small timescale as to be unnoticeable to the human ear. This lets us separately alter the duration and pitch of the output speech.

As with sampling for the STFT, we want each sectioned window to overlap with adjacent samples slightly. However, sampling will not be performed at windows of constant width. Instead, we wish to sample around the centre of each impulse response – this being the pitch synchronous aspect of the algorithm. This position can be determined using a peak detection algorithm in the time domain.

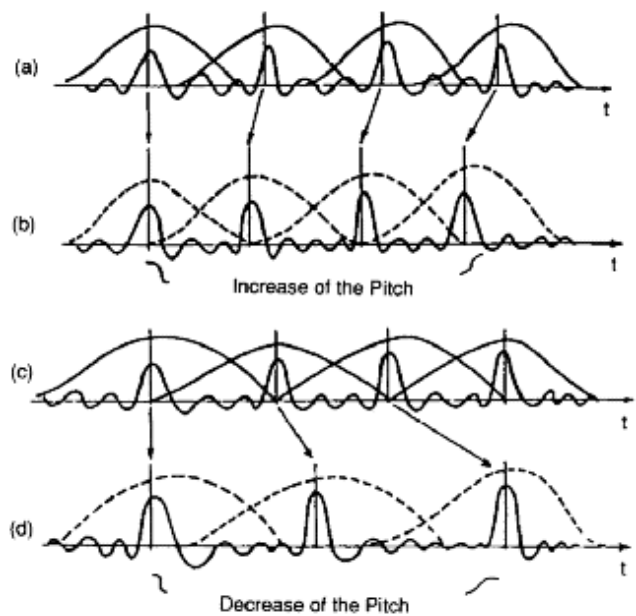

*Figure 22: PSOLA increasing and decreasing the pitch of a sound [32]*

Of the approaches to pitch manipulation discussed here, PSOLA is the most widely used in modern implementations. By operating purely in the time domain, no transformations into the frequency domain are needed to implement it, making it exceptionally fast and computationally cheap in comparison. One of the downsides is that this approach can result in undesirable phase interference between adjacent samples. By shifting a sample through an adjacent one and adding them, we might find frequencies at which constructive and destructive interference occur.

Another problem is that there is no particular consideration of the unvoiced sections of voice in PSOLA's design. In practice, the result is often acceptable despite these abnormalities. Most plosives are captured as peaks, and therefore sampled around; provided that they are not repeated or cut in time manipulation operations, the plosives of the output will sound similar to their equivalents in the unmodified waveform. Due to its widespread use and effectiveness, as well as computational efficiency, this approach seems very appealing for our objectives.

## 4.2. Generative Synthesis

Generative synthesis approaches, unlike sample-based approaches, produce waveforms which are not based on recordings from real life. They generate sounds purely programmatically, either by modelling physical speech production processes with varying accuracy, or by using simplified models which approximate the acoustic character of natural speech sounds. As speech sounds are generated at runtime rather than being stored in memory and played back, generative synthesis approaches often use less storage space than sample-based synthesis techniques, since there is no need for a large database. The downside is that the algorithms and models involved are more complex than simple concatenation, requiring a faster processor to produce speech in real time.

### 4.2.1. Articulatory Synthesis

Articulatory synthesis is a generative synthesis approach based on the knowledge that an accurate software simulation of the physical processes within the human vocal tract should produce an accurate approximation of human speech. Articulatory synthesis models are quite complex, encapsulating advanced biomechanical models and fluid dynamics simulations of air as it propagates through the vocal tract. This usually means that the models are difficult to implement in a computationally efficient way. Indeed, the computational complexity of articulatory synthesis approaches typically makes running them in real time impossible. Furthermore, a large amount of manual fine-tuning is required to make the waveforms generated by the model sound natural.

As articulatory synthesis emulates the way in which a real vocal tract behaves, captured images and video of real-world vocal tracts pronouncing different parts of speech usually form the basis of the dimensional and articulatory data in the model. Recording real world data lets us see the varying stricture of articulators in regular speech. This data is usually captured with an X-ray video camera and then traced, and these parameters are then used by the simulation. For example, Yale's Haskins Laboratories have developed a Configurable Articulatory Synthesizer (CASY) which uses an articulatory synthesis model based on real-world scans of the vocal tract. Their model constructs curves between key points, as shown in Figure 23, which then match real-world articulation movements as the virtual speaker produces different speech sounds. [53]

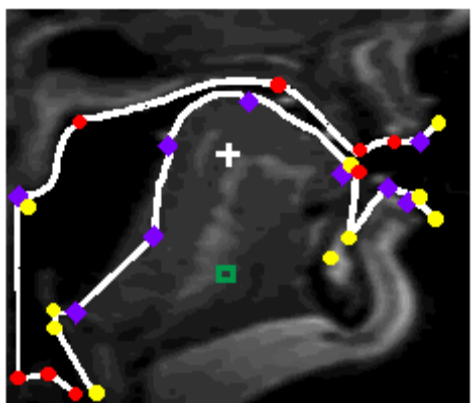

*Figure 23: Vocal tract trace from Haskins Laboratories' Configurable Articulatory Synthesizer (CASY) [53]*

While tracing scans is a reasonable method of data acquisition, it is often difficult to reconstruct all elements of the vocal tract from this captured data, as we are only seeing a two dimensional projection of the true three-dimensional surface. Even if information is captured from multiple perspectives, it is difficult to manually reconstruct a three dimensional representation. Further, it is very difficult to automate the process. It can also be very time-intensive to obtain the scanning data required, as we may wish to capture a great deal of transitional data on speech production – the greater the body of data collected, the more robust, accurate, and detailed the simulation can be.

One advantage of articulatory synthesis is that the same articulatory model can be adjusted to match multiple people's vocal tracts and articulatory behaviours in speech. This means that we only have to construct the technical aspects of the model once, and it can be applied to any particular person or speech pattern. This is especially useful since we are able to adapt the same model to languages other than English while using almost identical programming.

At the present time, articulatory synthesis is mostly of interest in research rather than application; it is a synthesis approach rarely used in practice for consumer electronics or software. This is because many models still have low naturalness and high computational cost; it is typically far easier to use a sample-based synthesis approach. However, with further research, the naturalness in such systems may greatly improve, and the flexibility of the model would allow users to extensively customise the synthesizer's voice simply by modifying system parameters. As available processing power on systems improves, and we develop technologies which are better able to capture three-dimensional internal biomechanical data, articulatory synthesis may find wider use in future applications.

## 4.2.2. Sinusoidal Synthesis

Where articulatory synthesis was based on modelling as closely as possible the articulation of the vocal tract, sinusoidal synthesis works by imitating acoustic properties of the waveforms produced in normal speech. This approach is far simpler to develop, requiring only frequency-domain analysis of speech waveforms to determine the desired frequencies. As previously discussed, vowel formant frequencies are the primary contributors to vowel quality. Therefore, by producing sinusoids at the first few formant frequencies (most typically three), we can approximate the acoustic properties of the waveform of vowels produced in natural speech production. Consonants may be modelled by using white noise. Haskins Laboratories have implemented this with their SineWave Synthesizer, which uses the first three formants to produce intelligible speech, as shown in Figure 24.

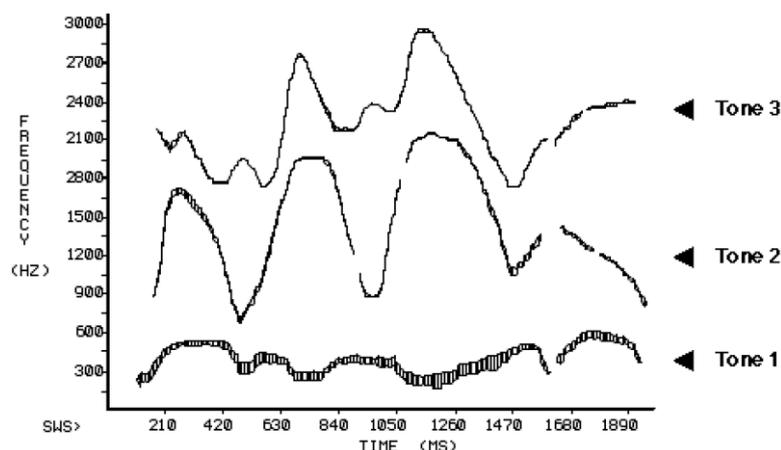

*Figure 24: Haskins SineWave Synthesizer (SWS) Frequencies for the phrase "Where were you a year ago?"*

Sinusoidal synthesis is the least computationally intensive generative synthesis method, since it only requires us to generate sinusoids and noise. The main problem is that the speech sounds that the approach produces have exceptionally poor naturalness, though still remaining somewhat intelligible. It is primarily of note in that such approaches are intelligible at all: despite the huge simplification of the model in ignoring many characteristics of human speech production, it is still possible for listeners to understand utterances produced by such systems. This helps to indicate the robustness of human speech comprehension.

By having an extremely small data footprint and being computationally simplistic, sinusoidal synthesis systems have historically found use in embedded systems [54]. However, due to their very poor naturalness, they should never be used on a modern system when an alternative is available.

## 4.2.3. Source-Filter Synthesis

Source-filter synthesis is essentially a compromise between the computationally expensive articulatory synthesis approach and the low naturalness of the sinusoidal synthesis approach. Rather than aiming to physically simulate the articulation processes as precisely as possible, formant synthesis uses a simplified source filter model of speech production.

The source filter model uses two different sources: a periodic source for the production of voiced speech, and a noise source for other speech elements. Unlike with sinusoidal synthesis, the shape of the periodic voiced sound source is designed to match real-world glottal excitation. Both of these sources are then fed into separate filters. For an input sequence of phonemes, the system alters the properties of the filters. This results in the appropriate change in formant frequencies that an articulatory synthesis approach would model, but with far less computational load.

The filters are designed in one of two ways. Their design can be acoustically-based, designed such that they correspond to the formant peaks and turbulence frequencies in the spectral envelope produced by natural speech. Alternatively, their design can be articulatory-based, where there is an additional layer of computation between the model and the filter. In such a design, the filter corresponds to an acoustic tube model of the vocal tract.

Such models make certain key simplifications in their simulation relative to articulatory synthesis models. First, they consider sound propagation within the vocal tract to be one-dimensional, such that sound is only travelling along the central axis of the vocal tract. For nasal consonants, we are able to simulate this by replacing a model of the oral cavity with the nasal cavity. Next, the cross section of the vocal tract is always considered to be circular. This assumption helps to make calculations substantially easier, and the approximation makes only a small change to the acoustic character of the system. Then, the vocal tract is either modelled as a sequence of cylinders of constant radius, as in Figure 25, or as a sequence of consecutive cone segments. The cone segment model can provide a more accurate approximation of the vocal tract, but increases the computational complexity of the model.

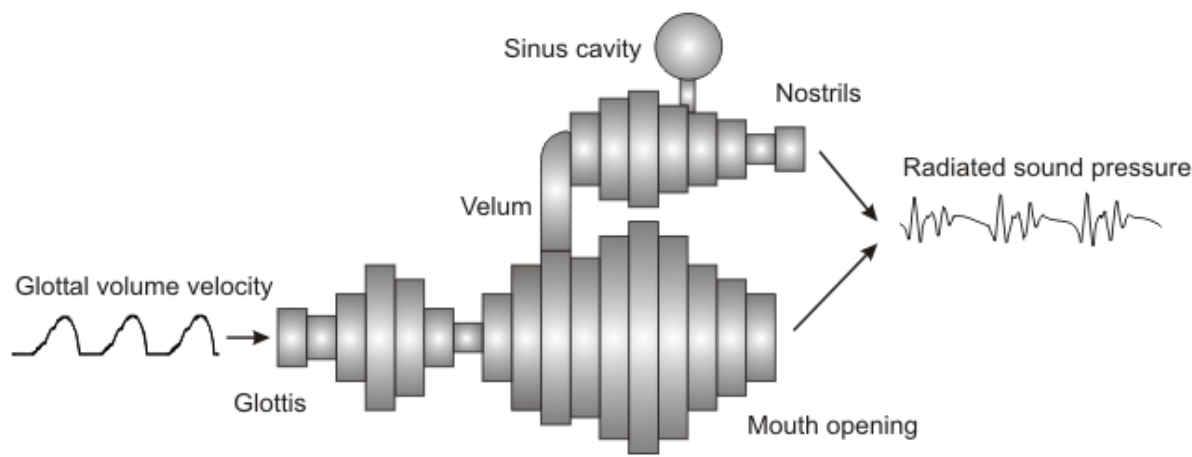

*Figure 25: Cylindrical Tube Speech Filter Model [55]*

More advanced models can also take into account how sections of the vocal tract slightly expand and contract according to the varying air pressure from sound propagation; this is another example of where a bioacoustic understanding of speech production can improve our design. Source-filter models can also take into account the contribution of viscosity loss within air on the produced speech waveform. These different kinds of losses reduce the power peaks of formant frequencies while widening their bands [56]. These acoustic models are then mathematically consolidated and implemented as filters, which allows us to easily change the articulation in our model over time.

The source-filter synthesis technique allows us to separately modify the model's glottal excitation and articulator positions. This lets us overlay any desired prosody in a natural way: we do not need to implement signal processing techniques for our output as with concatenative synthesis approaches. The changes in pitch and duration here can therefore sound more continuous than applying signal processing to a sample. This is especially true with slower speech: as most diphone samples are quite short in duration, applying algorithms such as PSOLA to extend their length will result in many repeated sections, which may sound unnatural.

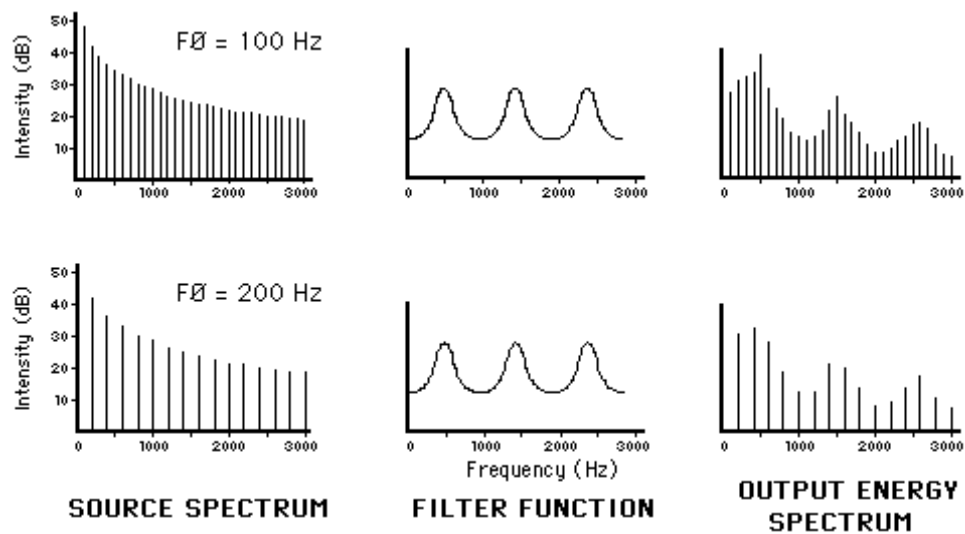

*Figure 26: Source filter model with identical articulation but different glottal excitation frequencies [57]*

The source-filter synthesis approach simply lets us move between different articulator positions as slowly as we desire, while keeping glottal excitation and the turbulence source the same. This can be seen in Figure 26, where the filter function denoting articulation remains the same while the source frequency changes. Source-filter models of speech production can make these articulatory transitions as arbitrarily fast or slow as desired, giving it exceptional flexibility without the computational complexity of articulatory synthesis.

## 4.3. Hidden Markov Model Synthesis

Where the previous sample-based and generative approaches were comfortably distinct from each other, Hidden Markov Model (HMM) synthesis effectively straddles the two. While the approach requires a large amount of recorded and tagged speech to construct the model, when synthesizing speech HMM synthesis is not concatenating different recordings, but generating speech sounds programmatically.

In general terms, a Hidden Markov Model is a statistical model where the system we are analysing is considered to be a Markov process (a process where future states are exclusively dependent on the current state) with hidden, unknown states. HMMs are a very useful tool in speech recognition, where we know a speaker is pronouncing a word composed of sequential phones: if each phone is considered a state, then a HMM can help us to determine what the underlying states that the system is transitioning between.

In HMM synthesis, we instead use the model to define prosodic aspects of speech such that they are similar to those in natural speech. To construct a suitable HMM, we use an algorithm which analyses sections of human speech, and then tries to determine which system configurations will result in similar outputs. There are various algorithms which are widely used to perform this task in a computationally efficient manner, such as the Viterbi algorithm and the Baum-Welch algorithm.

*4. Review of Speech Synthesis Techniques*The actual synthesis component of HMM synthesis may be produced using a source-filter model of speech which uses Hidden Markov Models to define the frequency of the glottal source and filter properties over time, as well as the duration and other prosodic aspects of speech. HMM synthesis can alternatively use an additional HMM model to synthesize the waveform itself in a more direct fashion.

The HMMs used in speech synthesis are designed to find a maximum likelihood estimation of state transitions within words or sentences. This is a very effective way of applying naturalness to synthesized speech, since prosodic changes can be individually modelled according to real-world speech patterns. Depending on how we train the system, we can therefore have the system imitate different prosodic speech patterns by training it in a different way.

The training of a HMM synthesis system requires extensive, annotated databases of comparable size to the ones used in unit selection synthesis. This database, being of natural human speech, ensures that the trained synthesizer will model human speech modulation patterns in an accurate fashion. HMM synthesizers not only offer intelligibility and naturalness comparable with or greater than unit selection synthesis, but the HMM approach requires far less memory and storage. Rather than storing the transitions as waveforms, we only include the HMM parameters which such waveforms would correspond to when resynthesized. This means that we can train a HMM system with an even larger corpus to improve its effectiveness, with no corresponding downside of a larger database size as with unit selection.

HMM synthesis is one of the newest and most effective techniques in speech synthesis, and a great deal of research is currently underway into optimising and improving it. Its advantages over other approaches come at a cost of increased complexity to implement.

## 4.4. Choosing a Technique for this Project

The objective of this project is initially to establish the intelligibility of our speech synthesis system, and then implement techniques to maximise naturalness. It is also important to consider the viability for a single person to construct any of these systems within the timeframe of the project. It is therefore desirable to reach a baseline of intelligibility as quickly as possible, so that we can progress to the investigation of more advanced challenges and techniques for applying prosodic overlay.

While generative synthesis approaches allow for extensive manipulation of different aspects of speech, they still suffer from needing either a large amount of real-world data or manual fine-tuning to sound reasonably intelligible or natural. As we wish to maximise both over a shorter timeframe, we should choose a sample-based approach rather than a generative one. Implementing a HMM synthesis system would also require an extensive amount of training speech to be recorded and tagged, which is similarly unreasonable for the timeframe available.

Considering the two options in concatenative synthesis which allow for a broad range of speech, Unit Selection and Diphone synthesis, we know that there are no freely available Unit Selection databases, and it would take a prohibitively long time to construct one from scratch. Diphone synthesis databases are still somewhat time-consuming to construct, but can be recorded and correctly formatted in a far more reasonable timeframe. Further, diphone synthesis gives us phonetic-level control over prosodic aspects of speech, where a unit selection database would require extensive tagging to allow us to arbitrarily change low-level prosodic characteristics. Using a diphone synthesis approach allows us to establish different degrees of success in our project, and lets us examine the more advanced research topics earlier.

As such, it was decided that a diphone synthesis system was the most suitable for this project.

047

# 5. Synthesizing Intelligible Speech

Now that we have determined the speech synthesis approach we intend to use, we will start by developing a system which reaches a basic level of intelligibility. After completing this, we can investigate more complex topics from a solidly established platform of operation. Thus, the aim of this section is to complete a relatively rudimentary but reasonably intelligible diphone text to speech synthesis system.

As we will be iteratively improving the synthesis system as we develop it, and we will later analyse the differences in capabilities between our system's stages of development, we should establish a reference name for our initial TTS system. We will therefore refer to the system developed in this section as the BAsic Diphone Speech synthesis system, or BADSPEECH.

## 5.1. Word to Phoneme Lookup

For this iteration of our speech synthesis system, we only want to establish basic functionality as a starting point for further work. A full Text to Speech system should be able to accept an arbitrarily formatted input, including punctuation and terms which are not necessarily known English words. For our implementation of BADSPEECH, we will assume that all input is in the form of space separated English words. As such, we only intend to implement a word-to-phoneme lookup function.

This will require the use of a machine-readable pronouncing dictionary. As this is primarily a research project, we wish to use a freely available database so that our results and findings can be freely reproduced and distributed. The most expansive such database for the English language is the Carnegie Mellon University Pronouncing Dictionary, or CMUdict. [58]

CMUdict is provided as a plaintext file from Carnegie Mellon University's website. It uses a reduced version of the Arpabet transcription system, using the same stress markers but only including 39 of the full Arpabet's 48 symbols. As with all Arpabet transcriptions, each transcription is a representation of the word in the General American English accent. Some example words from the file are shown in Table 14. In this project, CMUdict will be the primary source of pronunciation information; for this section, however, we will only be implementing it for lookup.

*Table 14: Example CMUdict Entries*

```
ARTICULATE  AA0 R T IH1 K Y AH0 L EY2 T
CUSTODIAL  K AH0 S T OW1 D IY0 AH0 L
DUBLIN  D AH1 B L IH0 N
KEGS  K EH1 G Z
LOTTERY  L AA1 T ER0 IY0
READ  R EH1 D
READ(1)  R IY1 D
THOUGHT  TH AO1 T
WACKY  W AE1 K IY0
ZIP  Z IH1 P
```

As discussed in the introduction of this report, heterophonic homographs are words which are spelled the same but pronounced differently based on context. This is another problem which will not be considered in this section, but postponed for the next iteration of our system. CMUdict includes multiple pronunciations for heterophonic homographs which are delineated using a numerical tag. This can be seen in Table 14, providing two distinct pronunciations for the word "read" (with pronunciation varying between the past tense "I have read" and future tense "I will read"). For BADSPEECH, we will simply take the first pronunciation that we encounter and use it to produce our speech, ignoring the numerical tag.

## 5.2. Constructing a Diphone Database

As we can now take an input in words and find a corresponding phoneme sequence, we wish to construct a database of diphones to use. With over a thousand diphones in the English language, manual extraction from recorded speech can take days. While this can give a better quality of recording, as all diphone transitions are manually checked and extracted, it is preferable to automate this process. There are various techniques for doing this.

One of the most common approaches is to perform speech recognition on a large amount of recorded speech. Distinct diphones are extracted and saved into the database with some redundancy. Where multiple samples of the same diphone are extracted, each sample is assigned a quality weighting dependent on desirable qualities of the sample. While this method is effective, it requires us to construct a general speech recognition system; this is well beyond our project's scope.

A simpler method (on behalf of the engineer, though perhaps not the speaker) is to prompt a speaker to produce every phonetic transition in isolation. As we will know the phones that the speaker is transitioning between, we are operating within a greatly restricted scenario relative to the general speech recognition system required for automatic extraction as above. All that is required for this methodology is the ability to tell when the phonetic transition has occurred. This is therefore the method that we will implement.

While many programming languages are capable of capturing and playing back audio signals through the use of a software library, it is far easier to use a language which automatically handles audio capturing and playback. MATLAB has a very simple system for capturing audio from a microphone input, automatically selecting any microphones connected to the system; similarly, we need only call the *sound* function to play back a waveform. This removes any potential interfacing difficulties which could introduce additional challenges. Further, there are many functions already implemented in MATLAB which are well suited for solving the kinds of signal processing problems we will be considering in this project. For these reasons, we will implement our program using MATLAB.

## 5.2.1. Initial Considerations

As we wish to construct a diphone database, the 5 Arpabet diphthongs AW, AY, EY, OW, and OY should be treated as a combination two distinct phones. While it is simple enough that OY becomes AO IH, the other Arpabet diphthongs use the IPA vowels /a/, /o/, and /e/, which do not exist in isolation within the Arpabet transcription system. We will therefore denote these new vowels in our database using the graphemes IPAA, IPAO, and IPAE respectively. We can now replace these symbols within CMUdict with new symbols as shown in Table 15.

*Table 15: Diphthong Replacements in our Database*

| Arpabet | IPA | Database |
|---------|-----|----------|
| EY | /eɪ/ | IPAE IH |
| AY | /aɪ/ | IPAA IH |
| OW | /oʊ/ | IPAO UH |
| AW | /aʊ/ | IPAA UH |
| OY | /ɔɪ/ | AO IH |

Now, after processing, each grapheme in our database only corresponds to one phone. We should note that we will only need to capture the phone transition going from IPAE to IH instead of IPAE to every other phone: it is impossible for IPAE to be followed by any phone other than IH, so we only need a variety of transitions going to the IPAE phone. This similarly applies to IPAA and IPAO, reducing the total number of possible phonetic transitions we need to capture for our system.

Now, we can consider some preliminary design choices. Our initial methodology can be broadly split into two stages. In the first stage, monophone extraction, we will prompt the speaker to first pronounce each possible phone within the database. The fundamental frequency of voiced phones should be kept as similar as possible, so that the pitch of any synthesized speech will remain close to constant. This recording also lets us characterise the speaker's pronunciation of each phone.

As part of monophone extraction, we also want to extract the transitions between each phone and silence. The phonetic representation of the absence of speech sounds will be denoted using X in our database. Therefore, the diphone X IH is the initial part of producing the IH phone, while IH X is the terminating section. Therefore, as part of monophone extraction, we need to capture three distinct sections: the transition from silence to the phone, the phone in persistence, and the transition from the phone to silence. After we have captured all monophones, we can move on to the second stage.

For the second stage of our method, diphone extraction, we wish to record the speaker producing every possible diphone. The current diphone we wish to capture is communicated to the speaker in two ways. Firstly, the phone transition to be captured is denoted using Arpabet notation. Secondly, we create an audio prompt by concatenating the recordings in persistence from the first stage together. For example, if we are capturing the transition IH AA, we take the persistent sections recorded for IH and AA from the first stage, reduce their duration to a smaller time period, and play them back in order.

This prompt will not sound like a natural phonetic transition; after all, we are capturing diphones in order to better capture that transition. This prompt mainly serves to communicate to the speaker the desired target phones at the start and end of the diphone. It is particularly important that each target phone be produced at close to the same pitch and articulation. This consistency will make our concatenated speech sound more natural, as the points of concatenation will sound as similar as possible. This will make the transition between any two diphone recordings sound smoother.

Automating monophone extraction should be reasonably easy: we need only to distinguish between the presence and absence of sound to automatically extract the phones. The automatic extraction of diphones will be more difficult, and require the implementation of more advanced techniques.

### 5.2.2. Implementation

Before discussing further, we should first create a distinction between certain phone groups. There are three categories of phones which we wish to consider: sonorants, obstruents, and stops. Our use of these terms here does not strictly match what the same terms in linguistics do. Instead, we define these groups in a way which is more useful while we are considering different aspects of diphone speech synthesis.

Sonorants are phones produced by voiced, non-turbulent airflow. This includes vowels, semivowels, liquids, and nasals. Sonorants can be continuously sustained. Acoustically, they therefore have a fundamental frequency from their voicing, and no turbulent production. This results in a smooth waveform when viewed in the time domain.

Partial obstruents are phones produced by partially obstructing airflow to produce high-frequency turbulence, which appears as noise in the time domain. While the formal linguistic category of obstruents also includes stops, in our categorisation here we only include phones which can be continuously sustained; in the remainder of this paper, the term "obstruent" will therefore only refer to partial obstruents. This category includes both voiced and unvoiced fricatives, as well as the aspirate phone HH. Together, sonorants and obstruents are the persistent phones.

Our category for stops includes stop consonants, as well as affricates, which begin as a stop consonant. Stops cannot be continuously sustained; that is, they are non-persistent. Further, their pronunciation requires complete occlusion of the vocal tract, stopping airflow entirely immediately prior to their pronunciation. For the remainder of this paper, the term "stop" shall be used to refer to both stop consonants and affricates.

This categorisation of phones into sonorants, obstruents, and stops will be important throughout this report, as each presents different challenges in the extraction and synthesis process.

### 5.2.2.1. Monophone Extraction

In the first stage of our system, the speaker is asked to pronounce monophones in isolation. If the phone can be continuously sustained, the speaker should produce the phone for one or more seconds. Then, we will extract and isolate waveforms capturing transitions in both directions between the phone and silence, as well as the phone itself. If the phone cannot be sustained, then it is a stop; it should be pronounced and terminated briefly, and we only need to isolate that sound.

The first step here is to determine when the speaker is silent. Any microphone will capture undesirable signal noise due to ambient air movement, electrical noise, or other, distant sounds. We must thus determine a minimum point, below which we assume sound produced to be silence. This is performed by simply opening the microphone channel and determining the maximum amplitude during silence; it is then multiplied by 2 to exclude sounds slightly louder than total silence. We also find the RMS value of this silence, which is used as a secondary check.

Once this silence threshold has been defined, we can identify and isolate the section containing the start, sustenation, and termination of the phone. We identify the first and final points in the waveform above both the amplitude and RMS silence threshold, to determine the space within which the phone is being produced. We then apply a linear window function to the surrounding sections over a 0.01 second timeframe; this means that the first and final points of the wave will be at exactly zero, minimising undesirable effects from concatenation.

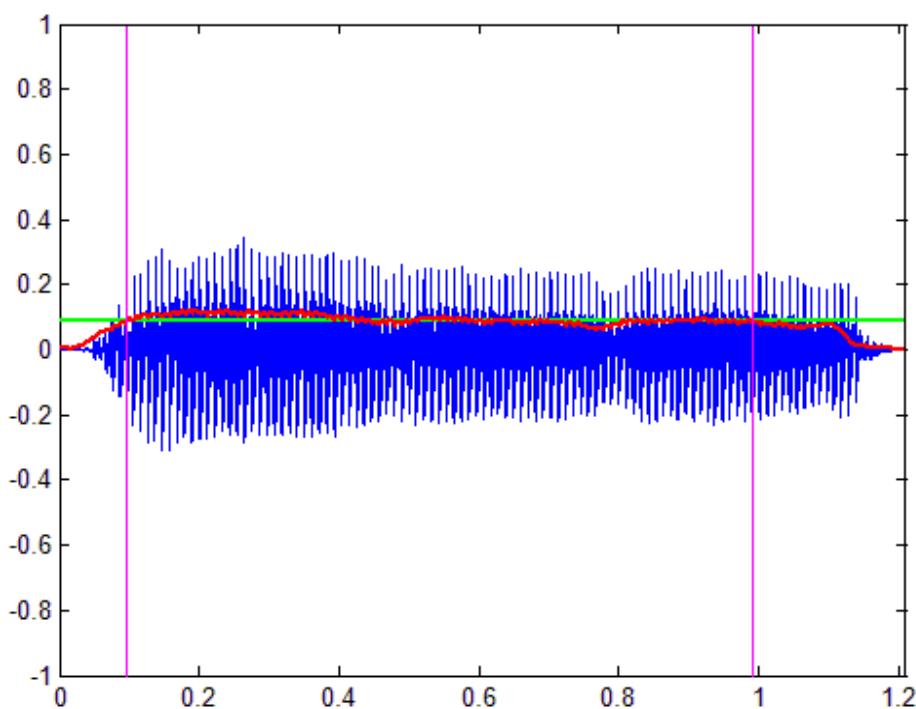

*Figure 27: AA Monophone being produced and automatically sectioned*

Figure 27 shows the process of recording and sectioning an AA phone; the silence in the recording surrounding this phone has already been removed. The blue line represents the relevant section of the recorded waveform, while the red line is the short-term RMS of the phone. The green horizontal line is the total RMS of the entire waveform; similarly, this is also the mean of the red line. If the phone is persisted for sufficient time, this value will be slightly less than the RMS of the wave when the phone is being produced at its maximum volume.

To determine the points at which we split the phone, we find the first positive edge crossing of the green line by the red line. This is the point at which the phone has mostly transitioned to a fully articulated state, and is marked by the first vertical pink line. The second vertical pink line occurs at the final negative edge crossing of the green line by the red line; the section of the phone occurring after this is the termination of the phone. By then storing each of these sections of the waveform separately, we have extracted the transition to the phone from silence, persistent production of the phone, and termination of the phone. For non-persistent phones, we simply remove the surrounding silence and store the produced phone.

### 5.2.2.2. Persistent Diphone Extraction

In recording diphones where both phones are persistent, there should be four sustained states from the speaker: initial silence, the production of the first phone, the production of the second phone, and a return to silence. We are interested in automatically capturing the transitioning between the two persistent phones. To do this, we need to be able to distinguish between different phones as pronounced. As has been previously discussed, it is easiest to tell the difference between two phones by looking at the waveform in the spectral domain using a STFT.

We wish to consider over time how spectrally similar our recording is in the short-term to the two monophones that compose it. Fortunately, we have recordings of each phone in persistence from the previous stage. We can therefore do this by defining a distance metric between two spectra. The log-spectral distance is a common distance measure used for this purpose [59]. It can be found using the formula:

$$D_{LS} = \sqrt{\frac{1}{2\pi}\int_{-\pi}^{\pi}\left(10\,log_{10}\frac{P(\omega)}{\hat{P}(\omega)}\right)^2 d\omega}$$

This is the distance measure as calculated for an analog signal; we need to make some modifications to it here, as we are dealing with a digital signal. Further, we can simplify the metric for computational efficiency without reducing its usefulness. The MATLAB code shown below is therefore used as our distance metric, to find the distance between two spectral profiles:

```
distance=0;
for k = 1:length(spectrum1)
    p1 = spectrum1(k)+0.1;
    p2 = spectrum2(k)+0.1;
    distance = distance + (max(p1,p2)/min(p1,p2))-1;
end
distance = log(1/(distance+1));
```

This sums the ratios of the larger to the smaller of the two spectra at each point subtracted by one, such that if the values are the same there is no net change to the distance sum. We add 0.1 to each value of the spectrum; this is small relative to the spectral peaks. This ensures that if both values at a given instant are very small from an absolute perspective, even if one is large relative to the other our distance metric does not change much. We then take the logarithm of 1/(distance+1); this results in our distance measure being negative and approaching zero with closeness, such that a peak of our distance measure represents relative closeness rather than a trough.

Using this metric, we can determine the acoustic similarity of a particular time frame in our STFT to the phone that the speaker should be pronouncing. Therefore, if we wish to capture a transition between two persistent phones, we can perform a STFT and plot the distance of each time frame's spectrum to the spectrum of one of the monophones in persistence. This gives us a distance value which is close to zero when the chosen monophone is being produced, but negative with greater amplitude when it is not.

Since we are capturing diphone transitions, it is useful for us to take the distance metric over time for both the first and second phone. This gives us two distances varying over time; one peaks during the production of the first phone in the diphone, while the other peaks during the production of the second phone. If we subtract the first distance from the second distance, we therefore have a single line which is low during the production of the first phone and high during the production of the second phone. We shall refer to this as the distance line.

The distance line is very useful in determining where within our recording the phonetic transition occurs. If we first only consider that section of our recording during which sound is being produced, and we make the assumption that the production of each phone persists for approximately half of the time, then we anticipate a distance line which starts reasonably constant at a lower value, which then increases, and becomes reasonably constant at a high value.

If each phone composing the diphone takes up approximately half of our recording, then if we get the mean of the distance line, then it should be approximately the average of the higher and lower values that we are considering. If we then get the mean of every point in the distance line which is below this overall mean, then we expect a value close to the lower constant value; similarly, if we get the mean of every value in the distance line above the overall mean, then we find a value close to the higher constant value.

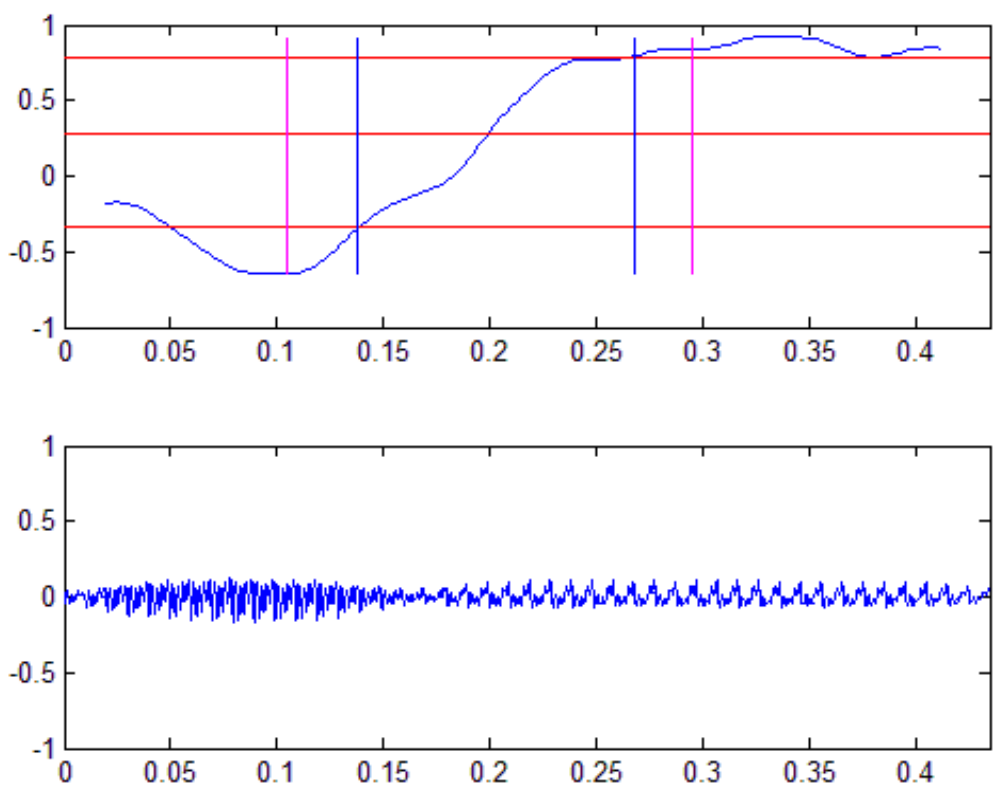

*Figure 28: L to N Diphone being produced and automatically sectioned*

This technique can be seen in Figure 28: the lower plot is of the waveform, while the blue line in the upper plot is the distance line (with some smoothing applied, to reduce short-term fluctuations from different time frames in our STFT capturing different sections of the wave). It is clear that the behaviour of the distance line is as expected: its value is low during the first half of the diphone, and high during the second half. The central horizontal red line is the mean of the distance line. The upper red line is the mean of all points in our distance line above the overall mean; the lower red line is the mean of all points in the distance line below the overall mean.

In finding a valid transition, we want to find a section of the distance line which passes through the lower, middle, and upper mean. In our sound wave, this corresponds to a phonetic transition. In our transition sectioning, we want the upper and lower means to only be passed through once, with the lower mean being crossed first and the upper mean being crossed last.

Figure 28 shows two vertical blue lines at the crossing of the lower and upper means for our transition. While we could section our wave at these points, we want our waveform to be as similar as possible to the monophones recorded in persistence. As such, we move backwards in time from the crossing of the lower mean until the previous value is greater than the current value: this finds a local minimum of our function, and therefore a local point of maximal spectral closeness to the first phone. We do similarly with the crossing of the upper mean, moving forwards in time until we find a local maximum. These two points are represented by the vertical magenta lines in the figure. We can then save only that section of the recording which lies between these two magenta lines, which gives us the transition between the two phones. This technique is very effective at finding and extracting the phonetic transition between any two persistent phones.

### 5.2.2.3. Stop Diphone Extraction
Non-persistent phones are more difficult to automatically extract, as their articulation occurs over a short timespan. Here, we will exclusively capture transitions going from non-persistent phones to persistent phones.

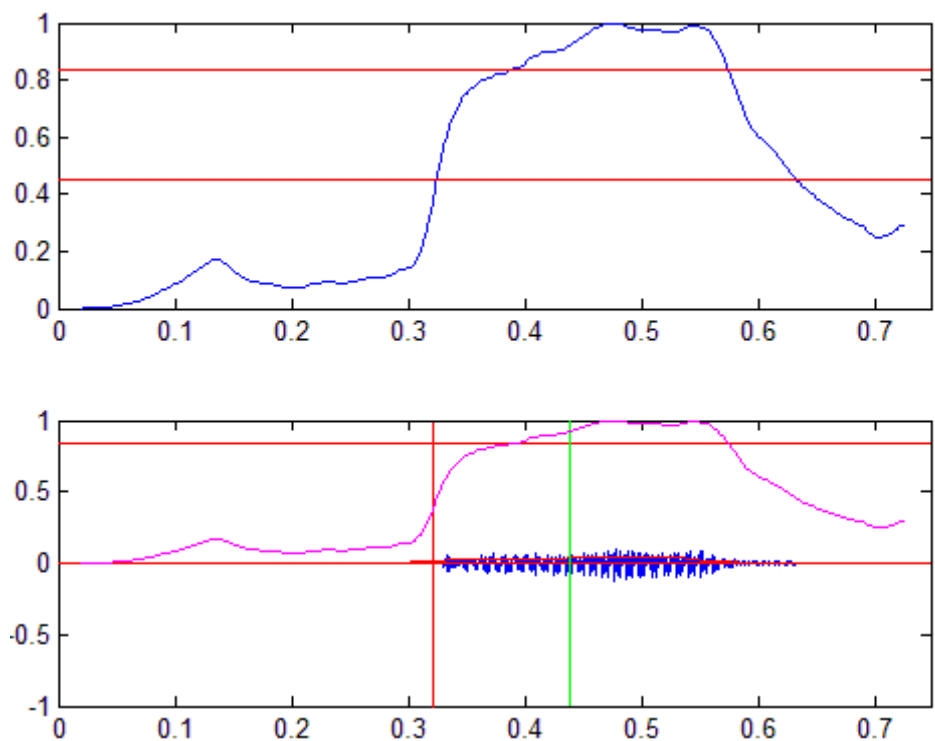

*Figure 29: G to R diphone being produced and automatically sectioned*

Figure 29 shows our methodology for extracting these transitions. The upper plot shows the spectral distance of our recording from the persistent phone, normalised between 0 and 1; then we perform a similar technique as was used for diphone transitions. First we find the mean of this distance line over time, then the upper mean, then we find the first positive crossing of the upper mean and move ahead to a local maximum. This is then marked in the lower plot by a vertical green line. The vertical red line is the closest point previous to the green line at which the short-term RMS of the wave was below the RMS silence threshold.

This methodology allows us to extract phonetic transitions leaving stop phones moving towards persistent phones. However, we will not capture transitions between two stop phones, nor will we extract transitions from persistent phones to stop phones.

First, when two stop phones are produced sequentially, there is no continuous transition between the two; unlike with transitions between persistent phones, there is no intermediary articulatory space. The first stop phone is produced, the vocal tract is again occluded, and the second stop phone is produced. Therefore, we do not need to capture stop/stop transitions, and instead simply play the first stop phone's recorded monophone, and then the second phone's monophone (or diphone, if the second phone is followed by a persistent phone).

As stop phones all include a similar occlusion of the vocal tract prior to articulation, we also do not need to extract transitions moving from persistent phones to stop phones. This is because the transition recorded from the persistent phone to a stop phone ends at the point of occlusion, which is acoustically almost identical to the transition from the persistent phone to silence. We have already captured this transition as part of our monophone extraction, so we can simply use that recording.

### 5.2.3. GUI and Automation

Now that we have a method for automatically extracting monophones and diphones from recorded speech, we can design an interface to automate the process, and make constructing a diphone database easier. First, we shall consider all of the constraints that we will place on what diphones must be captured.

As previously stated, there are only one or two possible terminating phones for our IPAA, IPAE, and IPAO diphones. Next, we do not wish to capture any diphone with a stop phone as its terminating phone. Finally, we also do not wish to try and capture diphones which transition between the phones M, N, or NG. As these three phones are nasal, in persistence they sound identical. This means that any transitions between them do not change the way that the phone sounds. As such, our method is not effective at telling the distinction between them; at the same time, their removal changes little about the system's intelligibility, since listeners cannot distinguish between those transitions either.

With these exclusions, our system must capture 37 monophones and 958 diphones. It is possible for the system to play a short prompt for diphone capture, record the speaker's response, and extract the desired diphone in about two seconds, while each monophone should be persisted for four seconds to best identify the monophone's spectrum. Adding this to the 10 seconds of silence that we must characterise before recording anything, this means that a full diphone bank can be produced in 2074 seconds, or about 35 minutes. Of course, speakers do not always produce diphones as quickly as possible, and may need short breaks. Further, not all speakers can respond to prompts as quickly as this; as such, we also want to permit the selection of a slower prompt speed. In practice, trained speakers can create a database in approximately 40 minutes, while untrained speakers take about 2 hours to record a full diphone database.

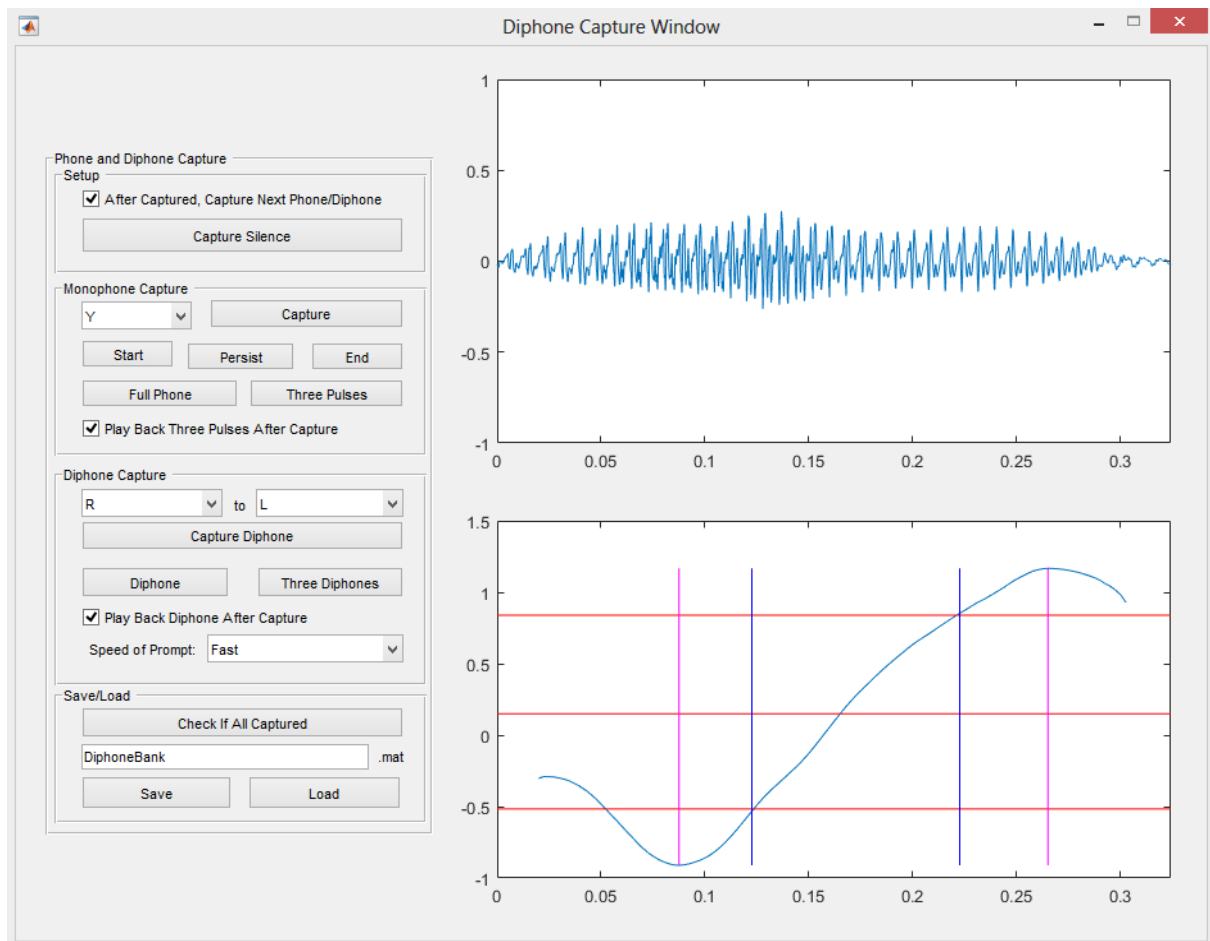

*Figure 30: GUI for capturing Monophones and Diphones.*

Figure 30 shows the user interface designed for capturing speech. The Phone and Diphone Capture panel is for user interaction, the upper plot shows the waveform recorded, and the lower plot shows the data used to separate monophone or diphone sections. In the Setup panel, a user records ambient background silence to determine the thresholds to use for the rest of the system. The checkbox in that panel also allows the system to immediately continue to the next phone or diphone once the present one has been recorded, permitting rapid recording and phone extraction.

The Monophone Capture and Diphone Capture boxes allow a user to select particular phones or diphones to capture, or after having been captured, play back the recordings. The user can choose to automatically play back the extracted diphone after recording, to confirm it has been extracted correctly. The speed of the prompt used for diphone capture is also selectable as Fast, Medium, or Slow, with each constituent phone being played back for 0.1, 0.2, or 0.3 seconds respectively.

Finally, the Save/Load panel lets users save recorded diphone banks, or import previously recorded ones to the workspace. Users can also check if all desired diphones have been captured in the current diphone bank, any missing phones being displayed in the MATLAB command window.

This interface makes it much easier to construct a diphone database than using the command line. The process can progress through monophones and diphones manually or automatically, and the speed of the prompt can be set according to the speaker. It is also easy to re-record if the initial recording was poor, or to capture the silence level again if conditions change. Of course, this software side should be complemented by high-quality hardware and good recording conditions: the best results are obtained by using a high quality microphone in an acoustically isolated environment.

## 5.3. Combining to Produce BADSPEECH

We can now use CMUdict to determine the pronunciation of English words within the dictionary, and then we can use a recorded diphone database to synthesize speech. However, our diphone sectioning method means that the beginning and endpoints of our diphone recordings are not usually at the same point within the phase of the phone, nor at the same amplitude. This means that playing these diphones sequentially results in audible clicks from the wave discontinuities. As such, we will need a way to smooth the connection between two concatenated waveforms.

### 5.3.1. Concatenation Smoothing

Our objective is to turn two separate diphone recordings into a single, continuous sound wave which smoothly transitions through the connective phone. For diphones connected by silence or a stop phone, we do not need to be concerned about smoothing, as our waveform should always be equal to zero at the point of concatenation. For persistent phones, we need to smoothly transition from one recording to the other.

To do this, we want to align the phase of the two diphones on their connective phone, then crossfade from one recording to the other over a short timeframe. An initial method considered was to find an alignment which minimises the average difference between the overlapping sections of each wave. However, this is computationally expensive, and could only be effective for sonorant connective phones. Using this approach fails for obstruent phones due to the wide band of high frequency noise, which increases the average difference between any two waves simply from random turbulence in speech articulation.

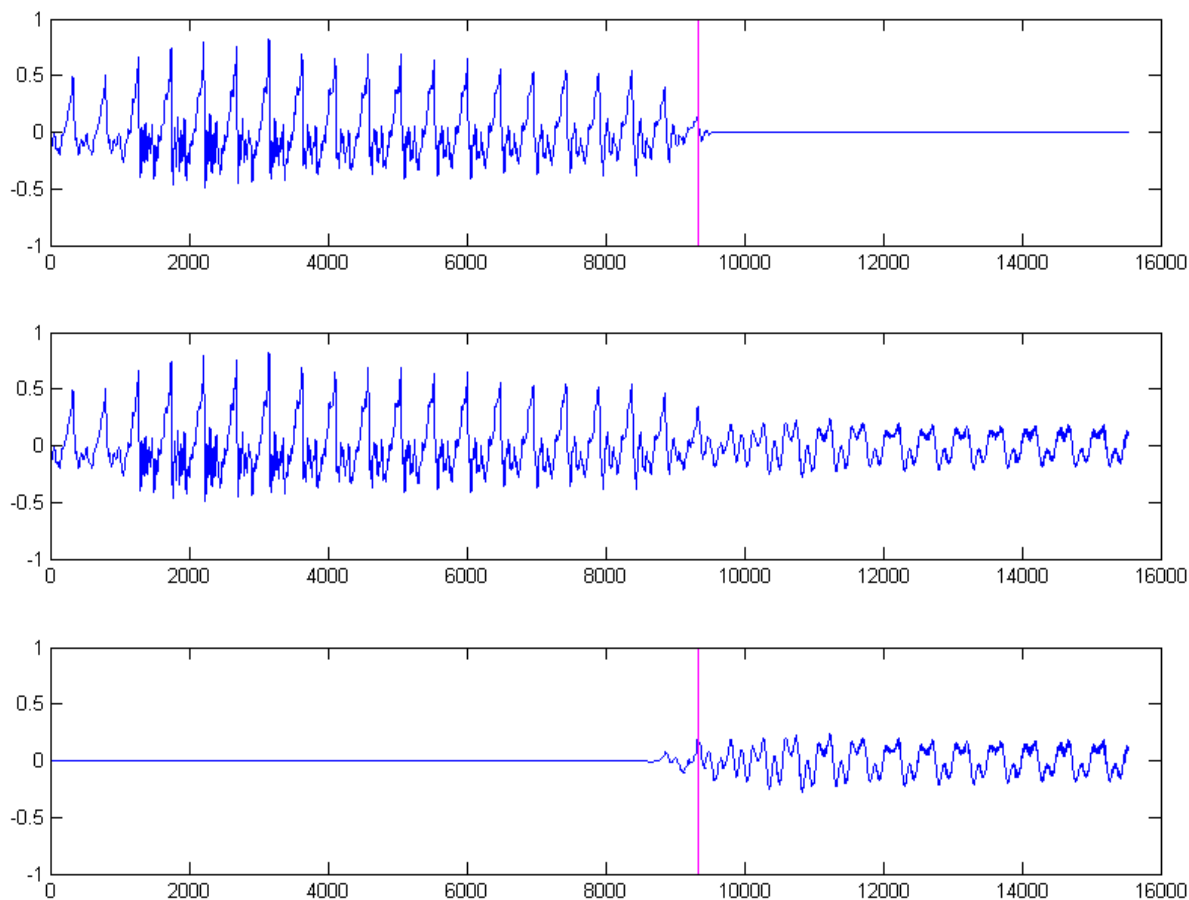

*Figure 31: Alignment and combination of the diphones W to Y and Y to AA.*

The approach used instead is to find the maximum point within the last 0.02 seconds of the first diphone's waveform, and the maximum point within the first 0.02 seconds of the latter diphone's waveform. The value of 0.02 seconds was chosen so that at least one full period of the wave is within that range, provided that the voiced component of speech is being produced at a higher frequency than 50 Hz, which human speech almost always is.

If at least one full period of voiced speech has been included in these regions, then the maximum points in each should occur at close to the same point within the phase of the signal. As such, by setting the two waves to overlap such that these maxima occur at the same point in time, then we have correctly aligned the waves. Then, for the amount that each wave overlaps with each other, we can multiply each by a ramp function such that the first wave fades out and the second wave fades in. Finally, we can add these two together, resulting in a smoothly concatenated wave.

This procedure is illustrated in Figure 31: the top plot is the recorded diphone from W to Y, and the bottom plot is the recorded diphone from Y to AA; their maxima are indicated by the vertical magenta lines, which have been aligned to the same point in time. These were then multiplied by ramp functions so that the W to Y diphone reduces in volume at the same time as the Y to AA diphone increases in volume. These waves are then added together, resulting in the plot in the centre. This gives a smoothly concatenated wave, with no abnormalities when played back.

This approach is very computationally cheap, and the same procedure works well when the connective phone is a voiced obstruent, as even with high-frequency signal noise the more prominent contribution of the voiced component of speech dominates, resulting in accurate phase alignments. Where the connective phone is an unvoiced obstruent, the entire signal is this noise: as such, crossfading between diphone recordings sounds reasonably natural, since the constituent noise signals are essentially random, meaning that their composition is also essentially random. If we run these through the same smoothing procedure, the alignment is an unnecessary part, but the crossfading means that it is still effective. With this completed, we simply combine these individual components to complete our implementation of BADSPEECH, as shown in Figure 32.

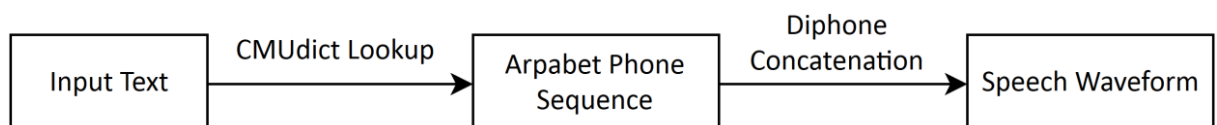

*Figure 32: Simplified BADSPEECH Software Flow Diagram*

## 5.3.2. Recorded Voices and Testing BADSPEECH

We wish to confirm the general effectiveness of our speech synthesis system; as such, it is important to use a number of diphone banks contributed by different speakers. This will help to demonstrate both the robustness of our diphone capturing method, in being able to successfully extract diphones for each speaker, and our synthesis method, in synthesizing speech from a range of diphone banks with different acoustic properties.

Informal testing confirms that BADSPEECH is reasonably intelligible for all of the recorded diphone banks. However, we will leave the results of formal testing until later in this report. After completing the more advanced iterations of our system, we intend to perform tests to determine their relative effectiveness. It is more useful to consider all such tests at the same time, so that we can evaluate our results as a whole. For this reason, we will similarly postpone the testing of future iterations of our system until they have all been completed.

Our analysis of different diphone voice banks is in Section 8.1.1. on Page 92.

# 6. Improving Our System

While BADSPEECH gives some intelligibility, it is still, relative to our goal of easily intelligible and reasonably natural synthesis… well, bad speech. In this section, we wish to optimise the intelligibility and efficiency of our system, as well as diversify both the number of words which can be synthesized by the system and the variety of pitches and speeds at which it can produce speech.

As before, we wish to use a label to refer to the synthesis system we develop in this section. Thus, we will refer to the system being developed here as the Optimised Diversified Diphone Speech system, or ODDSPEECH. A flow diagram of the final implementation is shown in Figure 33.

## 6.1. Broad Grapheme-Phoneme Conversion for English

With BADSPEECH, we could only produce words which were within CMUdict, having no way to pronounce words outside of that lookup table. Here, we will discuss how to determine the most likely pronunciation of an arbitrary input word composed of alphabetic characters.

Accurate Grapheme-Phoneme conversion is particularly important in GPS navigation systems, where synthesized audio instructions are a vital component of safe usage on the road. No unit selection database could feasibly contain a pronunciation of every street and town in the world. Even for those systems with a larger database, it is often necessary to fall back on the more flexible diphone synthesis approach. This must then be used in conjunction with an accurate conversion system from input words to their phonetic pronunciation.

While a more linguistics-focused approach would be to use a pre-existing understanding of phonetic rulesets to hard-code these rules, the engineering solution is data-driven. By using a large number of grapheme-phoneme correspondences as training data, we can determine the most likely pronunciation of a word not in that data. This requires a large dataset of words and pronunciations. Conveniently, as we were previously using CMUdict as our pronunciation dictionary, we have such a dataset readily available. However, CMUdict is only a dataset of word-scale grapheme-phoneme correspondences. To train our system, we need to determine correspondence on a smaller scale than whole words. Therefore, we need to align our grapheme data with our phoneme data, so we can determine which graphemes contribute to which phonemes within a word.

There are three main grapheme/phoneme alignment methodologies [60]. 1-to-n alignments map each grapheme to some number of phonemes, or no phonemes at all. 1-to-1 alignments map at most one grapheme to at most one phoneme. Finally, m-to-n alignments map groups of at least one grapheme to groups of at least one phoneme. Examples of each type are shown in Table 16.

*Table 16: Examples of 1-to-n, 1-to-1, and m-to-n alignments for the word "mixing".*

| m | i | x | i | n | g | m | i | x | - | i | n | g | m | i | x | ing |
|---|---|---|---|---|---|---|---|---|---|---|---|---|---|---|---|---|
| M | IH1 | K S | IH0 | NG | - | M | IH1 | K | S | IH0 | - | NG | M | IH1 | K S | IH0 NG |
| 1-to-n alignment ||||||| 1-to-1 alignment ||||||| m-to-n alignment ||||

Each of these alignment methodologies have distinct advantages and disadvantages, and each is preferable for different techniques. Different alignment methodologies may have varying usefulness for different languages, depending on the target language's morphological characteristics. General pronunciation systems for the Dutch and German languages, due to their consistent morphological structure, can use any of the 1-to-1 or m-to-n alignments with very little relative difference in accuracy – indeed, such systems commonly reach over 85% accuracy in their predictions [61]. However, the English language has many irregularly spelled words, due to various loanwords and linguistic influences from other languages, meaning that 1-to-1 predictive systems can only reach around 60% accuracy. By using an n-to-m alignment in English we see a notable improvement, bringing the accuracy closer to 65%.

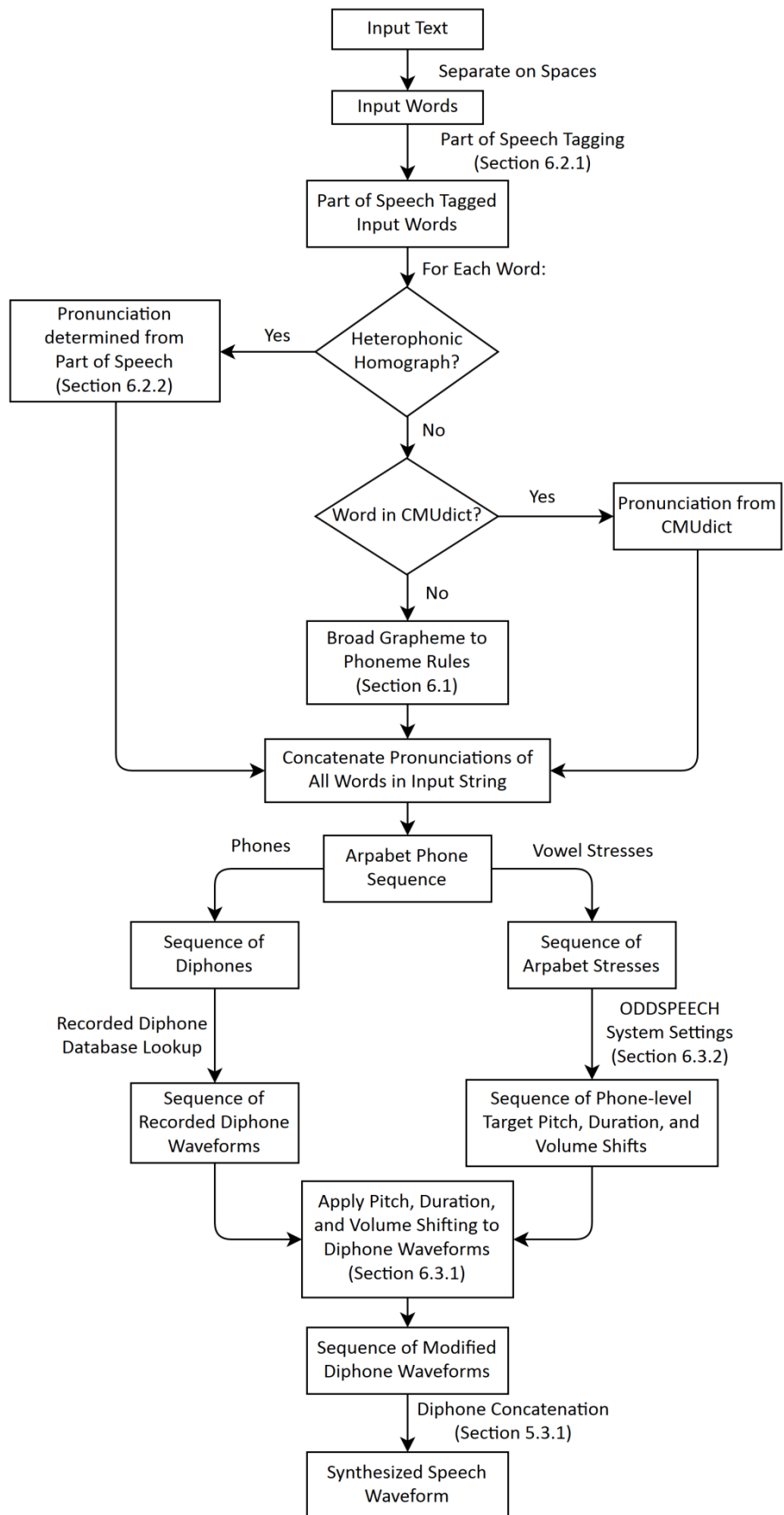

*Figure 33: ODDSPEECH Software Flow Diagram*

Performing an m-to-n alignment is also referred to as **co-segmentation**, as there are the same number of grapheme and phoneme groups. Each grapheme-to-phoneme correspondence in this kind of alignment is referred to as a **graphone**. m-to-n alignments are usually used for a family of techniques called Joint-Sequence models. In these models, the probability of a certain grapheme cluster corresponding to a particular phoneme cluster is determined based on statistical analysis of a known dataset. To consider an example (with these percentages taken from our completed implementation), of the times we see the grapheme combination "abs" in CMUdict, it most frequently corresponds to the phonemes "AE B S", and does so about 44% of the time.

For a given input word, we consider all possible graphone decompositions of the word. As each graphone has an associated probability, each decomposition has a joint probability associated with it. Continuing our example, for the word "absolve", we might break it into the two phoneme clusters "abs" and "olve"; we will denote this joint decomposition as "abs/olve". We know that the grapheme cluster "olve" corresponds to the phoneme cluster "AA L V" about 73% of the time, and we know "abs" corresponds to "AE B S" 44% of the time, so the decomposition of "abs/olve" into "AE B S AA L V" occurs with a confidence of 0.73 x 0.44 = 32.12%. On considering all possible decompositions (such as "a/bso/lve" or "ab/sol/ve"), we then pick the decomposition with the highest confidence probability, and use that as our pronunciation.

More advanced implementations of this approach can combine confidence probabilities of separate joint decompositions if they both correspond to the same pronunciation [60]. However, as there are many possible ways of combining the distinct confidence probabilities, and little research data on which should be preferred, this change does not provide a clear advantage over the simpler method.

There are some additions we can make to our dataset to improve the effectiveness of this algorithm. For example, we might note when the letter "x" is at the beginning of a word, it usually corresponds to the phoneme "Z", such as in the words "xenon" and "xylophone". If it is encountered in the middle or end of a word, such as in the words "mixing" or "lax", it most commonly corresponds to the phonemes "K S". In characterising the statistical likelihood of a certain grapheme/phoneme correspondence, we therefore append additional token graphones to indicate the start and end of a word. In this project, we will use the open bracket character, (, to denote the start of a word, and the close bracket character, ), to denote the termination of a word. This is shown in Table 17.

*Table 17: Alignments of the words "mixing" and "xenon" including start and end of word tokens*

| ( | m | i | x | i | ng | ) | | ( | x | e | n | o | n | ) |
|---|---|---|---|---|----|---|---|---|---|---|---|---|---|---|
| ( | M | IH | K S | IH | NG | ) | | ( | Z | IY | N | AA | N | ) |

Adding our start and end of word tokens, we can note that the graphone "(x" corresponds to the phoneme "Z" with 73% confidence, while overall "x" corresponds to "K S" 84% of the time, and "x)" almost always corresponds to "K S", occurring 99.54% of the time in CMUdict. Therefore, the observation made above can later be confirmed by our analysis of the dataset.

Including these tokens also improves the confidence in our previously considered example, "abs/olve"; this decomposition becomes "(abs/olve)". While the pronunciation remains the same, our confidence levels change: where "abs" had a confidence of 44%, "(abs" has a 56%, and where "olve" had a confidence of 73%, "olve)" has a confidence of 100%, as words ending on "olve" always correspond to the phonemes "AA L V" in CMUdict. This brings our joint confidence up to 56% where it was previously around 32%.

Most grapheme-to-phoneme decomposition systems are not concerned with determining the assignment of vowel stresses, as vowel stresses tend not to generalise well to unknown words [62] – indeed, the same word in English can often be pronounced with the stress on different vowels with no change in semantic meaning. As such, the numerical part of the Arpabet transcription system, used to denote vowel stress, will not be included when training our system, or determining the pronunciation of unknown words. Our training set will also not include words in CMUdict which include characters which are non-alphabetic, such as apostrophes or numerals.

Having given some examples of how the intended algorithm will work, we return to the problem of aligning our word level grapheme-to-phoneme database on the level of smaller graphones. Broad solutions to the database alignment problem often use a system which can perform general graphone inference through maximum likelihood estimation [60]. While this provides a robust solution and can be applied to any word-level pronunciation database, the method is very computationally intensive. It was determined that an analysis on CMUdict using this method would take prohibitively long to complete on the available computer hardware – especially if wanting to run the process multiple times to confirm results. Therefore, to expedite the alignment of our database, we will take advantage of some linguistic understanding of the English language.

We wish to perform an initial graphone alignment of our system on a subset of the database, and then use the information learned there to align the rest. We use our knowledge that, in general, clusters of vowel/semivowel graphemes of English (a, e, i, o, u, w, and y) will correspond to clusters of vowel/semivowel phonemes in Arpabet. Similarly, that clusters of consonant graphemes in English usually correspond to consonant phonemes in Arpabet. We also include the letter "r" in our list of vowel/semivowels, as the Arpabet monophthong ER is commonly co-articulated with an "r". Thus, a, e, i, o, u, w, y, and r are in one group of our graphemes, while the other group contains the remaining letters. Similarly, we group together the phonemes AA, AE, AH, AO, EH, ER, IH, IY, UH, UW, AW, AY, EY, OW, OY, W, Y, and R, with the other group containing the remaining phonemes.

Thus, we start by taking the word-level graphemes and phonemes and segmenting them into clusters as described above. If the number of grapheme clusters is the same as the number of phoneme clusters, then we can pair each cluster group to identify an initial graphone decomposition of the word. To see to what extent this rule holds true, we can find what percentage of words are broken down into the same number of grapheme clusters as phoneme clusters. This turns out to be 102685 of our 124568 words, telling us that about 82% of the words in CMUdict conform to this rule. Inspection of the words co-segmented in this way indicates that the segmentations are accurate, but not minimal; for example, in Table 18, we can see that the "nj" to "N J" and the "mpt" to "M P T" graphone clusters could be broken down further, as each individual grapheme clearly corresponds to exactly one phoneme.

*Table 18: Initial co-segmentations of the words "inject" and "empties".*

| i  | nj   | e  | ct  | e  | mpt   | ie | s |
|----|------|----|-----|----|-------|----|---|
| IH | N JH | EH | K T | EH | M P T | IY | Z |

On inspection of the words which this initial alignment did not work for, we find that they are typically those with silent vowel letters, such as the "e" at the end of the word "ate", which decomposes into "EY T". The grapheme grouping for this word gives us three distinct groups, while our phoneme grouping gives us only two; as such, we cannot successfully create an initial alignment.

We start our analysis with the words we successfully co-segmented. First, we count the number of times that each graphone appears. We also note the number of times that larger graphones, containing multiple smaller graphones, appear within the dataset. These larger clusters of graphemes contain greater context, giving correspondences which may have a higher percentage likelihood than the component graphones. Due to the constraints of physical computing, we must place a limit on the number of graphemes that can appear within each recorded graphone, or it takes a prohibitively long time to index our entire dataset. Here, we only record graphones with 4 or fewer corresponding graphemes. We also do not include the start or end of word tokens in our graphones at this point.

In counting these, we group together graphones which correspond to the same group of graphemes; for example, we group together all possible pronunciations of the grapheme cluster "hen". We then remove all pronunciations which occur less than 10% of the time relative to the most common pronunciation; this eliminates decompositions which are likely to be incorrect. Using this, we analyse our previously existing graphones, and see if we can further break them down.

To demonstrate how this procedure works, we use the example graphone between "mpt" and "M P T" as in the word "empties" from our initial alignment in Table 18. We take the largest grapheme cluster from the end of the word that does not contain the entire group – here, that is "pt" – and see if the end of our phoneme cluster "M P T" corresponds to one of the known pronunciations of "pt". Since we have seen elsewhere that "pt" is pronounced "P T", we can break the graphone up into two. This process is repeated for the "pt" to "P T" phoneme, breaking the graphone down into the minimal possible units. If we reduce a section of our graphone such that it contains only one grapheme or one phoneme, it cannot be further reduced, so we consider it completed. Once we have finished attempting to co-segment from the end of the graphone, we attempt to co-segment from the beginning; if neither can result in further decomposition, we consider the process complete. The result of this alignment on the words "inject" and "empties" is shown in Table 19.

*Table 19: Minimal co-segmentations of the words "inject" and "empties".*

| i | n | j | e | c | t | | e | m | p | t | ie | s |
|---|---|---|---|---|---|---|---|---|---|---|----|---|
| IH | N | JH | EH | K | T | | EH | M | P | T | IY | Z |

Using the same methodology, we can estimate the alignments for the words which could not initially be co-segmented. This allows us to identify new graphones which do not conform to the alignment of vowel-to-vowel and consonant-to-consonant, for example, we can now identify the "se" to "Z" graphone and the "le" to "AH L" graphone in Table 20, which we could not previously have isolated. This algorithm is performed on the remaining 18% of words which did not conform to our initial alignment rule.

*Table 20: Co-segmentations of the words "dispose" and "crumple".*

| d | i | s | p | o | se | | c | r | u | m | p | le |
|---|---|---|---|---|----|---|---|---|---|---|---|----|
| D | IH | S | P | OW | Z | | K | R | AH | M | P | AH L |

We have now co-segmented every alphabetic word in CMUdict, and can use this alignment to calculate our decomposition probabilities for use in the joint sequence model. In this count, we now want to include start and end of word tokens, so that we capture the separate probabilities of graphone correspondences when they occur at the start or end of words. As before, we must include a limit, so we include graphones containing up to 4 graphemes, or 5 if they contain a start or end of word token, and then count all of the graphones up to this size within each word in our database. We also never include graphones with both a start and an end of word token in them, since that contains the entire word: if an input word matches such a graphone, the pronunciation is already known to our lookup dictionary, so we should never need to estimate its decomposition.

For each grapheme cluster found, we record the different phoneme clusters which it can correspond to. After completing this count, we identify the most frequently occurring graphone for those graphemes, and record the percentage of the time that it occurred relative to the other pronunciations. Finally, we are left with a lookup table of graphones, where for a given sequence of letters, we are given a corresponding pronunciation and confidence percentage.

With this completed, we implement the previously discussed joint sequence model. For an input word, we create an array of all possible grapheme groupings (such as "(ab/sol/ve)" or "(abso/lve)"), calculate the corresponding confidence percentage for each, and then select the one with the highest likelihood as our pronunciation.

On inspection of the output produced by the system, this approach has some minor problems. First, we occasionally select a decomposition which repeats the same phoneme in sequence. For example, our decomposition for the word "paddle" is "pad/dle", corresponding to "P AE D D AH L"; in the correct pronunciation, we only have the one "D" phoneme, while our joint sequence model gives us two. Again, returning to our linguistic understanding, we know that in English pronunciation the same phone cannot occur twice sequentially (or alternatively, from a data-driven perspective, we know that this behaviour does not occur within CMUdict). As such, this can be fixed by removing repeated phonemes in our determined pronunciation.

Another problem is that some graphones within our group will never be used in decomposition. For example, the most likely graphone corresponding the grapheme "arms" has 47% certainty, while decomposing it into "arm/s" gives us a 58% certainty. As such, when the "arms" grapheme occurs in decomposition, it will never be used, as a grapheme decomposition into "arm/s" always has a higher confidence percentage. Running a program to find graphones where this is the case, we find that 16514 of our 98366 previously determined graphones are redundant, or approximately 16.8%. These graphones can be removed from our database to reduce its size, and improve our operation speed as we no longer consider graphones which will never be selected.

A more significant problem is that this approach doubles the required memory and processing time for each letter added to the word we wish to decompose. This is because between each pair of letters within the word, we can choose to either split the word at that point or not. This means that if the number of letters in our word is L, the number of possible decompositions of that word is $2^{L-1}$. As such, the approach resolves in exponential time.

A word with twelve letters in it takes approximately one second to complete using this method. This means that particularly long words will take an excessive amount of time, such as the 21-letter word "incomprehensibilities", which we would need over eight and a half minutes to find a pronunciation for – as well as using over 350 megabytes of memory. This is unacceptably slow and inefficient for general use.

The reason for this is that we calculate total joint probabilities for all possible decompositions, store them all in memory, and then choose the most likely one. Fortunately, we can reformulate how we implement this solution to make it more computationally simple to solve, turning it into a compact graph theoretic structure rather than a large, branching tree. If we consider each point between two characters in an input word to be a node on a graph, then we can create edges between nodes such that each edge represents a possible subsequent grapheme cluster.

This is illustrated in Figure 34 for the word "hello", including its start and end of word markers. We can see that Node 1 is the start of the word, while Node 6 is the end of the word. Each possible path from Node 1 to Node 6 corresponds to a possible grapheme grouping of the word, with each node along the pathway corresponding to a split in the word. For example, the path (1,3,6) corresponds to the grouping "(he/llo)". If each edge is weighted by the probability of its corresponding graphone's confidence, then if we multiply along the path as we move through, we can determine the overall probability.

As such, the multiplicative path from Node 1 to Node 6 with the largest probability is the most likely path, and therefore the one we choose. Unfortunately, due to having to multiply rather than simply add along the path, this is still computationally difficult. Fortunately, it is possible to further adjust our model to turn this into a shortest path problem.

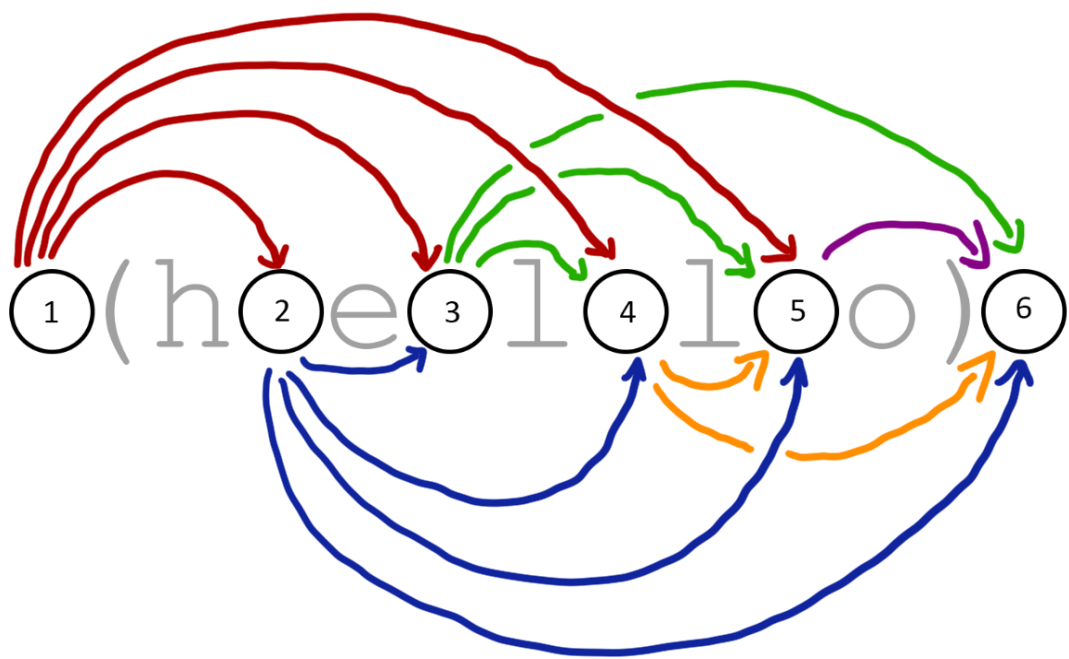

*Figure 34: Directed graph corresponding to decomposition of the word "hello". Weightings not pictured.*

We know that addition in logarithmic space corresponds to multiplication in real space; as such, if we get the logarithm of each percentage and add along the path, then the maximum total path length from Node 1 to Node 6 will correspond to the maximum likelihood. We can also note that since these percentages are always less than or equal to 1, the logarithms of these percentages will always be less than or equal to zero, meaning all edge weights are negative or zero. If we instead use the negative of the logarithm of the percentages, then all edge weights will become positive, and the preferred decomposition corresponds to the shortest path from Node 1 to Node 6.

Shortest path problems are easy for computers to solve, with various algorithms existing which are optimised for use with different types of graphs [63]. Since version R2015b, MATLAB includes native support for graph theoretical structures, and so there is a built-in function to do this for us. We can improve efficiency by noting that the graph produced for any given word is a directed, acyclic graph; that is, our edges are directional, and there is no way to return to a node that has previously been visited by moving along edges. Knowing this is the case, we can tell MATLAB to use an algorithm with better performance than the more general Dijkstra algorithm it would use by default.

With this reformulation, for each new letter in our input, we only need to add one new node and at most 4 new paths to our graph, meaning that the complexity of the algorithm is bounded above by 4L. We are now solving the problem in linear time rather than exponential, making its operation substantially faster: we can now determine a pronunciation for the word "incomprehensibilities" in 5.927 milliseconds rather than several minutes, which is a far more acceptable speed.

With the efficiency of our grapheme to phoneme algorithm now greatly improved, we should attempt to determine its accuracy. To do this, we use the algorithm to find pronunciations for all of the alphabetic words in CMUdict, and count the number of times that our determined pronunciation is the same as the pronunciation given. We know that our system will not be able to identify the correct pronunciation all of the time, as it is based simply on the most common pronunciations. As such, many pronunciations will have minor errors which do not reduce intelligibility when synthesized. We therefore also tally minor errors; with One Off referring to when only one of the phones in our pronunciation was incorrect, Missing Phone meaning that our pronunciation was missing a phone, and Extra Phone meaning our pronunciation had an additional phone.

*Table 21: Accuracy of our Grapheme to Phoneme algorithm in CMUdict.*

|  | Correct | One Off | Missing Phone | Extra Phone | Incorrect | Total |
|---|---|---|---|---|---|---|
| Count | 57809 | 25932 | 4395 | 7096 | 29336 | 124568 |
| Percentage | 46.41% | 20.82% | 3.53% | 5.70% | 23.55% | 100.00% |
|  |  | 30.04% |  |  |  |  |

The results of this count are shown in Table 21. We can see that our system finds a perfectly accurate pronunciation of a word 46.41% of the time, with 30.04% of pronunciations only containing minor errors, and 23.55% of pronunciations being incorrect to a greater degree. Overall, our system has a lower accuracy than might be desired, as many modern English-language systems can achieve above 65% accuracy on their training sets [61]. This can be partially attributed to CMUdict's database including the pronunciations of some abbreviations and acronyms, which interfere with the more phonetically normal data that our system is trying to predict pronunciations from. It is also probable that our initial alignment of the data was inaccurate in some places, due to our assumed decomposition rules not always generalising. We could also likely achieve greater accuracy by using a larger maximum grapheme cluster size, but this in turn would increase required memory and reduce speed.

We should again note that the technique described in this section can be generalised to any language and any phoneme set. If we already have a database of m-to-n alignments between words and phonemes, we can determine a maximally likely pronunciation of words in that language which are not in the original dataset. For languages conforming to regular phonological rules, this approach would find decompositions with greater accuracy. Unfortunately, English has been altered and added to by the influences of many other languages, resulting in pronunciation irregularities that make the problem more difficult.

For our purposes, the level of accuracy we have achieved should prove sufficient. First, we are only using this method for when we cannot find our word in CMUdict, which already contains the pronunciations for most English words; as such, this procedure will be used infrequently, making its relative inaccuracies less apparent. It is also likely that many of the minor errors tallied in Table 21 are insufficient to alter word-level intelligibility when spoken within a sentence. If we assume that a listener can still recognise a word which only has these minor errors in pronunciation, then our system should be able to intelligibly produce about 76% of unknown words. By using this algorithm, ODDSPEECH is now capable of attempting a pronunciation for any word composed of alphabetic characters, making it far more versatile in general usage than BADSPEECH was.

## 6.2. Heterophonic Homograph Disambiguation

Homographs are words which are spelt in an identical way but mean different things depending on context. This is distinct from homonyms, which are single words with multiple meanings but only one pronunciation, and homophones, which are different words which are pronounced the same. As seen in Figure 35, a heteronym is a homograph which is not a homophone, and a heterograph is a homophone which is not a homonym. However, in the literature of speech synthesis, the term homograph is often used to refer exclusively to heteronyms and other heterophonic homographs; we shall adopt this usage from here onwards.

Homonyms, always being spelt and pronounced the same way, are of no concern to us in designing text to phoneme rules: they retain a one-to-one correspondence between their encoding in graphemes and their pronunciation in phones. Heterographs can cause problems in speech recognition, especially for automatic transcription; from input speech, the correct transcription of a word can be unclear if there are multiple possible written words that the phones produced could correspond to. For speech synthesis, homophones present no issue.

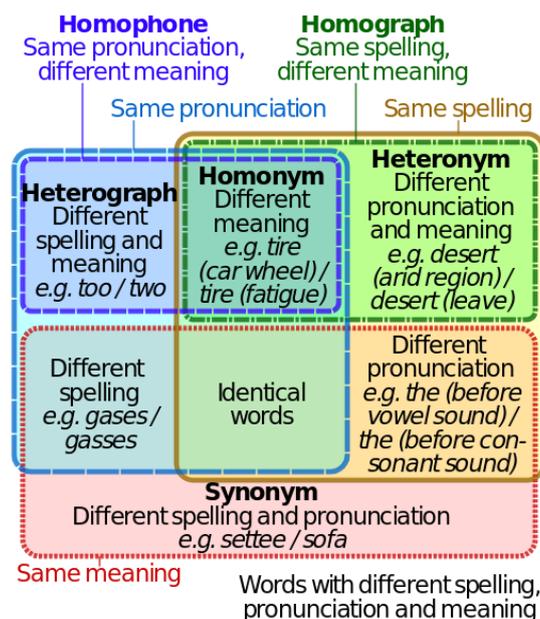

*Figure 35: The difference between Homographs and Homophones [64]*

Homographs are the inverse problem to homophones, and are the primary category which concerns us in text to speech synthesis [65]. For homographic input words, there are multiple possible pronunciations; it is unclear without context which pronunciation is correct. While with our development of BADSPEECH we ignored this problem, simply choosing the pronunciation that occurred first in order within CMUdict, for ODDSPEECH we will attempt to determine the correct pronunciation of a homographic input word based on context.

For some of these, determining the correct pronunciation is relatively straightforward. For example, in the word "the", we can determine the correct pronunciation based on whether the sound immediately following it is a vowel or a consonant. This is simple to implement, as we can simply find decompositions for the other words within the sentence, and then determine the preferred pronunciation of "the" from the following phone.

With other words, determining the correct pronunciation is more difficult; some examples of such words are shown in Table 22. For example, the word "project" as used in the phrase "We will undertake this project" is pronounced with different vowel stress from in the phrase "You need to project your voice better". The former pronunciation is used when the word is being used as a noun, while the latter is for when the word is being used as a verb.

Similarly, in the sentence "I dove towards the dove", the word "dove" is pronounced differently when referring to the bird as a noun, or the action of diving as a verb. For many words, if we can determine that word's lexical category, then we will be able to correctly determine the pronunciation of the word.

*Table 22: Words with multiple possible pronunciations in CMUdict.*

```
PROJECT    P R AA1 JH EH0 K T
PROJECT(1)   P R AA0 JH EH1 K T
OBJECT   AA1 B JH EH0 K T
OBJECT(1)   AH0 B JH EH1 K T
LEAD   L EH1 D
LEAD(1)   L IY1 D
DOVE   D AH1 V
DOVE(1)    D OW1 V
```

## 6.2.1. Part of Speech Tagging (POS Tagging/POST)

Part of speech tagging aims to identify the lexical categories of words within a sentence. For many words, this is a simple task; a particular grapheme combination might only corresponds to one particular word meaning, so we can simply consult a database of these correspondences to identify such words. For example, the word "bird" is always a noun.

In establishing a part-of-speech tagging system, homonyms are an issue: the same homonym, even if it is pronounced identically, can be a member of multiple word categories depending on context. For example, the word "walk" can be a verb, as in "I walk every morning", or a noun, as in "I took a walk". While these words are pronounced identically, meaning that we do not need to disambiguate their pronunciation, it will be useful to know which lexical category a word is a member of in the future, as we will wish to overlay prosody on sentences in later iterations of the speech synthesizer. Thus, at this stage, we aim to identify the lexical category of every word within a sentence, even if we do not need to use that data to determine its correct pronunciation.

There are several different categorisation systems for tagging lexical categories. For example, the Brown University Standard Corpus of Present-Day American English, developed in the 1960s, uses 79 distinct lexical categories and 6 distinct tags for grammatical elements such as brackets, colons, commas and dashes [66]. Many of these are categories including only one word, such as "am" or "do"; others are specific conjugations or categorisations of verbs or adverbs. While this specificity may be suitable for some natural language processing tasks, for our purposes this is needless complexity when simpler solutions suffice.

The University of Sheffield's Institute for Language Speech and Hearing makes available various word lists as part of the Moby Project. A sub-project within this, Moby Part-of-speech (or MPOS), specifically lists words and their corresponding lexical categories. MPOS only uses 15 different lexical categories, and includes all categories that a word may belong to, as illustrated in Table 23.

*Table 23: Lexical Categories used in MPOS, and sample entries for certain words*

| Noun | N |
|---|---|
| Plural | p |
| Noun Phrase | h |
| Verb (usu participle) | V |
| Verb (transitive) | t |
| Verb (intransitive) | i |
| Adjective | A |
| Adverb | v |
| Conjunction | C |
| Preposition | P |
| Interjection | ! |
| Pronoun | r |
| Definite Article | D |
| Indefinite Article | I |
| Nominative | o |

```
articulate×AVti
custodial×AN
Dublin×N
keg×N
lottery×N
read×VtNiA
thought×VN
zip×Nit
```

It should be noted that in MPOS certain phrases larger than a single word can be tagged as a singular part of speech; for example, the three words in "a la mode" can be tagged together as an adjective(A), and the term "mandelic acid" can be tagged as a noun phrase(h), where both words identify the noun. For now, however, we will not be grouping these words together.

To start working on part of speech tagging, we will design a function which expects some number of space-separated words in the form of a sentence as its input. This function should then output a cell array where one of the rows contains the words (or groupings of words) in the sentence, and the other contains the possible lexical categories corresponding to each word (or grouping) as found in MPOS. If a word exists in the sentence that is not found in MPOS, we will assign it to a sixteenth category of Unknown Words, denoted by a question mark (?).

Before implementing this, we must address a property of the MPOS list, that being that entries are case-sensitive. For example, the word "Ring", when the first letter is capitalised, has an entry N, but the word "ring" in all lowercase has entry NVt. This can introduce problems; for example, a sentence such as "Ring the bell" capitalises the word "ring", but it is used as a verb, not a noun. Thus we will pre-process MPOS to identify redundant entries so that only a single, case-insensitive entry remains.

Thus, we search through the list finding entries with the same spelling but differing case. If the entries have the same part of speech (such as with the entries "Yen" and "yen", both being nouns) then we remove the first entry. If one of the entries' part of speech categories are entirely a subset of the other (as with "Ring" and "ring" previously, where "ring" already includes N), then we remove the entry that is a subset of the other. If both entries have distinct elements not in the other (such as in the entry for "Wobbly", as a noun (N), and "wobbly", as an adjective(A)), then the elements not overlapping are added to the latter entry (the decision of which entry to keep is arbitrary). When this methodology was implemented and run, it identified and fixed 4607 redundant entries.

After this has been completed, the cell array containing the text corresponding to MPOS is turned into a hash table for increased efficiency, using the uppercase versions of the entries as keys. Now we can simply retrieve our part of speech data by using a call of the form *posMap('WORD')*, which returns the corresponding entry for WORD. If there is no corresponding entry; that is, if the word is not in the MPOS list, an error is thrown by this process, and we instead assign a question mark (?).

To illustrate how this version operates, we will use sentence Number 8 of List 24 of the Harvard Psychoacoustic Sentences as an example:
**Jerk the rope and the bell rings weakly.**
This is input to our system as space-separated words without the period at the end. The output is a cell matrix as shown in Table 24.

*Table 24: Initial POS Tagging of "Jerk the rope and the bell rings weakly".*

| JERK | THE | ROPE | AND | THE | BELL | RINGS | WEAKLY |
|------|-----|------|-----|-----|------|-------|--------|
| VtN  | Dv  | Nti  | CN  | Dv  | NVt  | ?     | Av     |

From this example, it is clear that we cannot simply use the MPOS list to break a sentence down into its parts of speech: none of the words in the sentence were reduced to a single, definitive lexical class, and the word "rings" was not even found in the MPOS list. We could include some general conjugation rules to widen the scope of MPOS; for example, the word "kegs" is not included, while "keg" is. Since we know that "keg" is a noun(N), and the general rule of addending the suffix -s to a word makes it plural, we would know the word "kegs" is likely a plural noun(p). Similarly, the word "ring" has possible classes NVt, resulting in the conjugation "rings" giving the possible classes pVt.

While implementing something like this would in many cases be useful, it is a linguistics-focused solution, and requires a large number of hard-coded possible conjugation rules, where we would (as engineers) prefer a data-driven solution to the problem. Instead, we wish to identify the most likely lexical class based on the context of the possible adjacent lexical classes. This can be done using a probabilistic approach based on a large corpus.

If we have a large corpus of data with annotated part of speech information, we can determine the frequency of certain groupings of part-of-speech elements. For example, adverbs should occur more frequently next to verbs when used within a sentence, and similarly adjectives should occur more often next to nouns. As such, we wish to determine the frequency with which words in certain lexical classes are adjacent to each other. The term for this technique is an n-gram model, where n is the number of items in our group; in particular, the 2-gram case is referred to as a bigram model, and the 3-gram case is referred to as a trigram model.

As an example, here we illustrate how a bigram model can help us to determine the lexical class of words within a sentence. We can consider a subset of the previously used sentence, including only the phrase "jerk the rope". This is shown again in Table 25.

*Table 25: Initial POS Tagging of "Jerk the rope".*

| JERK | THE | ROPE |
|---|---|---|
| VtN | Dv | Nti |

A bigram model can incorporate the relative probabilities of each word's lexical class being adjacent to another. We need to consider all possible assignments of lexical class: we have three possible lexical classes for the words "jerk" and "rope", while the word "the" can only belong to two. As such, there are 2*3*3 = 18 possible lexical class assignments for the entire sentence. To compare which is most likely, we compare the probabilities of each combination; as this example is with a bigram model, we do this in groups of two words each.

We will initially consider the probability of the lexical classes being, in order, V/D/N. First, we look at the first pair of words, "jerk" and "the", given the assignments V/D. We already know the number of times that the V/D combination occurs within our corpus. We then consider the number of times that the other possible lexical bigram assignments for these two words occurred in our corpus; those being V/v, t/D, t/v, N/D, and N/v. We sum together the number of times that these occurred with the frequency of V/D, giving the number of times that any valid assignment of these words occurred in our database. Finally, we can turn this into a percentage, by dividing the number of times V/D occurred by this total found. This gives us the statistical likelihood of the first two words being assigned as V/D.

We then must consider the likelihood of the next assignment in a similar way; though here, we treat the previously found assignments as invariant, and only consider the possible assignments for the following word. We want the probability that, given a fixed assignment of D to the word "the", our assignment of N to the word "rope" is correct. This is done in the same way as the first bigram, but only considering variation in the latter assignment: we consider the frequency with which D/N occurs in our corpus, and divide by the sum of the number of times that D/N, D/t, and D/I bigrams occur within our corpus, so that we find a percentage value.

Once we have the percentage chance of V/D being assigned, and the probability of N being subsequently assigned, we can multiply these two values to find the total likelihood that the overall assignment of V/D/N is correct. Then, we wish to find the probability of each of the other 17 possible assignments being correct, and choose the one with the highest probability. However, the computational complexity of a naïve implementation of this algorithm increases exponentially with the introduction of additional words. Fortunately, we can use a similar graph-theoretical method to the one we have used for grapheme to phoneme decomposition, so that the complexity of the algorithm increases linearly instead.

A trigram model works in a similar way to the bigram model; however, rather than finding the probability of the next word's lexical class from the lexical class of the word immediately prior, a trigram model uses likelihoods based on two prior values. A trigram model can give more accurate results than a bigram model, as it incorporates more of the sentence's prior context. It also uses more nodes in the graph theoretic structure, since each position within the sentence corresponds to a number of possible states equal to the number of possible assignments to the previous two words. However, as there are a finite number of possible states, there are a maximum number of possible combinations of two states, so the approach is still bounded above linearly.

Now that we understand how this approach works, we will consider how to implement it. To do this, we will need a dataset of n-grams as training data. Brigham Young University makes available data for the million most commonly occurring n-grams in the Corpus of Contemporary American English (COCA) with part of speech tags already applied [67].

This dataset has the capacity to be extremely useful to us; however, the part of speech tags used are the CLAWS7 tagset rather than the simplified dataset used in MPOS. The CLAWS7 tagset uses 137 distinct tags; we wish to map these to the 14 distinct tags which are used in MPOS so that the data is useful (the tag h, identifying an entire noun phrase, will not apply, as CLAWS7 tags on the single word level). However, CLAWS7 makes no distinction between transitive and intransitive verbs (the tags t and I respectively); as such, we will need to consolidate all such verbs under the singular verb identifier, V. Similarly, the nominative case of a noun (denoted as o in MPOS) is not separately identified in CLAWS7, so will be consolidated under the primary noun identifier N. Finally, we must consolidate the less common indefinite article, I, into the grouping of definite articles, D, to be able to create a meaningful, surjective mapping from the CLAWS7 tagset to our own tagset. This gives us a reduced tagset of 11 tags and one unknown, as shown in Table 26. The table of conversion from CLAWS7 tags to our tagset is in the Appendix on Page A-17.

*Table 26: Lexical Categories used in ODDSPEECH*

| Noun | N |
|---|---|
| Plural | p |
| Noun Phrase | h |
| Verb | V |
| Adjective | A |
| Adverb | v |
| Conjunction | C |
| Preposition | P |
| Interjection | ! |
| Pronoun | r |
| Article | D |
| Unknown | ? |

After converting the CLAWS7 tags to our custom tagset, we have approximately one million rows in our table of information, in the form shown in Table 27. As we are only interested in the counts for each part of speech 3-gram, we can consolidate this information into $10^3$, or 1000 rows of data (since CLAWS7 doesn't map to h, and trigrams including ? are not useful), each with a corresponding frequency count. The form of this consolidated data is shown in Table 28.

*Table 27: Initial rows of COCA Corpus particular trigram data*

| Frequency | Words | | | CLAWS7 Tags | | | Custom Tagset | | |
|---|---|---|---|---|---|---|---|---|---|
| 48 | *a* | *B.A.* | *degree* | at1 | nn1 | nn1 | D | N | N |
| 56 | *a* | *B.A.* | *in* | at1 | nn1 | ii | D | N | P |
| 41 | *a* | *B.S.* | *in* | at1 | np1 | ii | D | N | P |
| 33 | *a* | *BA* | *in* | at1 | nn1 | ii | D | N | P |
| 28 | *a* | *babble* | *of* | at1 | nn1 | io | D | N | P |

*Table 28: Initial rows of consolidated COCA Corpus lexical trigram data*

| 3-gram | Count |
|---|---|
| NNN | 289770 |
| NNp | 66052 |
| NNV | 142513 |
| NNA | 17081 |
| NNv | 38436 |
| NNC | 198507 |
| NNP | 290077 |
| NN! | 215 |
| NNr | 21543 |
| NND | 7103 |

Now that we have our data from MPOS to give us a list of the possible lexical categories that a word might be a member of, and n-gram data for predictions from COCA, we have the datasets required to construct an n-gram model to determine part of speech within a sentence. Before continuing, we will further discuss the n-gram technique which was previously outlined.

This technique is an example of a Hidden Markov Model, or HMM. In a HMM, we assume that the underlying system is a Markov process. This is a system where the likelihood of a transition from one state to another is entirely determined by the current state. Here, our "current state" is the lexical classes of the last two words in order and the possible lexical classes of the following word, and our possible future states are the most recent word paired with the prediction for the next word. The probabilities of moving from one state to another are defined by our COCA corpus data. If, for example, we had a Noun and a Noun, and the following word could possibly be either a Noun or a Verb, we can look at Table 28 to see that NNN occurs 289770 times, while NNV occurs 142513 times in the corpus; together, these sum to 432283. Therefore using this method, we predict the probability of the next word being a noun is 289770/432283, or about 67%, leaving the probability of the next word being a verb as 33%.

In our HMM, we observe the probabilities of each path through our data, and then find the maximally probable path. This is the "hidden" component of the Hidden Markov Model: we have the complete sentence as the "output" of the Markov process, and wish to determine the underlying sequence of system states that it passed through. Those states then correspond to the sequence of lexical classes through the sentence. This technique is known as a Most Likely Explanation HMM, as it gives the most likely state sequence for the overall sentence. While this is similar to the technique we used in grapheme to phoneme conversion, in that system each grapheme cluster could only possibly correspond to one graphone. A HMM model would include multiple possible graphones for each grapheme cluster, and find the maximum likelihood based on adjacent clusters. While this would offer improved effectiveness, it drastically increases database size.

Returning to part of speech tagging, we must consider what to do if a sentence is smaller than three words long. We know the data in MPOS gives us the most likely lexical class first in the order of possible lexical classes; if our sentence is one word long, we simply assume the lexical class to be the first in the list. If we have a sentence with two words in it, we can determine the most likely sequence using bigram data which is also drawn from COCA.

We must also consider how to handle unknown words, currently tagged as "?". To them, we instead assign the possibilities "NpVav!". These categories are what are known as open classes, meaning that new, unknown words can be a member of them [68]. Closed classes, meanwhile, are classes which new words cannot be a member of. Conjunctions, prepositions, pronouns, and articles are all closed classes in English; they are composed of connective or relation-describing words such as "the", "to", and "me", which there a known and finite number of. However, it is always possible to add a new word to an open class and use it meaningfully by analogy, since these open classes communicate new semantic content. Closed classes in English mostly perform grammatical functions.

With these final adjustments, we are able to estimate the most likely part of speech assignments for an input sentence. Some example sentences and their assignments are shown in Table 29.

*Table 29: Examples of part of speech assignments from ODDSPEECH.*

| I | WANT | TO | PROJECT | MY | PROJECT | ONTO | THE | WALL |
|---|------|----|---------|----|---------|------|-----|------|
| r | V | v | V | D | N | P | D | N |

| I | DOVE | TOWARDS | THE | DOVE | TO | CATCH | IT |
|---|------|---------|-----|------|----|-------|----|
| r | V | P | D | N | P | N | N |

| I | OBJECT | TO | THE | USE | OF | THIS | OBJECT |
|---|--------|----|-----|-----|----|------|--------|
| r | V | v | v | V | P | D | N |

| THE | DOVE | DOVE | AWAY | FROM | ME |
|-----|------|------|------|------|----|
| D | N | N | N | P | N |

We can see for the first three sentences used, the correct lexical classes for the different uses of the words "project" "dove", and "object" were determined. Using this information, we could determine their correct pronunciations in those scenarios. However, in the last sentence, we can note that incorrect lexical class assignments are given; both uses of the word "dove" are assigned nouns, where one should be considered a verb.

It is important to note that the approach we are using is fundamentally limited. We are exclusively attempting to find the lexical class of individual words based on context, without considering the semantic relationship between words in a sentence. Our approach gives us enough information to estimate a word's lexical class (with some inaccuracies), but it is an insufficient method to derive meaning from a sentence; because of this, our system does not recognise if certain assignments are nonsensical, or understand what is being communicated.

To use an example, in the sentence "The man at the end of the street is angry", our tagging system can identify the word "angry" as an adjective, but there are multiple nouns in the sentence – house, end, and street – and our system does not tell us which the adjective is referring to. More advanced natural language processing systems can determine specific information about a query, or determine properties of an object as described in a sentence.

Another advantage such a system would have over our own is that in some cases knowing the lexical class of a word is insufficient to identify its pronunciation. Taking an example from Table 22, the word "lead" can either be pronounced as "L EH1 D" or as "L IY1 D". In Table 30, we can see that the word can be pronounced in either fashion, but that lexical class alone is insufficient to determine the pronunciation. For verb usage, we would need to determine the intended chronological tense; this is something which could be done by determining the tense of known words, such as "was" or "to", and deducing the intended tense from context. While this would require the use of a system with additional information on words and an expanded tagset, the problem could be solved with a similar approach (using bigram or trigram likelihoods) to what we have already used.

*Table 30: Different pronunciations of the word "lead".*

| Sentence | Lexical Class | Pronunciation |
|---|---|---|
| I was in the lead. | Noun | `L IY1 D` |
| The ball was made of lead. | Noun | `L EH1 D` |
| I dropped the dog's lead. | Noun | `L IY1 D` |
| I was lead off the path. | Verb | `L EH1 D` |
| He helped to lead me. | Verb | `L IY1 D` |

For usage of the word "lead" as a noun, however, we would need to use a more advanced technique. Rather than using lexical class trigrams, we could use word level trigrams; for example, if we see the word "lead" after the words "the dogs", we might know that the phonemes "L IY1 D" most commonly correspond to it. This would still not be perfect, for example, in the sentence "The dogs lead me to my destination", this trigram approach would still determine an incorrect pronunciation.

The database used for this would also need to index every word-level trigram in text to a corresponding pronunciation; no such database seems to exist. Creating such a database would be exceptionally time-consuming, as a large corpus of written text would have to have its corresponding word-level pronunciations manually transcribed. (This process could not be automated, since a system which can identify the pronunciations is already performing the task we wish to solve – if we had a computer that can do it for us, we would already have a solution!)

Another alternative approach is to determine not only the lexical class of a word, but its underlying meaning, associating it with a more concrete object from which we can define a specific pronunciation. A system which can identify semantic meaning could also use that information to define more useful prosodic behaviours. From the sentence "The man at the end of the street is angry", we could determine that "angry" is an adjective relating to "man", and could emphasise those words, as they are central to the semantic meaning being communicated by the sentence. The nouns "end" and "street" are less semantically important, and such a system could therefore put less stress on them.

Large databases such as the Prague Dependency Treebank [41] describe the dependency structure within a large number of sentences, which can be used to train a system. Where this helps to determine a sentence's linguistic and grammatical structure, semantic meaning can be derived using a resource such as the public domain-licensed Wikidata, which indexes over 20 million objects as structured data [69]. For example, in its entry for "Norway", with index Q20 in the database, there are entries for its population, head of state, and even IPA transcription of the word in different accents. Many of these entries are themselves objects in the database; for example, we know Norway's current monarch is Harald V, who occupies index Q57287, from which we can find his various relatives, awards he has received, his place of birth, and so on. This gives a vast, heavily interconnected semantic network, which can be extremely useful for many natural language processing tasks.

While ideally we could use these approaches in this project, they introduce substantially greater computational complexity, requiring much larger datasets, and more advanced models to determine these deeper properties of sentences. Such implementations could take months or years for a large cross-disciplinary team to complete. Further, while we would find an improvement for our system in specific instances, most of this information would not be useful to our objective of producing a speech synthesis system. As such, the simpler HMM approach has been used for this project.

Unfortunately, unlike with word decomposition, we cannot test the accuracy of our procedure on the same dataset that it was trained on. The COCA database is only searchable via web interface, and while the n-gram data from it is available for download, the database itself is not. As such, we shall test the data on the Brown corpus. Similar to our reduction of the CLAWS7 tagset, we need to reduce the Brown tagset to our own limited tagset (the table for this conversion is in the Appendix on Page A-19). Then, we split the Brown corpus up at the sentence level, and apply our tagging methodology to it. Finally, we compare the manual tagging to our automatic system.

This process gives us an overall word-level accuracy of 53%. Relative to modern systems, this accuracy is very poor, as they often accomplish word-level accuracies of 97% or higher [70]. This poor result can be partially attributed to having eliminated most of the detail in the low-level connective tags to make our system work at a reasonable speed in MATLAB. We should also note that our initial tagging uses MPOS, our trigram data is from COCA, and we have evaluated the system effectiveness with the Brown corpus: all of these use different tagging systems. Our mapping from each system to the simplified tagset has likely compounded the errors in the system.

To try and improve this, we introduce an additional step to our system. Prior to the construction of our graph, we check to see if any specific word (rather than lexical) trigrams within the sentence exist in the COCA corpus trigram data. If they do, we define those part of speech assignments as-is, not permitting any alternative assignments. While this introduces some additional computational demand and requires more system memory, it also improves our tagging accuracy to 76.8%. This is still quite low relative to the state of the art; however, we are still only including the million most common specific trigrams in the COCA. Including further trigrams could push this percentage even higher, but the larger COCA trigram databases are not freely available.

While this degree of accuracy leaves much to be desired, our objective for this project is not to manufacture an outstanding part of speech tagger, but to implement an effective speech synthesizer, of which part of speech tagging is only a small section. Additional time spent on this particular part of the problem will give diminishing returns, so we must settle for this accuracy rate. Despite our relatively poor results, this research has led to a better understanding of the approaches most commonly used to solve the problem.

## 6.2.2. Using POS Tagging for Correct Homograph Pronunciation

Now that we have a part of speech tagger, in many cases we can use the part of speech tags found through analysis of the sentence to determine the correct pronunciation of a word. This is a process which must be performed for each word: we index its spelling along with the corresponding pronunciations for when used with different parts of speech. As this must be done manually for each word (as the CMUdict database does not contain this information, only indexing the possible pronunciations), in this project it has only been performed for the most common English words which can be pronounced in different ways.

Our system can now choose the correct pronunciation for words in phrases such as "please do not desert me in the desert" and "I want to advocate for the advocate". Using this in conjunction with our general grapheme to phoneme rules, we are now able to input an arbitrary alphabetic sentence into ODDSPEECH and find an Arpabet pronunciation for all the words within it; from this, we can split the pronunciation string up into its diphones. Finally, this information can be used with our existing diphone databases to produce speech in the same way that BADSPEECH did.

However, though we can now distinguish between pronunciations of the same word with different phonetic stresses, ODDSPEECH as-is does not have any system in place to alter its prosodic characteristics. Even though we can identify the correct vowel stress in the word "project", our system still pronounces it in the same way, since the underlying phone sequence is identical. We must therefore now give ODDSPEECH a way of altering prosodic aspects of speech over time.

## 6.3. Arbitrary Volume, Pitch, and Duration Modification

There are four primary prosodic variables: in order of most to least important for their contribution to speech naturalness, they are pitch, duration, volume, and timbre [20] [38]. Of these, timbre is the most difficult to alter in existing speech recordings, as a different timbre corresponds to a different vocal articulation. As such, to alter timbre, we would need to determine articulator configuration from our waveform, alter that articulation, and then resynthesize; this is a prohibitively difficult task. Fortunately, it is the least important prosodic variable for naturalness, so it is not vital for us to be able to change it. We will therefore only consider pitch, duration, and volume.

As discussed in Section 4.1.3., various signal processing techniques exist to change the pitch and duration of a speech waveform. The three techniques we considered there were Phase Vocoders, Spectral Modelling, and Pitch Synchronous Overlap Add (PSOLA). In our considerations, we wish to consider the disadvantages of each approach. Phase Vocoders alter the unvoiced sections of speech in an undesirable way, while Spectral Modelling synthesis deconstructs and resynthesizes speech which results in some loss of the original waveform's detail; further, both of these methods require Short Time Fourier Transforms, making them computationally expensive.

The PSOLA algorithm allows for effective pitch shifting of voiced speech in the time domain, making it computationally efficient; it also keeps most of the fine detail of the original waveform. On the downside, the algorithm cannot usefully modify unvoiced speech, and its effectiveness is dependent on our ability to section the waveform into separate partitions for each glottal excitation. An implementation might also repeat or eliminate short-term speech components such as stop consonants, which should occur exactly once in the adjusted waveform.

In considering what we want from our system, we can note that we only actually wish to change the pitch of voiced speech. While we may want the duration of unvoiced speech to change, its acoustic character is exclusively determined by turbulence from articulator positions in the mouth, and therefore should not change in pitch over time in the same way that voiced speech does. Therefore, if we use PSOLA to vary the pitch and duration of voiced sections of speech, and a different method to change the duration of unvoiced speech, we can have the benefits of PSOLA and compensate for its disadvantages.

If we wanted to use this approach to vary the pitch and duration of an arbitrary speech waveform, detecting the difference between voiced and unvoiced speech would be the initial and most difficult problem to solve. Fortunately, this is not a concern for us. As we are synthesizing speech from a database of pre-recorded diphones, we know which two phones any diphone is transitioning between. If we perform our pitch and duration shifting on the diphone's waveform before adding it to the overall speech waveform, then we know exactly which two phones are represented in any waveform we wish to analyse.

Therefore, if we are transitioning between two voiced phones, we need only use PSOLA. If we are transitioning between two unvoiced phones, we know that PSOLA will be ineffective. If the first phone is a stop, we will know that is the case, and can make adjustments so that our approach does not remove the character of the stop phone from the adjusted waveform. Finally, if we are transitioning between a voiced and an unvoiced phone, we will know that we want to use PSOLA on only half of the waveform, and use an alternative approach for the unvoiced speech.

So what approach will we use on unvoiced speech? We know that unvoiced speech is composed of high-frequency broadband noise, so the rate at which the speaker changes their vocal articulation is very slow in comparison to the changing of our waveform. We also know that we only wish to change the duration of unvoiced speech, without modifying its frequency.

In contemplating this, we can consider how the PSOLA algorithm acts to change the duration of a wave without changing its pitch: sections of the original waveform are repeated or omitted in such a way that the frequency does not change. As the frequencies contributing to unvoiced speech are over a large range, this may seem particularly applicable. However, if we extract a section of the wave that is large relative to its constituent periods and either repeat or remove this section, then the discontinuities at concatenation points will be infrequent relative to the frequency of the wave. Further, as the source signal is mostly noise, we already have irregularities in the wave; minimal smoothing at the concatenation points should be sufficient to make this change inaudible.

To illustrate this, if the average frequency of the unvoiced speech signal's band is 2000 Hz, and we repeat or omit sections with widths of 0.05 seconds, then the concatenation points will occur every 0.05 seconds, with an overall rate of 20 Hz. This means that only one out of every hundred periods of a 2000 Hz wave will contain a concatenation point; our modification is on average only interfering with 1% of the cycles in the overall waveform, while changing the overall wave duration as desired. The larger that the chosen section length is, the less proportional signal discontinuity will occur.

If we wanted to change the length of a waveform where the speech articulators are in steady state, the problem is solved; we pick the largest length possible (the entire waveform) and either repeat it entirely or remove sections from the end. Unfortunately, since we are using diphones, the waveform we are modifying is not in steady-state: the speaker's articulators are moving in their vocal tract.

In a short time section of a diphone waveform, the speaker's vocal articulators are in a very similar position at the start of the section as they are at the end. In a longer section, the positions of the articulators change over the duration. This means that if we repeat a longer section, then the connective sections will have different spectral profiles, making the point of concatenation more noticeable to a listener. As such, the smaller that the chosen section length is, the fewer of these large scale artefacts will occur.

Therefore, we have a tradeoff in choosing the length of sections: smaller sections will retain the large-scale smoothness of the diphone recording while reducing short-time signal quality; larger section sizes retain signal quality but deteriorate the smoothness of the phonetic transition. As a tradeoff between these two, a section size of 0.01 seconds was chosen.

This choice means that concatenation interferes with 5% of wave periods on average, but is also small relative to the length of our system's diphone recordings, as most diphones in our banks are between 0.2 and 0.4 seconds long. As we want to name this approach, so we can refer to it for the rest of this report, we will call this technique Unvoiced Speech Duration Shifting, or USDS.

### 6.3.1. Implementation

Now that we have decided to adjust volume, pitch, and duration on the diphone level, we should consider the details of this implementation. First, we wish to use PSOLA for diphones which only contain voiced phones, or transitions from stops to voiced phones. Second, we wish to use USDS for diphones which only contain unvoiced phones, or which are transitions from stops to unvoiced phones. Finally, for diphones between a voiced and an unvoiced phone, we will want to use a combination of PSOLA on the voiced section and USDS on the unvoiced section.

Each phone is assigned a certain multiplicative factor for volume, pitch, and duration: a 2 in duration means we wish to double duration, while a 2 in pitch means we want to double the frequency. This means that each diphone has two target volumes, pitches, and durations. Pitch and duration are handled by PSOLA or USDS, while the change in volume is applied at the end by simply multiplying the final waveform by a line between the two target volumes over the duration of the phone.

## 6.3.1.1. PSOLA for Voiced Diphones

For our implementation of the PSOLA, we first need to be able to separate the wave into its distinct glottal excitations. This can be done with a peak detection algorithm; MATLAB already includes the *findpeaks* function, which lets us define various properties of the peak detection. However, it is likely that any given speech waveform has multiple peaks corresponding to each single glottal excitation. We therefore need to filter our signal such that each glottal excitation only corresponds to one peak.

This can be done by smoothing the overall waveform to reduce the higher frequency contributions of vocal tract resonance. The degree to which we want to smooth the waveform is dependent not only on the fundamental frequency, but also on the harmonic frequencies of the recorded voice, which will change depending on vowel character. Too little smoothing and we will still find multiple peaks from a single glottal excitation, while too much smoothing can smooth multiple peaks together, appearing like one glottal excitation. The most useful smoothing value can be best found empirically for each diphone bank.

As each glottal excitation then corresponds to the maxima of the waveform, we then clip this waveform below at 0 so that we only find peaks above the x-axis, corresponding to glottal exhalations. We can finally perform peak detection on this modified waveform to determine peaks corresponding to glottal excitation. Figure 36 shows this procedure, with the top plot showing the original speech waveform, and the lower plot showing the processed waveform and the peaks which are found. These are then also shown on the upper plot as red circles and numbered.

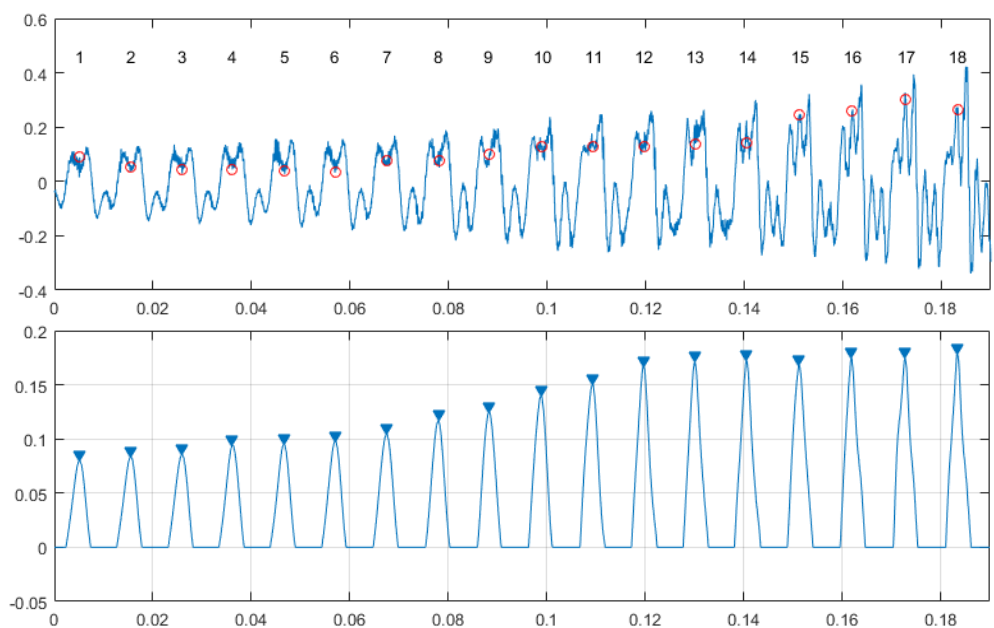

*Figure 36: Glottal Excitation Detection for Y to EH Diphone*

Now that we can automatically find these points of glottal excitation, we want to isolate each glottal excitation from the source waveform. In our implementation, this is done by using asymmetric Hann/raised cosine windows. The centre of each window is the point of glottal excitation, and each side of the window extends to the point of the previous or following glottal excitation (except the first and last excitations, which extend to the start and end of the waveform). This choice of window means that we can perfectly reconstruct the original waveform by adding these sections together.

Once we have separated each glottal excitation, we need to determine new locations to place excitations at. This is made somewhat more complicated due to having two target durations and pitches: one set for the first phone and another for the second.

In our implementation, we linearly alter both the pitch and duration of the waveform over the course of the diphone, moving from the first phone's target pitch and duration to the second. We will initially consider the desired duration shifting. We determine this by calculating the lengths of the gaps between glottal excitations in the original waveform. Then, we consider the points at the centres of these gaps, and determine the desired multiplicative change to duration at that point. We then multiply the gap lengths by these amounts. These shifts determine our excitation ranges: the space between the start of the wave and the first glottal excitation is Range 1, between excitations 1 and 2 is Range 2, between 2 and 3 is Range 3, and so on. The final excitation's range extends out infinitely after the end of these ranges.

Now that we have defined these ranges, we will insert glottal excitations sequentially according to our desired pitch shift. This starts by placing the first glottal excitation where it was in the original waveform. We then consider the length of the gap between excitations 1 and 2, and multiply it by the inverse of our target shift to frequency at the centre of that gap; this gives us our target shift to wave period. By adding this number to the position of the first excitation of the wave, we get the position of the next glottal excitation. To determine which extracted excitation we place at that point, we look at which Range we are currently in; if we are in Range 2, we put Excitation 2 from the original waveform at that point, in Range 3, we place Excitation 3 at that point, and so on. After we have an excitation placed within the final Range, we end this process and output the modified waveform.

Figure 37 shows the waveform from Figure 36 being shifted in duration by a constant amount over the waveform, with pitch kept the same: the upper plot shows a duration shift of 2, doubling waveform length, while the lower plot shows a duration shift of 0.5, halving waveform length. We know the expected behaviour of each; the upper should include each excitation twice, while the lower should include every other excitation (with the exceptions being the first and last excitations, which occur only once). This behaviour is approximately as expected, with minor differences due to our implementation. For example, in doubling the duration of the wave, we see Excitations 4 and 16 appear only once, while Excitations 5 and 17 appear three times each. This is due to the slightly different spacing between different glottal excitations over the course of the original waveform. The lower plot, however, behaves exactly as expected, including only odd-numbered excitations.

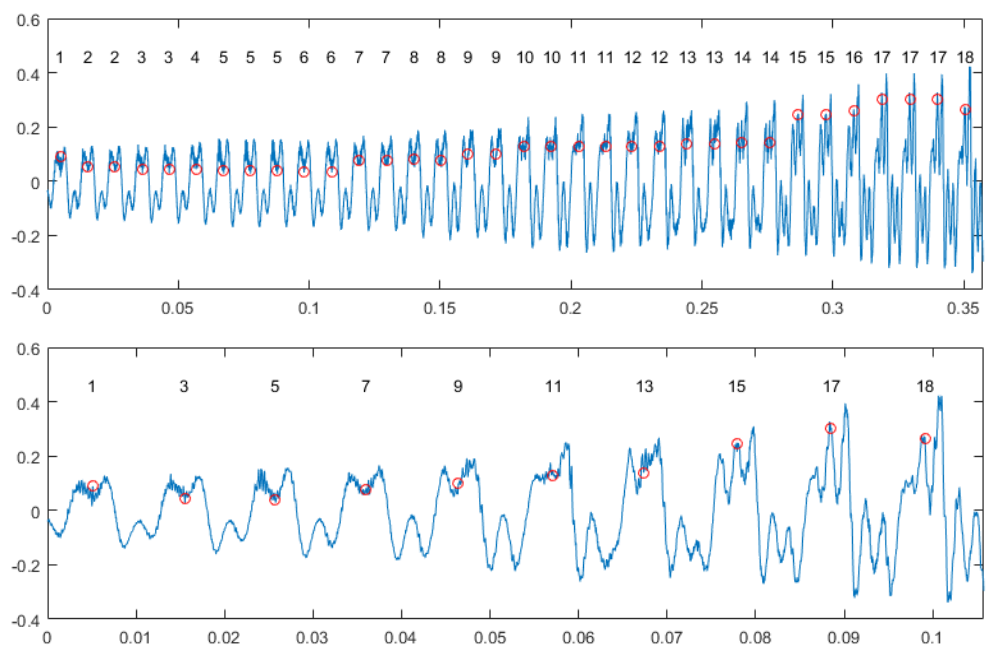

*Figure 37: PSOLA Constant Duration Shifting for Y to EH Diphone*

Figure 38 shows the waveform from Figure 36 with duration kept constant and pitch varying between 0.5 at the start to 1.5 at the end. We expect at the start of the wave every our frequency to be halved, while at the end the frequency is multiplied by a factor of 1.5. This matches the observed behaviour, again confirming the effectiveness of our implementation.

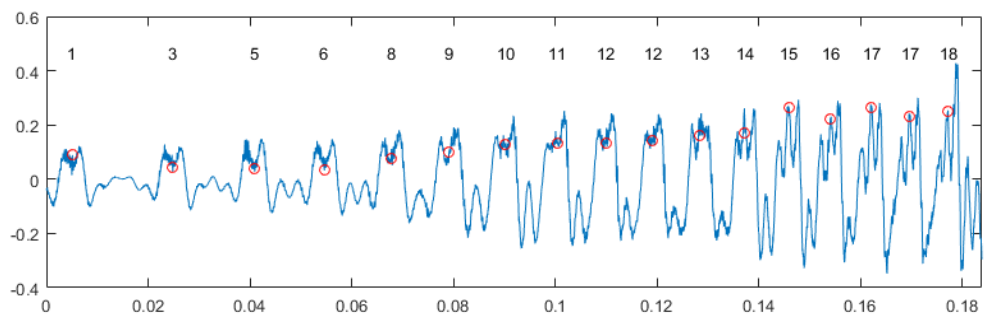

*Figure 38: PSOLA Varying Pitch Shifting for Y to EH Diphone*

This approach will also work for voiced diphones leaving stop phones: as our approach retains the first impulse's location in the waveform without modifying it, and stop phones are articulated as short-term impulses, then the stop component of the diphone will always be retained in the modified waveform. We also find this approach to be computationally effective, as it operates in the time domain. In summary of our approach here, we smooth the waveform, find peaks, apply window functions, determine new peak locations, and then place the glottal impulses where desired. All of these can be performed quite quickly in MATLAB.

## 6.3.1.2. USDS for Unvoiced Diphones

The procedure for USDS has already been explained: we wish to section up our diphone into 0.01 second frames, and then repeat or omit those frames as desired to scale the duration of the phone; we then perform slight smoothing between frames.

As with PSOLA, we want the duration shifting on the diphone to change continuously over time. This has been implemented through the use of a *count* variable which increases or decreases depending on the desired duration shift of the current frame, as linearly interpolated between the start and the end of the diphone. In each new frame, we add *dur-1* to the *count* variable, where *dur* is the desired multiplicative duration shift of that frame; then, if *count* is greater than 1, we subtract 1 and repeat the frame until it is less than 1. If *count* is less than or equal to -1, we remove that frame and add 1 to *count*. The way that this works is best understood through examples where we are duration shifting the entire phone by a constant amount.

If we want to shift the phones duration by a factor of 0.5, we want to remove 1 out of every 2 frames. Each frame, we add (0.5-1) = -0.5 to *count*. If *count* starts at 0, this means that after 2 frames, *count* will be equal to -1, and we will not include the second frame. Then we add 1, returning *count* to 0. This then repeats, removing 1 of every 2 frames as desired.

If our factor is 0.2, then we add (0.2-1) = -0.8 to *count* for each frame. Our first frame is included, and *count* becomes -0.8; then *count* becomes -1.6 after 2 frames, so we remove that frame and add 1, reaching a *count* value of -0.6. The next frame also adds -0.8, reaching -1.4; we remove that frame and add 1, reaching -0.4. In the same way, the following 2 frames are also removed, and *count* returns to 0, after which this behaviour repeats. This means we remove 4 out of every 5 frames: for our reduction to 0.2 times the length of the original wave, this is the desired behaviour.

*6. Improving Our System*

If our factor is 1.1, then we want to duplicate every tenth frame. We add (1.1-1) = 0.1 to *count* each frame. After 10 frames, *count* will be equal to 1; at this point, we repeat the frame and subtract 1, returning *count* to 0 and starting the process again. As such, one of every 10 frames is repeated; this indeed scales our wave's duration by a factor of 1.1. If our factor is 5, we add (5-1) = 4 to *count* each frame, repeating each frame 4 times after its initial incidence; this repeats each frame 5 times, again giving behaviour as expected. Implementing duration scaling in this way lets us vary the desired scaling over the course of the diphone.

Figure 39 shows this procedure for a HH to SH diphone; the top plot is the original waveform, while the centre plot is duration scaled at a factor of 2 and the lower plot is scaled by a factor of 0.5. We can see from the different horizontal axis scales for each that the behaviour is as expected. Further, we can note that the large-scale behaviour in each wave is similar. This indicates how each wave sounds like the same diphone, but with the articulatory change at a faster or slower speed.

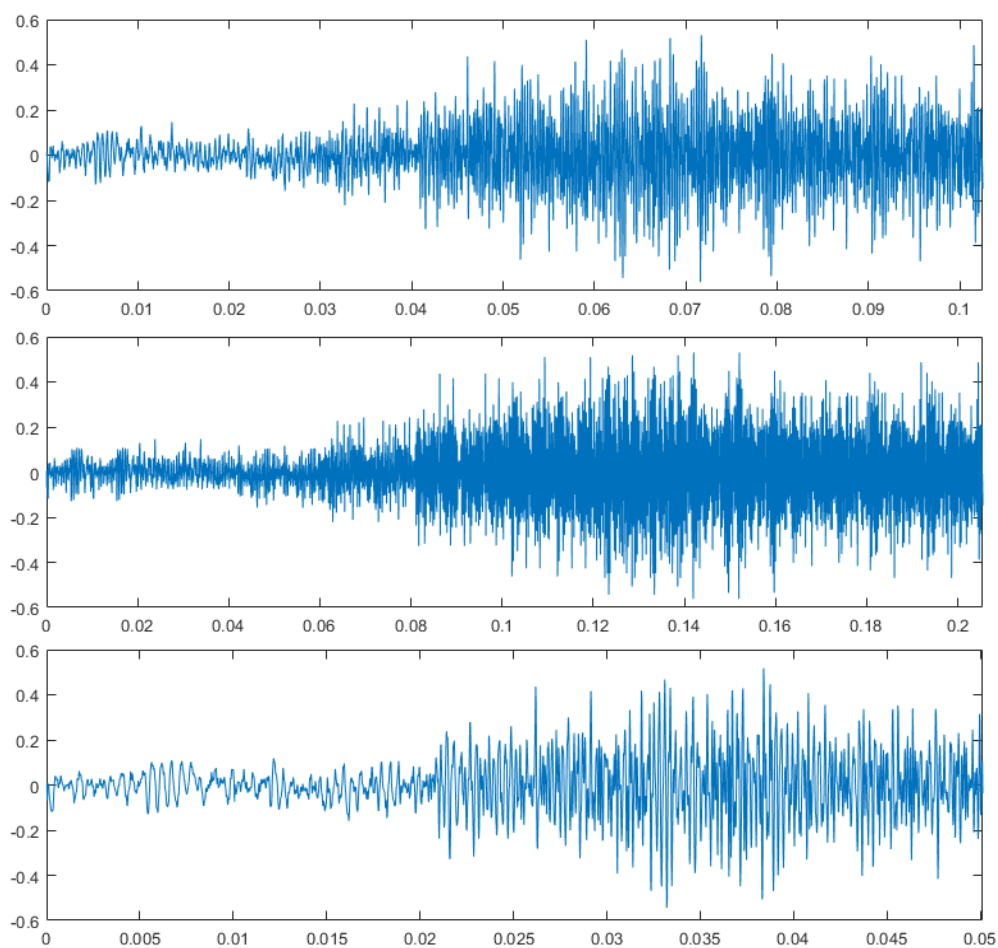

*Figure 39: USDS Constant Duration shifting for HH to SH Diphone*

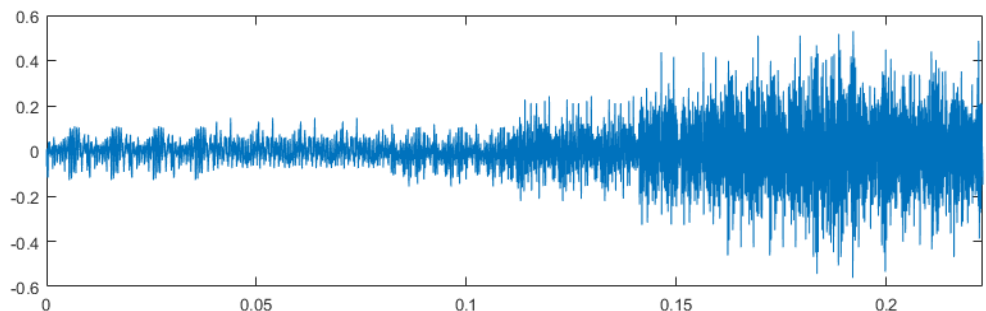

*Figure 40: USDS Variable Duration Shifting for HH to SH Diphone*

In Figure 40, we can see this procedure for a duration shift which is not constant over the course of the diphone; here, we start at a duration scaling of 5 and end with a duration scaling of 0.1. We can see visually that this prolongs the start of the phone while making the termination of the phone much briefer. When played back as audio, this increases the duration of the HH phone while reducing the duration of the SH phone. However, acoustic abnormalities are noticeable at the start of the phone from this large duration scaling. This is an artefact of our tradeoff between retaining the large-scale character of the phonetic transition and ensuring small-scale acoustic smoothness.

Different approaches, such as vocoders and spectral modelling, would suffer less from this problem; however, both involve a computationally expensive STFT transformation, performing operations on the resultant spectrogram, and then re-synthesis. However, as discussed on Page 39, these approaches then also have the problem of vertical and horizontal incoherence. USDS is extremely computationally cheap, and provided that we do not scale the duration excessively, results in minimal acoustic abnormalities.

### 6.3.1.3. Combining PSOLA/USDS for Voiced/Unvoiced Transitions

If a transition moves from an unvoiced phone to a voiced phone, the first step in our method is to flip it in time before proceeding. This means that we can perform the same operations as we would on a transition from a voiced to an unvoiced phone, and then flip it again after completion. Thus, in considering our procedure for transitions from voiced phones to unvoiced phones in this section, it will apply equally to transitions which go in the other direction.

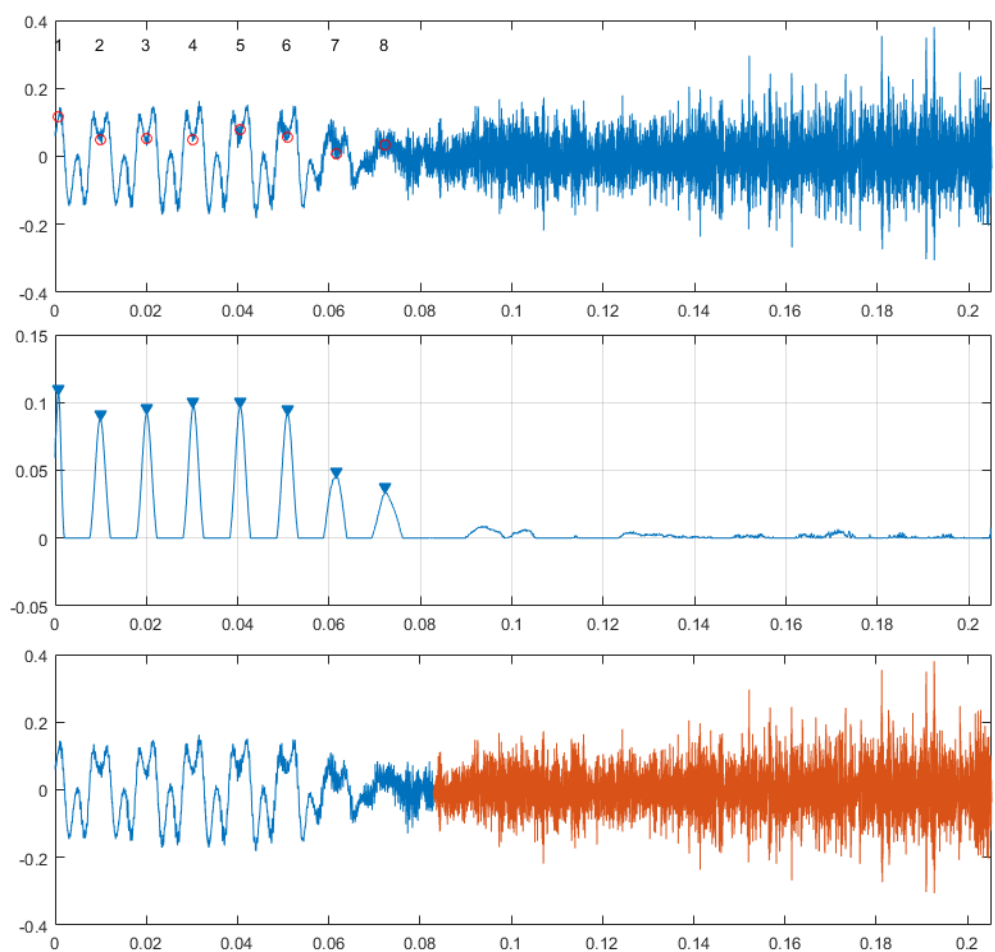

*Figure 41: Splitting Y to F Diphone into Sections*

The upper two plots in Figure 41 show our PSOLA implementation operating on the diphone from Y to F, with some minor differences. Here, we choose to also exclude peaks less than 20% of the height of our maximum peak. We also then find the median gap distance between peaks, and remove any peaks occurring more than twice this distance apart from the end of the waveform. This means that the small bumps during unvoiced speech in the centre plot (which are not entirely smoothed out) are excluded from our peak detection algorithm, ensuring we are only detecting peaks which correspond to glottal excitations.

We then consider the final PSOLA peak and find the distance between it and the second last PSOLA peak. We then move this distance forward in time from the final peak, and this is the point at which we choose to cut the wave into two sections. This is illustrated in the lowest plot in Figure 41, where the blue part of the plot denotes the voiced section of the diphone and the orange section of the plot denotes the unvoiced section. We can then pitch and duration shift the first part of the wave with our PSOLA algorithm, while duration shifting the latter part of the wave with our USDS implementation.

## 6.3.2. Using Tagged Stress Markers Within Words

We can now assign particular volume, pitch, and duration scalings to each phone, and use our combination PSOLA/USDS approach to apply the shifts to our recorded diphones. Now, all we need to do is use the tagged stress markers from CMUdict to determine how we modify the speech waveform over time. The different Arpabet vowel stresses are 0, for no stress, 1, for primary stress, and 2, for secondary stress. By assigning different volume, duration, and pitch values to these, ODDSPEECH output will overlay some prosody onto the sentence, helping it to sound less monotonous than BADSPEECH. Of course, there are many possible assignments of these variables; determining a preferable setting is empirical and subjective based on the listener's preferences.

## 6.4. Combining to Create ODDSPEECH

Figure 42 shows the GUI for synthesizing input text using ODDSPEECH. For a text input of space separated alphabetic words, we first determine the lexical class of each input word. Next, we find the pronunciation of any words where pronunciation varies depending on lexical class. After this, we attempt to find a pronunciation of words in CMUdict; if a word is not in CMUdict, then we determine a pronunciation with our grapheme to phoneme rules. We combine these into a sequence of phones, and then each diphone's volume, pitch, and duration is varied based on the settings of the system. The effectiveness of different ODDSPEECH configurations will be analysed in Section 8.1.

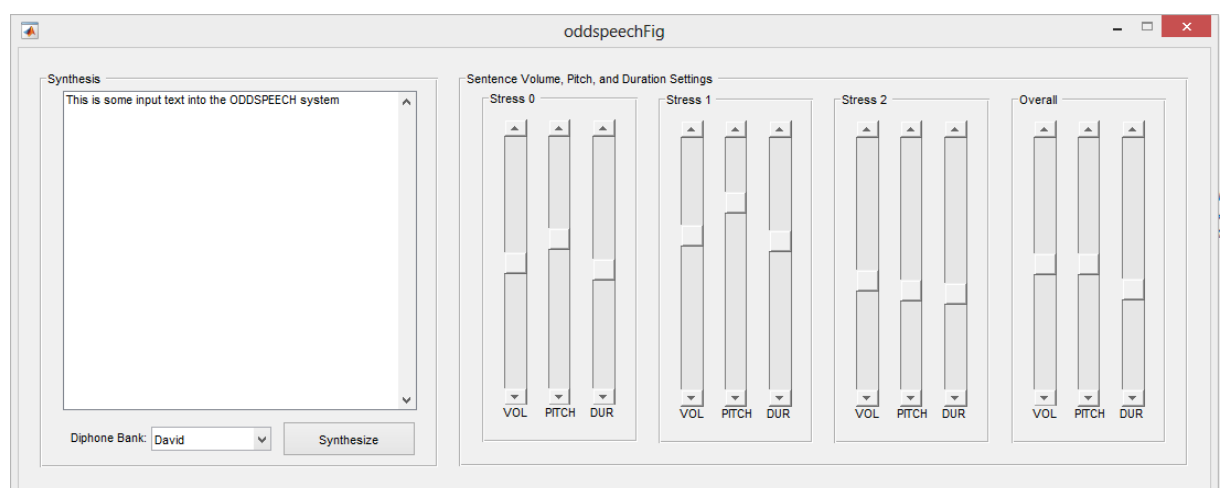

*Figure 42: ODDSPEECH GUI*

# 7. Speech Synthesis with Prosody

In having completed ODDSPEECH, our system is now able to vary volume, pitch, and duration on the phone level over the course of a sentence. While in ODDSPEECH we have only assigned prosodic overlay to accentuate vowel stress, we should consider what can be done to introduce prosody on a larger scale in our system. We will also implement a text preprocessing step, so that rather than requiring our input be space-separated alphabetic words, the system can handle arbitrary input.

As before with BADSPEECH and ODDSPEECH, we need a label for the more advanced synthesis system we are developing here. As we wish to overlay prosody in an aesthetically tasteful fashion that is typical of human speech, we will refer to it as PRosodic Enunciation of Tasteful TYpical Speech, or PRETTYSPEECH. A flow diagram of PRETTYSPEECH is shown in Figure 43.

## 7.1. Text Preprocessing

Our objectives in text preprocessing are quite simple: for an arbitrary input string, we want to separate that string into separate tokens, which can be composed of words, punctuation, digits, or other characters. We will then want a way of processing each token to find appropriate part of speech tags and pronunciations for them.

### 7.1.1. Tokenisation

In BADSPEECH and ODDSPEECH, our only step in "tokenisation" was to split the input string wherever there were spaces. In PRETTYSPEECH tokenisation, splitting an input string along spaces is still a good initial segmentation of an input string. After this, we will then need to separate out any sentence punctuation from these tokens.

*Table 31: Tokenisation for the Input String "Hello, everyone!"*

| Input String | Hello, everyone! | | | |
|---|---|---|---|---|
| Splitting on Spaces | Hello, | everyone! | | |
| Punctuation Separation | Hello | , | everyone | ! |

Table 31 illustrates how this is done in our system. For an input string of "Hello, everyone!", an initial splitting along the space gives us the two tokens "Hello," and "everyone!". In our implementation, we then process each input token and separate any non-alphanumeric characters which occur at the start or end of a token into a token of their own. This separates the comma in "Hello," and the exclamation mark in "everyone!", giving us the tokenisation in the final row in Table 31.

After this is completed, we want to find part of speech tags for each of the words within the input string. The input string might be multiple sentences long; in taking inputs for part of speech tagging, we therefore split on any sentence-terminating punctuation. In the English language, these are the characters ".", "?", and "!". We then apply our previously existing part of speech tagging approach from ODDSPEECH to all non-punctuation tokens within a sentence. The one adjustment we make is always tagging tokens which are entirely numerical as nouns. Strictly speaking, numbers should be tagged as "quantifiers", a lexical class which can also behave like adjectives. Unfortunately, our simplified tagset does not include this lexical class, so for simplicity's sake we will always consider them to be nouns. Table 32 shows this tagging process as applied on a punctuated input string.

*Table 32: Tokenisation and Tagging for the Input String "Yes, I'm going to buy 10 apples."*

| Input String | Yes, I'm going to buy 10 apples. | | | | | | | |
|---|---|---|---|---|---|---|---|---|
| Tokenisation | Yes | , | I'm | going | to | buy | 10 | apples | . |
| Part of Speech Tags | N | | N | V | v | V | N | p | |

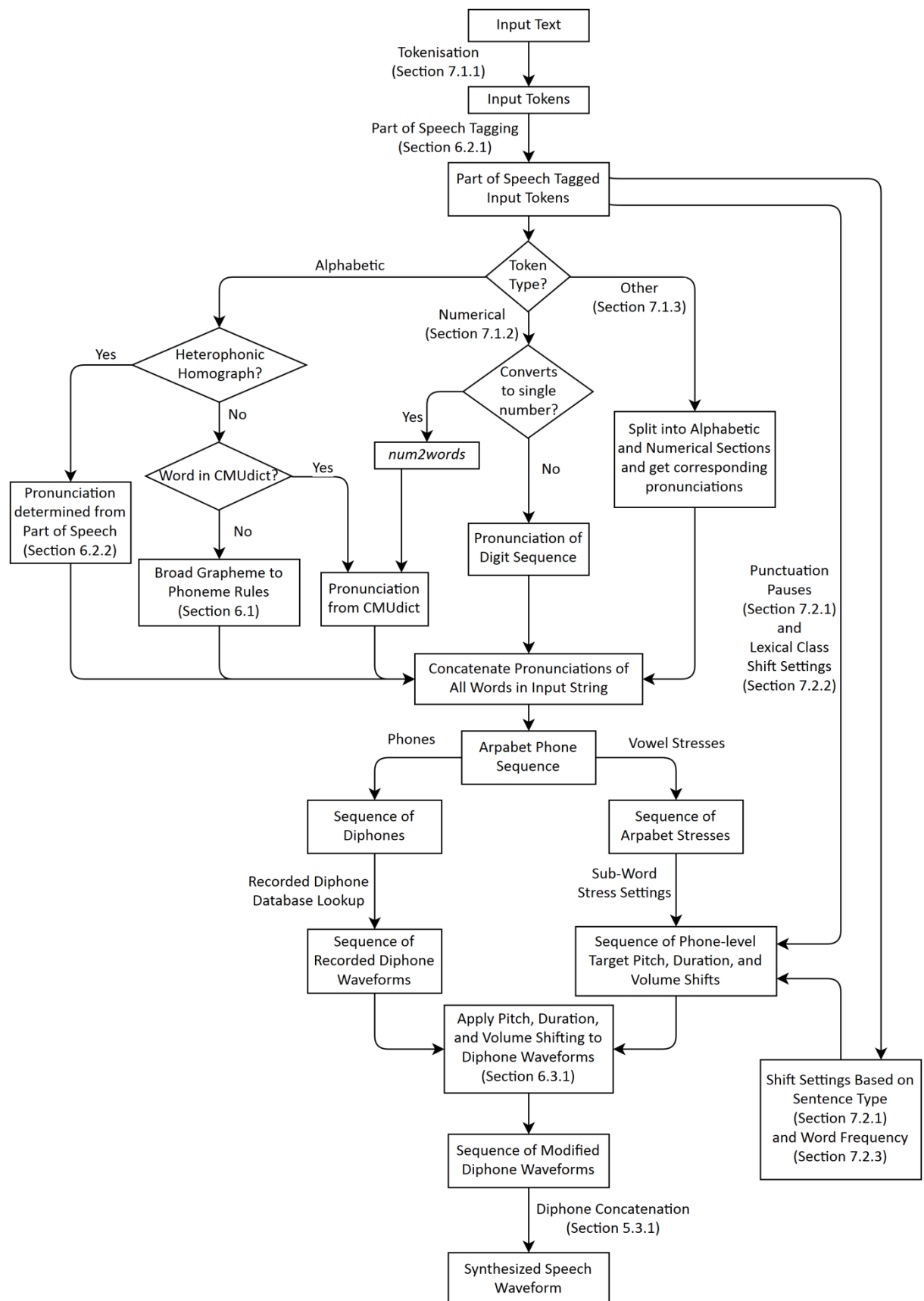

*Figure 43: PRETTYSPEECH Software Flow Diagram*

As before, we are then able to find pronunciations for heterophonic homographs according to their part of speech, and find pronunciations for known English words and alphabetic tokens. We must now design a method to pronounce non-alphabetic tokens: we categorise these tokens into being either numerical, or some combination possibly including alphabetic, numerical, and other non-alphanumeric characters.

### 7.1.2. Pronouncing Numerical Tokens

When we have an input number, we want to be able to find an appropriate pronunciation. In our implementation, we have used a file from the MathWorks File Exchange, *num2words*, which takes an input number value and finds corresponding string of words for that number [71]. This file works correctly for numerical token strings that can be successfully converted into a single number, and then outputs the input number in words. For instance, a call of *num2words(1024)* returns the string "one thousand and twenty-four". We can then find the appropriate corresponding pronunciations for these output words, which gives us a pronunciation for an input number.

There are then numerical input strings that may not convert into real numbers. For instance, IP addresses are given in a format like "127.0.0.1", which cannot directly be turned into a numerical value. In cases such as this, we pronounce each digit in order, and pronounce intermediary punctuation in an appropriate way. With "127.0.0.1", we want to pronounce each "." as "point", rather than treating it like a punctuation period. (This is why in our tokenisation step, we only remove punctuation from the beginning and end of words, without separating on punctuation within words. This also means that numbers formatted with place-marking commas, such as an input of "410,757,864,530", will be pronounced as a single number.)

One problem with this implementation is that, depending on context, there are some numbers that we might want to pronounce differently. One example is when numbers are used as years: in the phrase "Yuri Gagarin went to space in 1961", we would most commonly want to pronounce "1961" as "nineteen sixty one" rather than "one thousand nine hundred and sixty one". Telephone numbers are another example: we want to pronounce each component digit in order, rather than pronouncing it as a single very large number. While we already pronounce digits sequentially for nonstandard numerical tokens such as IP addresses, there are no clear delineators that instruct us to do this for phone numbers.

A simplistic approach would be to select certain ranges to pronounce numbers differently; for example, we might say that any number between 1900 and 2100 should be assumed to be referring to a year, or that any number with 8 digits should be assumed to be a phone number. While in many cases this works, it does not generalise flawlessly: ideally, we would want to know which pronunciation is suitable based on context.

Unfortunately, determining when to use these exceptions in this fashion would require more advanced semantic analysis of the input string. As was stated in our discussions on part of speech tagging, the complexity of semantic analysis makes it impossible to implement in the scope of this project. In our implementation, we have therefore chosen not to consider such cases, permitting more useful pronunciation behaviour in general.

### 7.1.3. Handling Other Tokens

Now, we will need a way to handle punctuation tokens. Fortunately, the Arpabet transcription system preserves sentence punctuation characters when converting from a sentence to its pronunciation. As such, any punctuation character's transcription in Arpabet is the character itself; at this stage, we are not concerned about what this will actually mean in our synthesized speech.

In allowing arbitrary input, we will also need a way to pronounce tokens which are not entirely alphabetic, numerical, or punctuation, but a combination of those and other characters. To do this, we will take the token and split it up into separate sections of alphabetic, numeric, punctuation, and other chunks. Then, we find a corresponding pronunciation for each alphabetic and numeric chunk, ignoring punctuation. Some non-alphanumeric and non-punctuation characters, such as &, are given pronunciations. Finally, the resulting pronunciations are concatenated together. This is done without placing any intermediary silence characters between them, meaning that the words will flow together rather than being pronounced separately. This procedure is illustrated in Table 33.

*Table 33: Finding a Pronunciation for the input token "quill12brigade".*

| Input Token | quill12brigade | | |
|---|---|---|---|
| **Splitting** | quill | 12 | brigade |
| **Pronunciations** | K W IH1 L | T W EH1 L V | B R AH0 G EY1 D |
| **Concatenation** | K W IH1 L T W EH1 L V B R AH0 G EY1 D | | |

## 7.1.4. Discussion of Advanced Text Preprocessing

Our implementation of text preprocessing, while sufficient for the scope of this project, is still quite primitive compared to modern systems. More advanced text preprocessing techniques could pronounce inputs like "$20" as "twenty dollars", or an input such as "5th" and pronounce it as "fifth". Depending on the implementation, it might also be desirable to distinguish between letter case when finding pronunciations; for example, the tokens "us" and "US", where the latter refers to the United States of America, and should be pronounced as an initialism instead.

While it would be nice for our system to handle all of these scenarios, the issue is that there are a great number of specific exceptions such as these. Rules for how to handle them have to be manually implemented, and no database for these rules appears to be freely available. Identifying these and similar scenarios is a challenge which is more grounded in linguistics than engineering. As this is an engineering-focused project, we will limit the depth of our investigation to what has already been discussed.

## 7.2. Methods to Introduce Sentence-Level Prosody

After tokenisation of an input string and finding Arpabet transcriptions of each token, we wish to produce our speech waveform in a similar fashion to ODDSPEECH. However, we also want to improve on it by including new behaviours to usefully handle punctuation characters, as well as to overlay sentence-level prosodic behaviours.

## 7.2.1. Prosodic Effects from Sentence Punctuation

When sentences are pronounced, tokens such as commas, semicolons, colons, and ellipses often correspond to breaks of varying lengths, while sentence terminating punctuation such as periods, question marks, and exclamation marks can correspond to longer breaks, as well as changing the way that a sentence is pronounced. For example, declarative sentences in English typically have falling intonation at the end, where interrogative yes-no questions typically terminate with rising intonation [72].

As with most questions of prosody, there is no universal value which is the "perfect" solution to these questions, simply what people subjectively assess to be the more aesthetically preferable decision. Rather than determining specific solutions, we therefore want our implementation to be flexible, permitting different values to be input according to a user's tastes. Addressing the first issue, we want a user to be able to define the length of pauses in our synthesized speech corresponding to each of the types of punctuation above, which is quite simple to implement.

We also want a user to be able to customise how volume, pitch, and duration change over the course of a sentence. This is done by recording user-defined curves for each prosodic variable over the course of a sentence. Separate curves should be defined for when the sentence ends in a period, exclamation mark, or question mark. The prosodic contribution from these curves are then applied on the word level: for a defined input curve, a number of points equal to the number of words within the sentence are equally spaced across the curve's domain, and prosodic attributes are determined based on the curve values at those points. The details of how users define this curve in our implementation will be discussed in Section 7.3.1.

## 7.2.2. Prosody Based on Lexical Class

We also want a user to be able to define additional prosodic characteristics on the word level based on the lexical class of the word being considered. For example, nouns and verbs usually communicate the most semantic meaning, and should be emphasized within a sentence. By contrast, lexical classes such as articles and conjunctions do not impart much meaning, being mostly syntactic. Pronouncing them more quickly or with less emphasis could improve system prosody.

In our implementation, a user picks a target volume, pitch, and duration for each lexical class. They can then also choose to include some random fluctuation. By using a small amount of randomness, we can prevent the system from sounding excessively monotonous. If all words of a certain lexical class are pronounced at the same pitch, this may become apparent and jarring to a listener. By adding some randomness, the target prosodic values will vary randomly while still being close together, potentially mitigating this problem.

In addition, we permit the user to determine whether the synthesized speech will be produced with exactly the target prosodic characteristics set, change a set amount relative to the prosodic character of the previous word, or move towards that target point a certain percentage relative to the previous word. This permits a wide variety of user-definable behaviours.

## 7.2.3. Prosody Based on Word Frequency

Another theory to improve prosody is to change prosodic variables based on the frequency of a word within a large corpus. This would allow commonly occurring words to be pronounced with less emphasis, and rarer words to be pronounced with greater emphasis. This is useful, as it is likely that listeners are more familiar with the more common words, and do not need as much time to understand what word is being said. A less common word is more likely to communicate meaning within a sentence, so should be accentuated so that a listener does not mishear it.

In our implementation, we have used a database of word frequencies derived from the British National Corpus [73] [74]. Words which occurred 10 or fewer times were ignored, and the Base 10 logarithm of all frequency counts was calculated. This assigns each word in the database a numerical value between 1 and 7; words not in this database are then assigned a value of 1. We then allow users to define prosodic curves over this domain; as with sentence-level prosody curves, the details of how this is implemented are further discussed in Section 7.3.1.

## 7.2.4. Discussion of Advanced Prosodic Overlay

There are many further additions we could make to improve the prosody of our system. If we knew the syntactic or semantic structure, we could likely determine better, more general prosodic behaviours. Another addition which could improve prosody would be for the system to produce inhalation sounds during sentences, emulating inhalation breaks in natural human speech. As with text preprocessing, finding techniques to improve prosody is more of a linguistics problem than an engineering one, so we will end our discussion of prosodic overlay here.

## 7.3. Combining to Create PRETTYSPEECH

Now that we have discussed our implementation of text preprocessing, and considered what methods of improving sentence prosody we want to include in PRETTYSPEECH, we must design a method for the user to define prosodic curves, as well as constructing a graphical user interface.

### 7.3.1. Customisable Curves

As previously discussed, the user must be able to define curves for prosodic overlay from sentence prosody and changes due to word frequency. Therefore, we want to design a way in MATLAB for users to intuitively create such curves.

To do this, when a user clicks within an existing *axes* object, our code creates a new *impoint* object. These are objects represented as a point on the screen which a user can manually reposition by clicking and dragging. We store each *impoint* created in this way inside of an array. Then, we consider only *impoint* objects within the bounds of the axes, and sort these points in order of their x coordinate. This gives us a sequence of user-defined points on the plot. To remove an *impoint* from the plot, it simply needs to be dragged off the axes.

We then want to interpolate between each of these points to create a curve. These sections can be defined by the user as sections of quintic curves (creating a quintic spline curve), sinusoidal curves, or straight lines between each point. Examples of each type are shown in Figure 44. This allows for a wide range of prosodic behaviours to easily be implemented by a user.

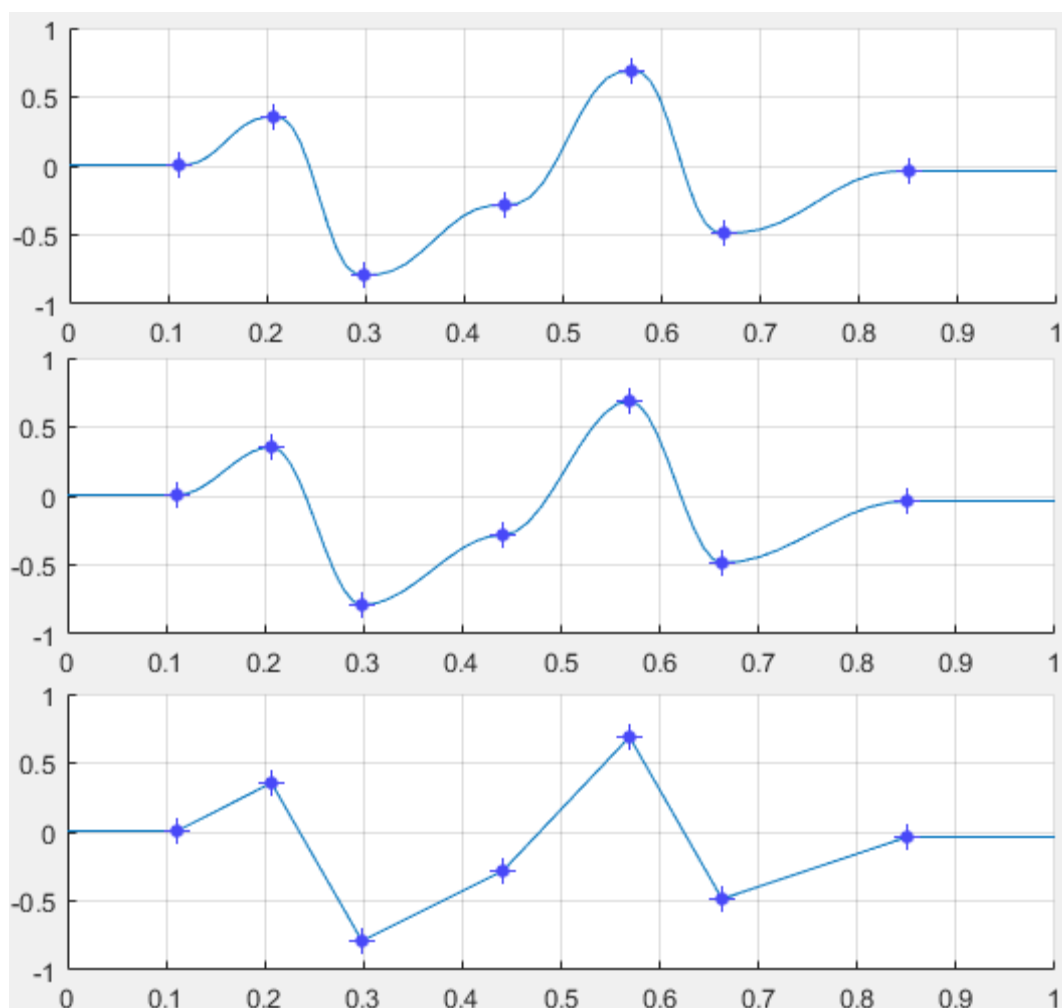

*Figure 44: Custom Curves with Quintic (upper), Sinusoidal (centre), and Linear (lower) Interpolation*

## 7.3.2. Final Design Decisions and Adjustments

Before continuing on to finalise PRETTYSPEECH, we wish make a change to the way that the system handles target pitches. Previously, the system has considered a multiplicative frequency shift. Here, we will instead apply these shifts on a logarithmic scale, allowing us to match certain behaviours in music. For example, a pitch which is one octave higher than another has double the frequency, while one octave lower corresponds to half of the original's frequency. By moving along logarithmic twelfths, we move along twelve-tone equal temperament, which is the standard musical tuning of piano keys; this correspondence is shown in Table 34.

*Table 34: Table of Shifts of Frequencies in Twelve-Tone Equal Temperament*

| Musical Note | Multiplicative Shift from a' | Logarithmic Shift from a' |
|---|---|---|
| a'' | 2 | 1 |
| g#''/a♭'' | 1.88775 | 11/12 |
| g'' | 1.78180 | 10/12 |
| f#''/g♭'' | 1.68179 | 9/12 |
| f'' | 1.58740 | 8/12 |
| e'' | 1.49831 | 7/12 |
| d#''/e♭'' | 1.41421 | 6/12 |
| d'' | 1.33484 | 5/12 |
| c#''/d♭'' | 1.25992 | 4/12 |
| c'' | 1.18921 | 3/12 |
| b' | 1.12246 | 2/12 |
| a#'/b♭' | 1.05946 | 1/12 |
| a' | 1 | 0 |

By defining our system in a way that easily corresponds to standard piano key pitches, we can more easily produce synthesized speech matching a desired musical pitch or rhythm. As such, our interface is defined such that a +1 pitch shift gives a shift up by one twelve-tone key; a -1 pitch shift gives a shift down by one twelve-tone key, and so on. This should help to let us determine more acoustically pleasing synthesized speech, as our sentence intonation can more easily match musical intonation. If we assign pitches manually, we can even have the speech synthesizer sing, by assigning each word the desired pitch.

## 7.3.3. GUI Design

Now that we have determined our last design decision, we need to design the remainder of our user interface. The layout of this interface is shown in Figure 45.

In the Presets pane, a user can easily access preset input text, as well as predefined prosodic settings and curves. In the Synthesis pane, a user inputs text and chooses the diphone bank to be used in synthesizing speech. Pressing the "Preprocess" button processes the input text and then outputs it as tokenised speech with corresponding Part of Speech tag and pronunciation in the Processed Text pane. If desired, a user can then press the Synthesize button in the Synthesis pane to produce this processed input text as speech. However, the speech does not have any prosodic overlay applied.

A user can apply prosodic overlay manually by setting the volume, pitch, and duration shift in the Processed Text pane. Alternatively, these values can be determined based on the predetermined rules defined in the various parts of the Settings pane by pressing the Overlay Prosody button.

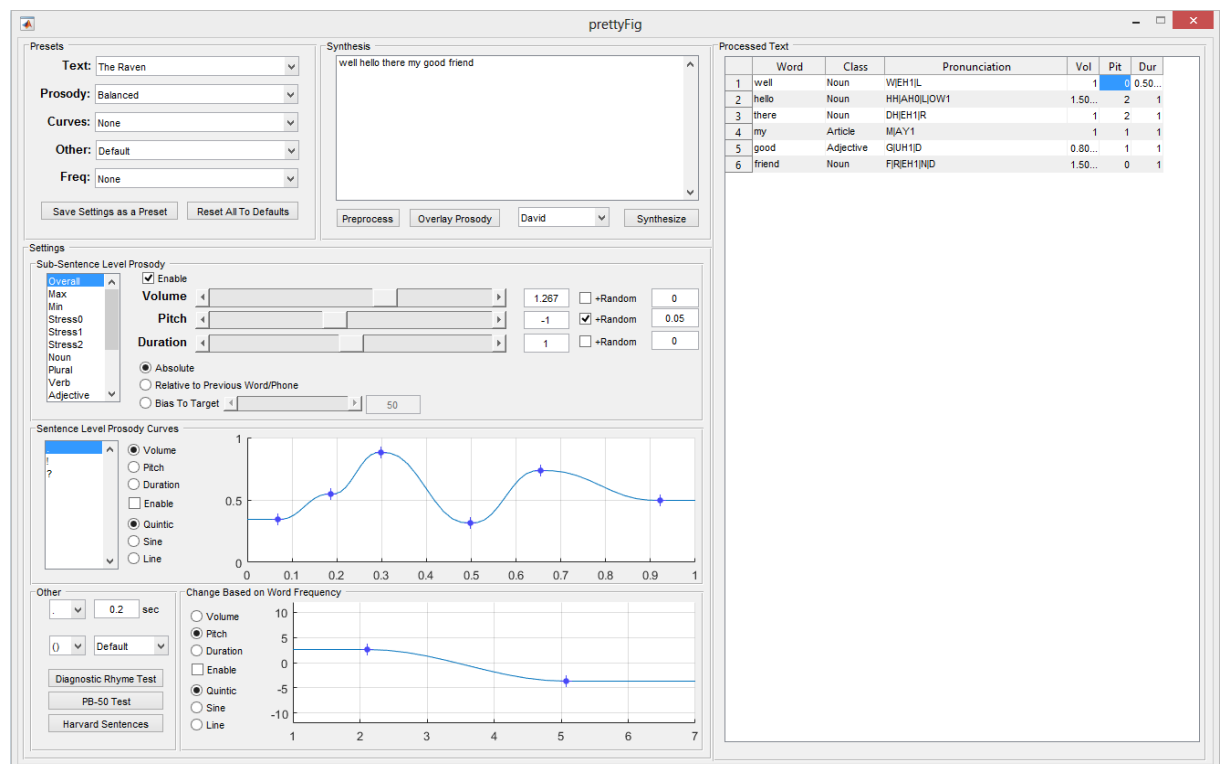

*Figure 45: PRETTYSPEECH GUI*

Users can manually determine the target volume, pitch, and duration of differently stressed phones, different parts of speech, or the overall waveform. Maximum and minimum values can also be defined here. Sentence prosody curves can be defined in the manner previously discussed, and particular curves selected and enabled through their associated radio buttons and a checkbox. The pause for each punctuation mark can be set in the Other pane, where we can also define a different diphone bank to be used for text within different kinds of brackets. This pane also contains buttons which let us set the input text to run a Diagnostic Rhyme Test, a PB-50 test, or one of the Harvard sentence sets.

This GUI permits an exceptionally wide range of prosodic behaviours to easily be defined by the user's input. If we do not overlay any prosody, we are essentially producing BADSPEECH, though with the added expansion of accepting arbitrary punctuated input. Similarly, if we only define behaviours for different stresses, the output operates in much the same way that ODDSPEECH does.

While the design of the PRETTYSPEECH GUI is mostly finalised, and the technical backend is complete at the present time, the connection between the interface and the underlying synthesis code has not yet been fully programmed. This task is not difficult from a technical standpoint, but substantially time consuming, so it has been left until after the completion and submission of this report. It is expected that this should be completed by demonstration day.

# 8. Testing, Future Research, and Conclusions

Now that PRETTYSPEECH has been completed, we have three systems with increasing complexity, each of which should provide an improvement in some way over the last. In this section, we will validate the effectiveness of our system experimentally.

## 8.1. Experimental Testing

Throughout the development of our speech synthesis systems, many informal tests were conducted to determine characteristics such as naturalness and intelligibility. These were enough to determine that our system was progressively improving. Here, we wish to perform some more formal tests, which can be evaluated and analysed in a more definite manner.

### 8.1.1. Comparison of Different Diphone Banks

As has been discussed earlier in the paper, the intelligibility of a diphone speech synthesis system is highly dependent on the diphone recordings that it is concatenating to produce the speech waveform. In this project, four separate diphone banks were recorded, as shown in Table 35.

*Table 35: Recorded Diphone Banks*

| Name | Sex | Accent |
|---|---|---|
| **David** | Male | Australian |
| **Alice** | Female | British |
| **Josh** | Male | American |
| **Megan** | Female | American |

These diphone banks were all recorded from native English speakers, and obtained using our automatic diphone extraction system at the medium speed setting. Each recording took place over the course of approximately two hours, including breaks for rest. We have recorded a male and a female voice from speakers with non-American sociolinguistic accents, David and Alice; we have also recorded male and female voices from speakers with American accents, Josh and Megan. All recordings were made using a Blue Yeti USB microphone in conjunction with a cloth mesh pop filter, as shown in Figure 46.

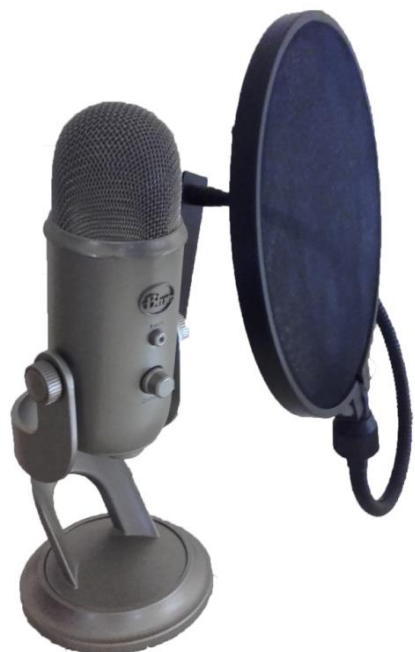

*Figure 46: Blue Yeti Microphone and Cloth Mesh Pop Filter used for recording*

Before continuing, we will make a few notes on the conditions under which these different banks were recorded, as well as some of the notable characteristics of each.

The David diphone bank's speaker has a General Australian English sociolinguistic accent and a lower vocal pitch. The recording was performed under reasonably quiet conditions, with a consistent level of background noise. All diphones were recorded continuously without breaks in recording. The recorded diphones are similar in pitch, and diphones including the same phone are close to the same acoustic character, resulting in a consistent synthesized speech waveform.

The Alice diphone bank's speaker has an Estuary English sociolinguistic accent and a higher vocal pitch. The recording was performed in a house which was located near a main road. The level of background noise was therefore inconsistent, requiring some re-recordings of some phones due to interference. The diphones are similar in pitch and the targets are close in acoustic character. However, some of these target phones, particularly vowels, were recorded at a different phonetic target than the actual desired target due to the speaker's accent. Interestingly, listeners could notice the speaker's sociolinguistic accent through the synthesized voice, despite the American word pronunciations being used.

The Josh diphone bank's speaker has a North-Central American sociolinguistic accent and a lower vocal pitch. The recording was performed under quiet conditions with consistent background noise. The speaker has a background in linguistics, so was able to consistently produce the desired target phones. However, the speaker was inconsistent in their production of each phone's target pitch. This causes the synthesized speech's pitch to change over time, even with no pitch shifting applied.

The Megan diphone bank's speaker has a Californian English sociolinguistic accent and a higher vocal pitch. The recording was performed under reasonably quiet conditions, and the recorded diphones are consistent in pitch, and reasonably consistent in articulation.

As determined in Section 3, our primary tools are the Diagnostic Rhyme Test (DRT), the Phonetically Balanced Monosyllabic Word Lists (PB-50), the Harvard Sentences, and the MOS-X. Since resources are limited, and only a short amount of time was available for testing, it was decided that performing a Harvard Sentences transcription test and then asking a subset of the MOS-X questions would be an effective tradeoff between speed of administering the test and the usefulness of the data then obtained. The MOS-X questions about social impression were not asked, and neither was the question about appropriate emphasis on the word level.

We want to test each diphone bank's effectiveness both with both BADSPEECH (no pitch or duration modification) and ODDSPEECH, with an overall duration shifting of 0.7 (so speech is said faster, taking 70% of the original time taken), and a pitch shifting of 1.5 on phones with primary stress. (As this is for ODDSPEECH's behaviour for pitch shifting, this means that the new frequency of stressed phones is the original's frequency multiplied by 1.5.)

Eight different tests were performed, with each using one of the first 8 sets of Harvard Sentences. A group of 5 listeners was used to determine the effectiveness of these. These listeners were all native Australian English speakers. Listeners were prompted to listen to each sentence and then transcribe it. Once all listeners had finished transcribing the sentence, the following sentence was played.

The detailed results of these tests are available in Appendix 4 from Page A-21. The consolidated results of the test are shown in Table 36; the average percentage accuracy of the semantic components of the Harvard transcription are given, as well as the average ratings of MOS-X intelligibility, naturalness, and prosody.

*Table 36: Harvard and MOS-X Results of Testing Diphone Banks*

|  | BADSPEECH | | | | ODDSPEECH | | | |
|---|---|---|---|---|---|---|---|---|
|  | David | Alice | Josh | Megan | David | Alice | Josh | Megan |
| **Harvard Transcription** | 62% | 57% | 57% | 59% | 46% | 40% | 73% | 27% |
| **MOS-X Intelligibility** | 2.9 | 2.8 | 1.8 | 2.6 | 1.5 | 1.6 | 3.6 | 1.4 |
| **MOS-X Naturalness** | 2.55 | 3.2 | 1.75 | 3.1 | 2.15 | 1.95 | 3.2 | 1.65 |
| **MOS-X Prosody** | 2.1 | 2.1 | 1.3 | 2.7 | 1.4 | 1.4 | 2.8 | 1.4 |

The transcription accuracies for BADSPEECH synthesis from the different diphone banks were quite similar. The Alice and Megan diphone banks had higher naturalness and prosody than the David and Josh recordings, though all evaluations of intelligibility, naturalness, and prosody were quite low. The highest BADSPEECH naturalness is produced from the Megan diphone bank, while the lowest naturalness was from the Josh diphone bank.

We also note that for three of the four diphone banks, the pitch and duration shifting applied in ODDSPEECH reduces the intelligibility, naturalness, and prosody of the speech produced. Yet when ODDSPEECH is used with the Josh diphone bank, it provides the expected improvement in intelligibility and naturalness: we find a transcription accuracy of 73%, which is by far the highest ranking. From just looking at the numerical results, it is not obvious why this should be the case.

As previously stated, the Josh diphone bank's speaker has a North-Central American sociolinguistic accent and a background in linguistics. They were able to consistently produce the correct phones as desired. As such, the pitch-shifted diphones should closely retain their correct acoustic character. For other diphone banks, where the speaker either did not have an American sociolinguistic accent or a background in linguistics, it is probable that the speaker's articulation was inconsistent or inaccurate over the course of recordings. The shifting process may have helped to accentuate these errors, resulting in the lower results.

Another contributor to the lower ODDSPEECH evaluations of the other voicebanks could be the empirically chosen smoothing settings in our PSOLA algorithm for pitch shifting. We want the smoothing step to identify each glottal pulse: a poorly chosen value may find multiple peaks where only one should be find, group together multiple peaks into one, or simply miss peaks entirely. These settings were based on what seemed intelligible to the developer of the system; based on the testing of other listeners, this was evidently less intelligible to a general listener.

We should also consider that the particular words used within the Harvard sentence sets may bias our results. While the Harvard sentences are phonetically balanced to English, they are not *diphonetically* balanced. As such, our measurements of intelligibility and naturalness may vary depending on which particular sentence set is being used. This hypothesis is supported by observing that there were certain sentences in our tests which all listeners were able to transcribe perfectly, while other sentences from the same speaker were difficult for anyone to transcribe.

This inconsistency is likely due to inconsistency in our recorded diphones. Some automatically extracted diphones may be produced by the speaker at a consistent pitch or articulation, while others are not. This problem could likely be addressed through greater manual curation of our diphone set, or using speakers who can more clearly articulate their phones and do so with consistent pitch.

Unfortunately, due to resource constraints, further intelligibility and naturalness testing of the system cannot be completed in the available time. However, the results here indicate that a diphone recording from a professional speaker should produce even more intelligible speech after shifting.

## 8.1.2. Evaluation of Computational Speed
In many speech synthesis applications, we want the system to produce synthesized speech as soon as the input is received. Unfortunately, stream processing in MATLAB is more complex than playing back a completed audio waveform. Because of this, our synthesis code constructs the complete waveform before playing it back over speakers, resulting in a pause before playback. This begs the question: could our system synthesize speech in real time if we used audio buffering?

To verify this experimentally, MATLAB's *tic* and *toc* functions were used to determine how long the system takes to produce a synthesized speech waveform for all of the Harvard sentences, and compare this to the duration of the waveform. The David diphone bank and ODDSPEECH settings were used, so this test also includes the computational load of pitch/duration shifting and part of speech tagging. This gives us a resulting waveform that is 5080 seconds long, which is produced in only 517 seconds; the waveform is generated in approximately 10% of the time it takes to play back the audio. As such, a modified version of our system could quite easily produce and play a synthesized speech waveform in real-time as soon as input is received.

This result was to be expected, due to our choice of computationally efficient time-domain signal processing techniques. An implementation in C rather than MATLAB would likely make this even faster. Sadly, making the change to streaming synthesis would require fundamentally altering almost all of the previously written code, making it infeasible to implement in the time available.

## 8.2. Possible Future Research
While we have only performed basic tests of our system due to resource and time constraints, these results indicate that it has not reached the level of intelligibility or naturalness which we initially set out to achieve. Given the open-ended nature of the problem, there are of course many further improvements that could be made to improve the system. Some examples of these are as follows:
- Our system has exclusively used freely available databases for pronunciation and training; it is possible the use of a more advanced proprietary database could improve our results.
- While this is a diphone synthesis system, natural speech is separated into individual syllables which have distinct pronunciations and effects on intelligibility and prosody. Determining appropriate syllabication of words could help improve the effectiveness of our system.
- Arpabet transcriptions make no distinction between adjacent phones where a hiatus transition should occur and where a diphthong transition should occur. Further, the limited nature of the Arpabet phone set makes no distinction between similar phones, such as the distinct dental /t/ and the alveolar /t̪/ both being the T phone in Arpabet. A switch to IPA phones would require a much larger database of diphones to be captured. However, it would let us synthesize English speech using any sociolinguistic accent, or even produce speech from other languages.
- Our diphone capturing methodology currently only captures one diphone at a time. A more advanced system operating in a similar way could extract multiple diphones from entire prompted words. Further, we perform no de-essing on the recorded speech; this was not found to be a problem with the speakers or microphone used, but would likely be a useful feature for general purpose speech capture.
- Our PSOLA smoothing parameters are determined from empirical judgement. An automatic approach could determine this smoothing length. Alternatively, a different technique might be more effective at isolating the contributing factors of each glottal impulse.
- Determining which ODDSPEECH or PRETTYSPEECH settings are most effective is left to the end user in this project; we have only facilitated the technical possibility of changing the pitch, duration, and volume. A linguist could likely tune this system in a more aesthetically pleasing manner to the average listener than an engineer.

## 8.3. Conclusions

For this project, our objective was to produce a speech synthesis system which produces intelligible and natural-sounding English language speech. Unfortunately, based on our testing, the system we have produced leaves much to be desired: we have only been able to achieve an evaluation of 73% intelligibility, where the bar for modern systems is typically at 95% intelligibility or more. Similarly, our evaluation of naturalness is also quite low.

However, in comparing our final product to other systems, we should consider the relative differences that set our implementation apart. Our diphone banks are not from recordings of professional speakers over the course of several days, but from everyday people over the course of only two hours. Our diphones were programmatically extracted, where the industry norm is for a more time-intensive manual extraction. We have only used freely available databases, whereas more advanced systems will often use more complete and higher quality resources. Considering these constraints, the level of intelligibility that we have achieved indicates that our system is built on a solid foundation.

Further, our system can reliably find acceptable pronunciations for arbitrary input text, as well as identifying the correct pronunciation for heterophonic homographs. For a speaker who consistently produces the correct phonetic articulation, our pitch and duration shifting techniques demonstrably improve both the intelligibility and naturalness of our synthesized speech.

While our final product is quite limited relative to top of the line systems, which are produced over longer periods by large teams, our implementation demonstrates the effectiveness of the techniques used. As a project by a single researcher, the level of progress accomplished in this project is promising, and would be a solid foundation for more advanced development. The primary factors limiting the naturalness and intelligibility of the current system are the ability of the speaker who is being recorded and the particular choice of emphasis settings in our system. Improving the former would require the contribution of a talented speaker; improving the latter would require fine-tuning from a linguistics expert.

As was stated at the beginning of this project, speech synthesis is a complex and cross-disciplinary problem. This project has addressed and provided a solution to all of the engineering challenges involved in the task, through the use of data-driven solutions and signal processing techniques. The engineering component has therefore been completed: further improvement is left as an exercise for the linguist.

# A. Appendices
## A.1. Appendix 1: IPA and Arpabet Tables

*Table 4: Arpabet and IPA Correspondence with Example Transcriptions in General American English*

| Arpabet | IPA | Example | Arpabet Transcription |
|---|---|---|---|
| AA | /ɑ/ | odd | AA D |
| AE | /æ/ | at | AE T |
| AH | /ə/ | hut | HH AH T |
| AO | /ɔ/ | ought | AO T |
| AW | /aʊ/ | cow | K AW |
| AY | /aɪ/ | hide | HH AY D |
| B | /b/ | be | B IY |
| CH | /tʃ/ | cheese | CH IY Z |
| D | /d/ | dee | D IY |
| DH | /ð/ | thee | DH IY |
| EH | /ɛ/ | Ed | EH D |
| ER | /ɝ/ | hurt | HH ER T |
| EY | /eɪ/ | ate | EY T |
| F | /f/ | fee | F IY |
| G | /g/ | green | G R IY N |
| HH | /h/ | he | HH IY |
| IH | /ɪ/ | it | IH T |
| IY | /i/ | eat | IY T |
| JH | /dʒ/ | gee | JH IY |
| K | /K/ | key | K IY |
| L | /ɫ/ | lee | L IY |
| M | /m/ | me | M IY |
| N | /n/ | knee | N IY |
| NG | /ŋ/ | ping | P IH NG |
| OW | /oʊ/ | oat | OW T |
| OY | /ɔɪ/ | toy | T OY |
| P | /p/ | pee | P IY |
| R | /r/ | read | R IY D |
| S | /s/ | sea | S IY |
| SH | /ʃ/ | she | SH IY |
| T | /t/ | tea | T IY |
| TH | /θ/ | theta | TH EY T AH |
| UH | /ʊ/ | hood | HH UH D |
| UW | /u/ | two | T UW |
| V | /v/ | vee | V IY |
| W | /w/ | we | W IY |
| Y | /j/ | yield | Y IY L D |
| Z | /z/ | zee | Z IY |
| ZH | /ʒ/ | seizure | S IY ZH ER |

*A. Appendices*## A.2. Appendix 2: Testing Datasets and Forms

## A.2.1. Diagnostic Rhyme Test

*Table 37: Diagnostic Rhyme Test Word List*

| Voicing | | Nasality | | Sustenation | | Sibilation | | Graveness | | Compactness | |
|---|---|---|---|---|---|---|---|---|---|---|---|
| veal | feel | meat | beat | vee | bee | zee | thee | weed | reed | yield | wield |
| bean | peen | need | deed | sheet | cheat | cheep | keep | peak | teak | key | tea |
| gin | chin | mitt | bit | vill | bill | jilt | gilt | bid | did | hit | fit |
| dint | tint | nip | dip | thick | tick | sing | thing | fin | thin | gill | dill |
| zoo | sue | moot | boot | foo | pooh | juice | goose | moon | noon | coop | poop |
| dune | tune | news | dues | shoes | choose | chew | coo | pool | tool | you | rue |
| vole | foal | moan | bone | those | doze | joe | go | bowl | dole | ghost | boast |
| goat | coat | note | dote | though | dough | sole | thole | fore | thor | show | so |
| zed | said | mend | bend | then | den | jest | guest | met | net | keg | peg |
| dense | tense | neck | deck | fence | pence | chair | care | pent | tent | yen | wren |
| vast | fast | mad | bad | than | dan | jab | gab | bank | dank | gat | bat |
| gaff | calf | nab | dab | shad | chad | sank | thank | fad | thad | shag | sag |
| vault | fault | moss | boss | thong | tong | jaws | gauze | fought | thought | yawl | wall |
| daunt | taunt | gnaw | daw | shaw | chaw | saw | thaw | bong | dong | caught | thought |
| jock | chock | mom | bomb | von | bon | jot | got | wad | rod | hop | fop |
| bond | pond | knock | dock | vox | box | chop | cop | pot | tot | got | dot |

A-2

## A.2.2. Modified Rhyme Test

*Table 38: Modified Rhyme Test Word List*

| | | | | | |
|---|---|---|---|---|---|
| went | sent | bent | dent | tent | rent |
| hold | cold | told | fold | sold | gold |
| pat | pad | pan | path | pack | pass |
| lane | lay | late | lake | lace | lame |
| kit | bit | fit | hit | wit | sit |
| must | bust | gust | rust | dust | just |
| teak | team | teal | teach | tear | tease |
| din | dill | dim | dig | dip | did |
| bed | led | fed | red | wed | shed |
| pin | sin | tin | fin | din | win |
| dug | dung | duck | dud | dub | dun |
| sum | sun | sung | sup | sub | sud |
| seep | seen | seethe | seek | seem | seed |
| not | tot | got | pot | hot | lot |
| vest | test | rest | best | west | nest |
| pig | pill | pin | pip | pit | pick |
| back | bath | bad | bass | bat | ban |
| way | may | say | pay | day | gay |
| pig | big | dig | wig | rig | fig |
| pale | pace | page | pane | pay | pave |
| cane | case | cape | cake | came | cave |
| shop | mop | cop | top | hop | pop |
| coil | oil | soil | toil | boil | foil |
| tan | tang | tap | tack | tam | tab |
| fit | fib | fizz | fill | fig | fin |
| same | name | game | tame | came | fame |
| peel | reel | feel | eel | keel | heel |
| hark | dark | mark | bark | park | lark |
| heave | hear | heat | heal | heap | heath |
| cup | cut | cud | cuff | cuss | cud |
| thaw | law | raw | paw | jaw | saw |
| pen | hen | men | then | den | ten |
| puff | puck | pub | pus | pup | pun |
| bean | beach | beat | beak | bead | beam |
| heat | neat | feat | seat | meat | beat |
| dip | sip | hip | tip | lip | rip |
| kill | kin | kit | kick | king | kid |
| hang | sang | bang | rang | fang | gang |
| took | cook | look | hook | shook | book |
| mass | math | map | mat | man | mad |
| ray | raze | rate | rave | rake | race |
| save | same | sale | sane | sake | safe |
| fill | kill | will | hill | till | bill |
| sill | sick | sip | sing | sit | sin |
| bale | gale | sale | tale | pale | male |
| wick | sick | kick | lick | pick | tick |
| peace | peas | peak | peach | peat | peal |
| bun | bus | but | bug | buck | buff |
| sag | sat | sass | sack | sad | sap |
| fun | sun | bun | gun | run | nun |

## A.2.3. Phonetically Balanced Monosyllabic Word Lists

*Table 39: Phonetically Balanced Monosyllabic Word Lists*

| List 1 | List 2 | List 3 | List 4 | List 5 | List 6 | List 7 | List 8 | List 9 | List 10 |
|---|---|---|---|---|---|---|---|---|---|
| are | awe | ache | bath | add | as | act | ask | arch | ail |
| bad | bait | air | beast | bathe | badge | aim | bid | beef | back |
| bar | bean | bald | bee | beck | best | am | bind | birth | bash |
| bask | blush | barb | blonde | black | bog | but | bolt | bit | bob |
| box | bought | bead | budge | bronze | chart | by | bored | boost | bug |
| cane | bounce | cape | bus | browse | cloth | chop | calf | carve | champ |
| cleanse | bud | cast | bush | cheat | clothes | coast | catch | chess | chance |
| clove | charge | check | cloak | choose | cob | comes | chant | chest | clothe |
| crash | cloud | class | course | curse | crib | cook | chew | clown | cord |
| creed | corpse | crave | court | feed | dad | cut | clod | club | cow |
| death | dab | crime | dodge | flap | deep | dope | cod | crowd | cue |
| deed | earl | deck | dupe | gape | eat | dose | crack | cud | daub |
| dike | else | dig | earn | good | eyes | dwarf | day | ditch | ears |
| dish | fate | dill | eel | greek | fall | fake | deuce | flag | earth |
| end | five | drop | fin | grudge | fee | fling | dumb | fluff | etch |
| feast | frog | fame | float | high | flick | fort | each | foe | fir |
| fern | gill | far | frown | hill | flop | gasp | ease | fume | flaunt |
| folk | gloss | fig | hatch | inch | forge | grade | fad | fuse | flight |
| ford | hire | flush | heed | kid | fowl | gun | flip | gate | force |
| fraud | hit | gnaw | hiss | lend | gage | him | food | give | goose |
| fuss | hock | hurl | hot | love | gap | jug | forth | grace | gull |
| grove | job | jam | how | mast | grope | knit | freak | hoof | hat |
| heap | log | law | kite | nose | hitch | mote | frock | ice | hurt |
| hid | moose | leave | merge | odds | hull | mud | front | itch | jay |
| hive | mute | lush | lush | owls | jag | nine | guess | key | lap |
| hunt | nab | muck | neat | pass | kept | off | hum | lit | line |
| is | need | neck | new | pipe | leg | pent | jell | mass | maze |
| mange | niece | nest | oils | puff | mash | phase | kill | nerve | mope |
| no | nut | oak | or | punt | nigh | pig | left | noose | nudge |
| nook | our | path | peck | rear | ode | plod | lick | nuts | page |
| not | perk | please | pert | rind | prig | pounce | look | odd | pink |
| pan | pick | pulse | pinch | rode | prime | quiz | night | pact | plus |
| pants | pit | rate | pod | roe | pun | raid | pint | phone | put |
| pest | quart | rouse | race | scare | pus | range | queen | reed | rape |
| pile | rap | shout | rack | shine | raise | rash | rest | root | real |
| plush | rib | sit | rave | shove | ray | rich | rhyme | rude | rip |
| rag | scythe | size | raw | sick | reap | roar | rod | sip | rush |
| rat | shoe | sob | rut | sly | rooms | sag | roll | smart | scrub |
| ride | sludge | sped | sage | solve | rough | scout | rope | spud | slug |
| rise | snuff | stag | scab | thick | scan | shaft | rot | ten | snipe |
| rub | start | take | shed | thud | shank | siege | shack | than | staff |
| slip | suck | thrash | shin | trade | slouch | sin | slide | thank | tag |
| smile | tan | toil | sketch | true | sup | sledge | spice | throne | those |
| strife | tang | trip | slap | tug | thigh | sniff | this | toad | thug |
| such | them | turf | sour | vase | thus | south | thread | troop | tree |
| then | trash | vow | starve | watch | tongue | though | till | weak | valve |
| there | vamp | wedge | strap | wink | wait | whiff | us | wild | void |
| toe | vast | wharf | test | wrath | wasp | wire | wheeze | wipe | wade |
| use | ways | who | tick | yawn | wife | woe | wig | with | wake |
| wheat | wish | why | touch | zone | writ | woo | yeast | year | youth |

| List 11 | List 12 | List 13 | List 14 | List 15 | List 16 | List 17 | List 18 | List 19 | List 20 |
|---------|---------|---------|---------|---------|---------|---------|---------|---------|---------|
| arc | and | bat | at | bell | aid | all | aims | age | ace |
| arm | ass | beau | barn | blind | barge | apt | art | bark | base |
| beam | ball | change | bust | boss | book | bet | axe | bay | beard |
| bliss | bluff | climb | car | cheap | cheese | big | bale | bough | brass |
| chunk | cad | corn | clip | cost | cliff | booth | bless | buzz | cart |
| clash | cave | durb | coax | cuff | closed | brace | camp | cab | click |
| code | chafe | deaf | curve | dive | crews | braid | cat | cage | clog |
| crutch | chair | dog | cute | dove | dame | buck | chaff | calve | cork |
| cry | chap | elk | darn | edge | din | case | chain | cant | crate |
| dip | chink | elm | dash | elf | drape | clew | chill | chat | did |
| doubt | cling | few | dead | fact | droop | crush | chip | chose | duke |
| drake | clutch | fill | douse | flame | dub | dart | claw | crude | eye |
| dull | depth | fold | dung | fleet | fifth | dine | claws | cup | fair |
| feel | dime | for | fife | gash | fright | falls | crab | dough | fast |
| fine | done | gem | foam | glove | gab | feet | cub | drug | flash |
| frisk | fed | grape | grate | golf | gas | fell | debt | dune | gang |
| fudge | flog | grave | group | hedge | had | fit | dice | ebb | get |
| goat | flood | hack | heat | hole | hash | form | dot | fan | gob |
| have | foot | hate | howl | jade | hose | fresh | fade | find | hump |
| hog | fought | hook | hunk | kiss | ink | gum | fat | flank | in |
| jab | frill | jig | isle | less | kind | hence | flare | fond | joke |
| jaunt | gnash | made | kick | may | knee | hood | fool | gin | judge |
| kit | greet | mood | lathe | mesh | lay | if | freeze | god | lid |
| lag | hear | mop | life | mitt | leash | last | got | gyp | mow |
| latch | hug | moth | me | mode | louse | ma | grab | hike | pack |
| loss | hunch | muff | muss | morn | map | mist | gray | hut | pad |
| low | jaw | much | news | naught | nap | myth | grew | lad | pew |
| most | jazz | my | nick | ninth | next | ox | gush | led | puss |
| mouth | jolt | nag | nod | oath | part | paid | hide | lose | quip |
| net | knife | nice | oft | own | pitch | pare | his | lust | ramp |
| pond | lash | nip | prude | pup | pump | past | hush | notch | retch |
| probe | laugh | ought | purge | quick | rock | pearl | lime | on | robe |
| prod | ledge | owe | quack | scow | rogue | peg | lip | paste | roost |
| punk | loose | patch | rid | sense | rug | plow | loud | perch | rouge |
| purse | out | pelt | shook | shade | rye | press | lung | raft | rout |
| reef | park | plead | shrug | shrub | sang | rage | lynch | rote | salve |
| rice | priest | price | sing | sir | sheep | reach | note | rule | seed |
| risk | reek | pug | slab | slash | sheik | ridge | ouch | sat | sigh |
| sap | ripe | scuff | smite | so | soar | roam | rob | shy | skid |
| shop | romp | side | soil | tack | stab | scratch | rose | sill | slice |
| shot | rove | sled | stuff | teach | stress | sell | sack | slid | slush |
| sign | set | smash | tell | that | suit | ship | sash | splash | soak |
| snow | shut | smooth | tent | time | thou | shock | share | steed | souse |
| sprig | sky | soap | thy | tinge | three | stride | sieve | thief | theme |
| spy | sod | stead | tray | tweed | thresh | tube | thaw | throat | through |
| stiff | throb | taint | vague | vile | tire | vice | thine | up | tilt |
| tab | tile | tap | vote | weave | ton | weep | thorn | wheel | walk |
| urge | vine | thin | wag | wed | tuck | weird | trod | white | wash |
| wave | wage | tip | waif | wide | turn | wine | waste | yes | web |
| wood | wove | wean | wrist | wreck | wield | you | weed | yield | wise |

## A.2.4. Harvard Psychoacoustic Sentences

*Table 40: Harvard Psychoacoustic Sentences*

### List 1
1. The birch canoe slid on the smooth planks.
2. Glue the sheet to the dark blue background.
3. It's easy to tell the depth of a well.
4. These days a chicken leg is a rare dish.
5. Rice is often served in round bowls.
6. The juice of lemons makes fine punch.
7. The box was thrown beside the parked truck.
8. The hogs were fed chopped corn and garbage.
9. Four hours of steady work faced us.
10. Large size in stockings is hard to sell.

### List 2
1. The boy was there when the sun rose.
2. A rod is used to catch pink salmon.
3. The source of the huge river is the clear spring.
4. Kick the ball straight and follow through.
5. Help the woman get back to her feet.
6. A pot of tea helps to pass the evening.
7. Smoky fires lack flame and heat.
8. The soft cushion broke the man's fall.
9. The salt breeze came across from the sea.
10. The girl at the booth sold fifty bonds.

### List 3
1. The small pup gnawed a hole in the sock.
2. The fish twisted and turned on the bent hook.
3. Press the pants and sew a button on the vest.
4. The swan dive was far short of perfect.
5. The beauty of the view stunned the young boy.
6. Two blue fish swam in the tank.
7. Her purse was full of useless trash.
8. The colt reared and threw the tall rider.
9. It snowed, rained, and hailed the same morning.
10. Read verse out loud for pleasure.

### List 4
1. Hoist the load to your left shoulder.
2. Take the winding path to reach the lake.
3. Note closely the size of the gas tank.
4. Wipe the grease off his dirty face.
5. Mend the coat before you go out.
6. The wrist was badly strained and hung limp.
7. The stray cat gave birth to kittens.
8. The young girl gave no clear response.
9. The meal was cooked before the bell rang.
10. What joy there is in living.

### List 5
1. A king ruled the state in the early days.
2. The ship was torn apart on the sharp reef.
3. Sickness kept him home the third week.
4. The wide road shimmered in the hot sun.
5. The lazy cow lay in the cool grass.
6. Lift the square stone over the fence.
7. The rope will bind the seven books at once.
8. Hop over the fence and plunge in.
9. The friendly gang left the drug store.
10. Mesh mire keeps chicks inside.

### List 6
1. The frosty air passed through the coat.
2. The crooked maze failed to fool the mouse.
3. Adding fast leads to wrong sums.
4. The show was a flop from the very start.
5. A saw is a tool used for making boards.
6. The wagon moved on well oiled wheels.
7. March the soldiers past the next hill.
8. A cup of sugar makes sweet fudge.
9. Place a rosebush near the porch steps.
10. Both lost their lives in the raging storm.

### List 7
1. We talked of the slide show in the circus.
2. Use a pencil to write the first draft.
3. He ran half way to the hardware store.
4. The clock struck to mark the third period.
5. A small creek cut across the field.
6. Cars and busses stalled in snow drifts.
7. The set of china hit, the floor with a crash.
8. This is a grand season for hikes on the road.
9. The dune rose from the edge of the water.
10. Those words were the cue for the actor to leave.

### List 8
1. A yacht slid around the point into the bay.
2. The two met while playing on the sand.
3. The ink stain dried on the finished page.
4. The walled town was seized without a fight.
5. The lease ran out in sixteen weeks.
6. A tame squirrel makes a nice pet.
7. The horn of the car woke the sleeping cop.
8. The heart beat strongly and with firm strokes.
9. The pearl was worn in a thin silver ring.
10. The fruit peel was cut in thick slices.

# A. Appendices

### List 9
1. The Navy attacked the big task force.
2. See the cat glaring at the scared mouse.
3. There are more than two factors here.
4. The hat brim was wide and too droopy.
5. The lawyer tried to lose his case.
6. The grass curled around the fence post.
7. Cut the pie into large parts.
8. Men strive but seldom get rich.
9. Always close the barn door tight.
10. He lay prone and hardly moved a limb.

### List 10
1. The slush lay deep along the street.
2. A wisp of cloud hung in the blue air.
3. A pound of sugar costs more than eggs.
4. The fin was sharp and cut the clear water.
5. The play seems dull and quite stupid.
6. Bail the boat, to stop it from sinking.
7. The term ended in late June that year.
8. A tusk is used to make costly gifts.
9. Ten pins were set in order.
10. The bill as paid every third week.

### List 11
1. Oak is strong and also gives shade.
2. Cats and dogs each hate the other.
3. The pipe began to rust while new.
4. Open the crate but don't break the glass.
5. Add the sum to the product of these three.
6. Thieves who rob friends deserve jail.
7. The ripe taste of cheese improves with age.
8. Act on these orders with great speed.
9. The hog crawled under the high fence.
10. Move the vat over the hot fire.

### List 12
1. The bark of the pine tree was shiny and dark.
2. Leaves turn brown and yellow in the fall.
3. The pennant waved when the wind blew.
4. Split the log with a quick, sharp blow.
5. Burn peat after the logs give out.
6. He ordered peach pie with ice cream.
7. Weave the carpet on the right hand side.
8. Hemp is a weed found in parts of the tropics.
9. A lame back kept his score low.
10. We find joy in the simplest things.

### List 13
1. Type out three lists of orders.
2. The harder he tried the less he got done.
3. The boss ran the show with a watchful eye.
4. The cup cracked and spilled its contents.
5. Paste can cleanse the most dirty brass.
6. The slang word for raw whiskey is booze.
7. It caught its hind paw in a rusty trap.
8. The wharf could be seen at the farther shore.
9. Feel the heat of the weak dying flame.
10. The tiny girl took off her hat.

### List 14
1. A cramp is no small danger on a swim.
2. He said the same phrase thirty times.
3. Pluck the bright rose without leaves.
4. Two plus seven is less than ten.
5. The glow deepened in the eyes of the sweet girl.
6. Bring your problems to the wise chief.
7. Write a fond note to the friend you cherish.
8. Clothes and lodging are free to new men.
9. We frown when events take a bad turn.
10. Port is a strong wine with a smoky taste.

### List 15
1. The young kid jumped the rusty gate.
2. Guess the results from the first scores.
3. A salt pickle tastes fine with ham.
4. The just claim got the right verdict.
5. These thistles bend in a high wind.
6. Pure bred poodles have curls.
7. The tree top waved in a graceful way.
8. The spot on the blotter was made by green ink.
9. Mud was spattered on the front of his white shirt.
10. The cigar burned a hole in the desk top.

### List 16
1. The empty flask stood on the tin tray.
2. A speedy man can beat this track mark.
3. He broke a new shoelace that day.
4. The coffee stand is too high for the couch.
5. The urge to write short stories is rare.
6. The pencils have all been used.
7. The pirates seized the crew of the lost ship.
8. We tried to replace the coin but failed.
9. She sewed the torn coat quite neatly.
10. The sofa cushion is red and of light weight.

### List 17
1. The jacket hung on the back of the wide chair.
2. At that high level the air is pure.
3. Drop the two when you add the figures.
4. A filing case is now hard to buy.
5. An abrupt start does not win the prize.
6. Wood is best for making toys and blocks.
7. The office paint was a dull sad tan.
8. He knew the skill of the great young actress.
9. A rag will soak up spilled water.
10. A shower of dirt fell from the hot pipes.

### List 18
1. Steam hissed from the broken valve.
2. The child almost hurt the small dog.
3. There was a sound of dry leaves outside.
4. The sky that morning was clear and bright blue.
5. Torn scraps littered the stone floor.
6. Sunday is the best part of the week.
7. The doctor cured him with these pills.
8. The new girl was fired today at noon.
9. They felt gay when the ship arrived in port.
10. Add the store's account to the last cent.

*A. Appendices*

### List 19
1. Acid burns holes in wool cloth.
2. Fairy tales should be fun to write.
3. Eight miles of woodland burned to waste.
4. The third act was dull and tired the players.
5. A young child should not suffer fright.
6. Add the column and put the sum here.
7. We admire and love a good cook.
8. There the flood mark is ten inches.
9. He carved a head from the round block of marble.
10. She has st smart way of wearing clothes.

### List 20
1. The fruit of a fig tree is apple-shaped.
2. Corn cobs can be used to kindle a fire.
3. Where were they when the noise started.
4. The paper box is full of thumb tacks.
5. Sell your gift to a buyer at a good gain.
6. The tongs lay beside the ice pail.
7. The petals fall with the next puff of wind.
8. Bring your best compass to the third class.
9. They could laugh although they were sad.
10. Farmers came in to thresh the oat crop.

### List 21
1. The brown house was on fire to the attic.
2. The lure is used to catch trout and flounder.
3. Float the soap on top of the bath water.
4. A blue crane is a tall wading bird.
5. A fresh start will work such wonders.
6. The club rented the rink for the fifth night.
7. After the dance they went straight home.
8. The hostess taught the new maid to serve.
9. He wrote his last novel there at the inn.
10. Even the worst will beat his low score.

### List 22
1. The cement had dried when he moved it.
2. The loss of the second ship was hard to take.
3. The fly made its way along the wall.
4. Do that with a wooden stick.
5. Lire wires should be kept covered.
6. The large house had hot water taps.
7. It is hard to erase blue or red ink.
8. Write at once or you may forget it.
9. The doorknob was made of bright clean brass.
10. The wreck occurred by the bank on Main Street.

### List 23
1. A pencil with black lead writes best.
2. Coax a young calf to drink from a bucket.
3. Schools for ladies teach charm and grace.
4. The lamp shone with a steady green flame.
5. They took the axe and the saw to the forest.
6. The ancient coin was quite dull and worn.
7. The shaky barn fell with a loud crash.
8. Jazz and swing fans like fast music.
9. Rake the rubbish up and then burn it.
10. Slash the gold cloth into fine ribbons.

### List 24
1. Try to have the court decide the case.
2. They are pushed back each time they attack.
3. He broke his ties with groups of former friends.
4. They floated on the raft to sun their white backs.
5. The map had an X that meant nothing.
6. Whitings are small fish caught in nets.
7. Some ads serve to cheat buyers.
8. Jerk the rope and the bell rings weakly.
9. A waxed floor makes us lose balance.
10. Madam, this is the best brand of corn.

### List 25
1. On the islands the sea breeze is soft and mild.
2. The play began as soon as we sat down.
3. This will lead the world to more sound and fury
4. Add salt before you fry the egg.
5. The rush for funds reached its peak Tuesday.
6. The birch looked stark white and lonesome.
7. The box is held by a bright red snapper.
8. To make pure ice, you freeze water.
9. The first worm gets snapped early.
10. Jump the fence and hurry up the bank.

### List 26
1. Yell and clap as the curtain slides back.
2. They are men nho walk the middle of the road.
3. Both brothers wear the same size.
4. In some forin or other we need fun.
5. The prince ordered his head chopped off.
6. The houses are built of red clay bricks.
7. Ducks fly north but lack a compass.
8. Fruit flavors are used in fizz drinks.
9. These pills do less good than others.
10. Canned pears lack full flavor.

### List 27
1. The dark pot hung in the front closet.
2. Carry the pail to the wall and spill it there.
3. The train brought our hero to the big town.
4. We are sure that one war is enough.
5. Gray paint stretched for miles around.
6. The rude laugh filled the empty room.
7. High seats are best for football fans.
8. Tea served from the brown jug is tasty.
9. A dash of pepper spoils beef stew.
10. A zestful food is the hot-cross bun.

### List 28
1. The horse trotted around the field at a brisk pace.
2. Find the twin who stole the pearl necklace.
3. Cut the cord that binds the box tightly.
4. The red tape bound the smuggled food.
5. Look in the corner to find the tan shirt.
6. The cold drizzle will halt the bond drive.
7. Nine men were hired to dig the ruins.
8. The junk yard had a mouldy smell.
9. The flint sputtered and lit a pine torch.
10. Soak the cloth and drown the sharp odor.

*A. Appendices*

**List 29**
1. The shelves were bare of both jam or crackers.
2. A joy to every child is the swan boat.
3. All sat frozen and watched the screen.
4. ii cloud of dust stung his tender eyes.
5. To reach the end he needs much courage.
6. Shape the clay gently into block form.
7. The ridge on a smooth surface is a bump or flaw.
8. Hedge apples may stain your hands green.
9. Quench your thirst, then eat the crackers.
10. Tight curls get limp on rainy days.

**List 30**
1. The mute muffled the high tones of the horn.
2. The gold ring fits only a pierced ear.
3. The old pan was covered with hard fudge.
4. Watch the log float in the wide river.
5. The node on the stalk of wheat grew daily.
6. The heap of fallen leaves was set on fire.
7. Write fast, if you want to finish early.
8. His shirt was clean but one button was gone.
9. The barrel of beer was a brew of malt and hops.
10. Tin cans are absent from store shelves.

**List 31**
1. Slide the box into that empty space.
2. The plant grew large and green in the window.
3. The beam dropped down on the workmen's head.
4. Pink clouds floated JTith the breeze.
5. She danced like a swan, tall and graceful.
6. The tube was blown and the tire flat and useless.
7. It is late morning on the old wall clock.
8. Let's all join as we sing the last chorus.
9. The last switch cannot be turned off.
10. The fight will end in just six minutes.

**List 32**
1. The store walls were lined with colored frocks.
2. The peace league met to discuss their plans.
3. The rise to fame of a person takes luck.
4. Paper is scarce, so write with much care.
5. The quick fox jumped on the sleeping cat.
6. The nozzle of the fire hose was bright brass.
7. Screw the round cap on as tight as needed.
8. Time brings us many changes.
9. The purple tie was ten years old.
10. Men think and plan and sometimes act.

**List 33**
1. Fill the ink jar with sticky glue.
2. He smoke a big pipe with strong contents.
3. We need grain to keep our mules healthy.
4. Pack the records in a neat thin case.
5. The crunch of feet in the snow was the only sound.
6. The copper bowl shone in the sun's rays.
7. Boards will warp unless kept dry.
8. The plush chair leaned against the wall.
9. Glass will clink when struck by metal.
10. Bathe and relax in the cool green grass.

**List 34**
1. Nine rows of soldiers stood in line.
2. The beach is dry and shallow at low tide.
3. The idea is to sew both edges straight.
4. The kitten chased the dog down the street.
5. Pages bound in cloth make a book.
6. Try to trace the fine lines of the painting.
7. Women form less than half of the group.
8. The zones merge in the central part of town.
9. A gem in the rough needs work to polish.
10. Code is used when secrets are sent.

**List 35**
1. Most of the new is easy for us to hear.
2. He used the lathe to make brass objects.
3. The vane on top of the pole revolved in the wind.
4. Mince pie is a dish served to children.
5. The clan gathered on each dull night.
6. Let it burn, it gives us warmth and comfort.
7. A castle built from sand fails to endure.
8. A child's wit saved the day for us.
9. Tack the strip of carpet to the worn floor.
10. Next Tuesday we must vote.

**List 36**
1. Pour the stew from the pot into the plate.
2. Each penny shone like new.
3. The man went to the woods to gather sticks.
4. The dirt piles were lines along the road.
5. The logs fell and tumbled into the clear stream.
6. Just hoist it up and take it away,
7. A ripe plum is fit for a king's palate.
8. Our plans right now are hazy.
9. Brass rings are sold by these natives.
10. It takes a good trap to capture a bear.

**List 37**
1. Feed the white mouse some flower seeds.
2. The thaw came early and freed the stream.
3. He took the lead and kept it the whole distance.
4. The key you designed will fit the lock.
5. Plead to the council to free the poor thief.
6. Better hash is made of rare beef.
7. This plank was made for walking on.
8. The lake sparkled in the red hot sun.
9. He crawled with care along the ledge.
10. Tend the sheep while the dog wanders.

**List 38**
1. It takes a lot of help to finish these.
2. Mark the spot with a sign painted red.
3. Take two shares as a fair profit.
4. The fur of cats goes by many names.
5. North winds bring colds and fevers.
6. He asks no person to vouch for him.
7. Go now and come here later.
8. A sash of gold silk will trim her dress.
9. Soap can wash most dirt away.
10. That move means the game is over.

# A. Appendices

**List 39**
1. He wrote down a long list of items.
2. A siege will crack the strong defense.
3. Grape juice and water mix well.
4. Roads are paved with sticky tar.
5. Fake &ones shine but cost little.
6. The drip of the rain made a pleasant sound.
7. Smoke poured out of every crack.
8. Serve the hot rum to the tired heroes.
9. Much of the story makes good sense.
10. The sun came up to light the eastern sky.

**List 40**
1. Heave the line over the port side.
2. A lathe cuts and trims any wood.
3. It's a dense crowd in two distinct ways.
4. His hip struck the knee of the next player.
5. The stale smell of old beer lingers.
6. The desk was firm on the shaky floor.
7. It takes heat to bring out the odor.
8. Beef is scarcer than some lamb.
9. Raise the sail and steer the ship northward.
10. The cone costs five cents on Mondays.

**List 41**
1. A pod is what peas always grow in.
2. Jerk the dart from the cork target.
3. No cement will hold hard wood.
4. We now have a new base for shipping.
5. The list of names is carved around the base.
6. The sheep were led home by a dog.
7. Three for a dime, the young peddler cried.
8. The sense of smell is better than that of touch.
9. No hardship seemed to keep him sad.
10. Grace makes up for lack of beauty.

**List 42**
1. Nudge gently but wake her now.
2. The news struck doubt into restless minds.
3. Once we stood beside the shore.
4. A chink in the wall allowed a draft to blow.
5. Fasten two pins on each side.
6. A cold dip restores health and zest.
7. He takes the oath of office each March.
8. The sand drifts over the sill of the old house.
9. The point of the steel pen was bent and twisted.
10. There is a lag between thought and act.

**List 43**
1. Seed is needed to plant the spring corn.
2. Draw the chart with heavy black lines.
3. The boy owed his pal thirty cents.
4. The chap slipped into the crowd and was lost.
5. Hats are worn to tea and not to dinner.
6. The ramp led up to the wide highway.
7. Beat the dust from the rug onto the lawn.
8. Say it slow!y but make it ring clear.
9. The straw nest housed five robins.
10. Screen the porch with woven straw mats.

**List 44**
1. This horse will nose his way to the finish.
2. The dry wax protects the deep scratch.
3. He picked up the dice for a second roll.
4. These coins will be needed to pay his debt.
5. The nag pulled the frail cart along.
6. Twist the valve and release hot steam.
7. The vamp of the shoe had a gold buckle.
8. The smell of burned rags itches my nose.
9. Xew pants lack cuffs and pockets.
10. The marsh will freeze when cold enough.

**List 45**
1. They slice the sausage thin with a knife.
2. The bloom of the rose lasts a few days.
3. A gray mare walked before the colt.
4. Breakfast buns are fine with a hot drink.
5. Bottles hold four kinds of rum.
6. The man wore a feather in his felt hat.
7. He wheeled the bike past. the winding road.
8. Drop the ashes on the worn old rug.
9. The desk and both chairs were painted tan.
10. Throw out the used paper cup and plate.

**List 46**
1. A clean neck means a neat collar.
2. The couch cover and hall drapes were blue.
3. The stems of the tall glasses cracked and broke.
4. The wall phone rang loud and often.
5. The clothes dried on a thin wooden rack.
6. Turn on the lantern which gives us light.
7. The cleat sank deeply into the soft turf.
8. The bills were mailed promptly on the tenth of the month.
9. To have is better than to wait and hope.
10. The price is fair for a good antique clock.

**List 47**
1. The music played on while they talked.
2. Dispense with a vest on a day like this.
3. The bunch of grapes was pressed into wine.
4. He sent the figs, but kept the ripe cherries.
5. The hinge on the door creaked with old age.
6. The screen before the fire kept in the sparks.
7. Fly by night, and you waste little time.
8. Thick glasses helped him read the print.
9. Birth and death mark the limits of life.
10. The chair looked strong but had no bottom.

**List 48**
1. The kite flew wildly in the high wind.
2. A fur muff is stylish once more.
3. The tin box held priceless stones.
4. We need an end of all such matter.
5. The case was puzzling to the old and wise.
6. The bright lanterns were gay on the dark lawn.
7. We don't get much money but we have fun.
8. The youth drove with zest, but little skill.
9. Five years he lived with a shaggy dog.
10. A fence cuts through the corner lot.

**List 49**
1. The way to save money is not to spend much.
2. Shut the hatch before the waves push it in.
3. The odor of spring makes young hearts jump.
4. Crack the walnut with your sharp side teeth.
5. He offered proof in the form of a lsrge chart.
6. Send the stuff in a thick paper bag.
7. A quart of milk is water for the most part.
8. They told wild tales to frighten him.
9. The three story house was built of stone.
10. In the rear of the ground floor was a large passage.

**List 50**
1. A man in a blue sweater sat at the desk.
2. Oats are a food eaten by horse and man.
3. Their eyelids droop for want. of sleep.
4. The sip of tea revives his tired friend.
5. There are many ways to do these things.
6. Tuck the sheet under the edge of the mat.
7. A force equal to that would move the earth.
8. We like to see clear weather.
9. The work of the tailor is seen on each side.
10. Take a chance and win a china doll.

**List 51**
1. Shake the dust from your shoes, stranger.
2. She was kind to sick old people.
3. The dusty bench stood by the stone wall.
4. The square wooden crate was packed to be shipped.
5. We dress to suit the weather of most days.
6. Smile when you say nasty words.
7. A bowl of rice is free with chicken stew.
8. The water in this well is a source of good health.
9. Take shelter in this tent, but keep still.
10. That guy is the writer of a few banned books.

**List 52**
1. The little tales they tell are false.
2. The door was barred, locked, and bolted as well.
3. Ripe pears are fit for a queen's table.
4. A big wet stain was on the round carpet.
5. The kite dipped and swayed, but stayed aloft.
6. The pleasant hours fly by much too soon.
7. The room was crowded with a wild mob.
8. This strong arm shall shield your honor.
9. She blushed when he gave her a white orchid.
10. The beetle droned in the hot June sun.

**List 53**
1. Press the pedal with your left foot.
2. Neat plans fail without luck.
3. The black trunk fell from the landing.
4. The bank pressed for payment of the debt.
5. The theft of the pearl pin was kept secret.
6. Shake hands with this friendly child.
7. The vast space stretched into the far distance.
8. A rich farm is rare in this sandy waste.
9. His wide grin earned many friends.
10. Flax makes a fine brand of paper.

**List 54**
1. Hurdle the pit with the aid of a long pole.
2. A strong bid may scare your partner stiff.
3. Even a just cause needs power to win.
4. Peep under the tent and see the clowns.
5. The leaf drifts along with a slow spin.
6. Cheap clothes are flashy but don't last.
7. A thing of small note can cause despair.
8. Flood the mails with requests for this book.
9. A thick coat of black paint covered all.
10. The pencil was cut to be sharp at both ends.

**List 55**
1. Those last words were a strong statement.
2. He wrote his name boldly at the top of tile sheet.
3. Dill pickles are sour but taste fine.
4. Down that road is the way to the grain farmer.
5. Either mud or dust are found at all times.
6. The best method is to fix it in place with clips.
7. If you mumble your speech will be lost.
8. At night the alarm roused him from a deep sleep.
9. Read just what the meter says.
10. Fill your pack with bright trinkets for the poor.

**List 56**
1. The small red neon lamp went out.
2. Clams are small, round, soft, and tasty.
3. The fan whirled its round blades softly.
4. The line where the edges join was clean.
5. Breathe deep and smell the piny air.
6. It matters not if he reads these words or those.
7. A brown leather bag hung from its strap.
8. A toad and a frog are hard to tell apart.
9. A white silk jacket goes with any shoes.
10. A break in the dam almost caused a flood.

**List 57**
1. Paint the sockets in the wall dull green.
2. The child crawled into the dense grass.
3. Bribes fail where honest men work.
4. Trample the spark, else the flames will spread.
5. The hilt. of the sword was carved with fine designs.
6. A round hole was drilled through the thin board.
7. Footprints showed the path he took up the beach.
8. She was waiting at my front lawn.
9. A vent near the edge brought in fresh air.
10. Prod the old mule with a crooked stick.

**List 58**
1. It is a band of steel three inches wide.
2. The pipe ran almost the length of the ditch.
3. It was hidden from sight by a mass of leaves and shrubs.
4. The weight. of the package was seen on the high scale.
5. Wake and rise, and step into the green outdoors.
6. The green light in the brown box flickered.
7. The brass tube circled the high wall.
8. The lobes of her ears were pierced to hold rings.
9. Hold the hammer near the end to drive the nail.
10. Next Sunday is the twelfth of the month.

*A. Appendices***List 59**
1. Every word and phrase he speaks is true.
2. He put his last cartridge into the gun and fired.
3. They took their kids from the public school.
4. Drive the screw straight into the wood.
5. Keep the hatch tight and the watch constant.
6. Sever the twine with a quick snip of the knife.
7. Paper will dry out when wet.
8. Slide the catch back and open the desk.
9. Help the weak to preserve their strength.
10. A sullen smile gets few friends.

**List 60**
1. Stop whistling and watch the boys march.
2. Jerk the cord, and out tumbles the gold.
3. Slidc the tray across the glass top.
4. The cloud moved in a stately way and was gone.
5. Light maple makes for a swell room.
6. Set the piece here and say nothing.
7. Dull stories make her laugh.
8. A stiff cord will do to fasten your shoe.
9. Get the trust fund to the bank early.
10. Choose between the high road and the low.

**List 61**
1. A plea for funds seems to come again.
2. He lent his coat to the tall gaunt stranger.
3. There is a strong chance it will happen once more.
4. The duke left the park in a silver coach.
5. Greet the new guests and leave quickly.
6. When the frost has come it is time for turkey.
7. Sweet words work better than fierce.
8. A thin stripe runs down the middle.
9. A six comes up more often than a ten.
10. Lush fern grow on the lofty rocks.

**List 62**
1. The ram scared the school children off.
2. The team with the best timing looks good.
3. The farmer swapped his horse for a brown ox.
4. Sit on the perch and tell the others what to do.
5. A steep trail is painful for our feet.
6. The early phase of life moves fast.
7. Green moss grows on the northern side.
8. Tea in thin china has a sweet taste.
9. Pitch the straw through the door of the stable.
10. The latch on the beck gate needed a nail.

**List 63**
1. The goose was brought straight from the old market.
2. The sink is the thing in which we pile dishes.
3. A whiff of it will cure the most stubborn cold.
4. The facts don't always show who is right.
5. She flaps her cape as she parades the street.
6. The loss of the cruiser was a blow to the fleet.
7. Loop the braid to the left and then over.
8. Plead with the lawyer to drop the lost cause.
9. Calves thrive on tender spring grass.
10. Post no bills on this office wall.

**List 64**
1. Tear a thin sheet from the yellow pad.
2. A cruise in warm waters in a sleek yacht is fun.
3. A streak of color ran down the left edge.
4. It was done before the boy could see it.
5. Crouch before you jump or miss the mark.
6. Pack the kits and don't forget the salt.
7. The square peg will settle in the round hole.
8. Fine soap saves tender skin.
9. Poached eggs and tea must suffice.
10. Bad nerves are jangled by a door slam.

**List 65**
1. Ship maps are different from those for planes.
2. Dimes showered down from all sides.
3. They sang the same tunes at each party.
4. The sky in the west is tinged with orange red.
5. The pods of peas ferment in bare fields.
6. The horse balked and threw the tall rider.
7. The hitch between the horse and cart broke.
8. Pile the coal high in the shed corner.
9. The gold vase is both rare and costly.
10. The knife was hung inside its bright sheath.

**List 66**
1. The rarest spice comes from the far East.
2. The roof should be tilted at a sharp slant.
3. A smatter of French is worse than none.
4. The mule trod the treadmill day and night.
5. The aim of the contest is to raise a great fund.
6. To send it. now in large amounts is bad.
7. There is a fine hard tang in salty air.
8. Cod is the main business of the north shore.
9. The slab was hewn from heavy blocks of slat'e.
10. Dunk the stale biscuits into strong drink.

**List 67**
1. Hang tinsel from both branches.
2. Cap the jar with a tight brass cover.
3. The poor boy missed the boat again.
4. Be sure to set the lamp firmly in the hole.
5. Pick a card and slip it. under the pack.
6. A round mat will cover the dull spot.
7. The first part of the plan needs changing.
8. The good book informs of what we ought to know.
9. The mail comes in three batches per day.
10. You cannot brew tea in a cold pot.

**List 68**
1. Dots of light betrayed the black cat.
2. Put the chart on the mantel and tack it down.
3. The night shift men rate extra pay.
4. The red paper brightened the dim stage.
5. See the player scoot to third base.
6. Slide the bill between the two leaves.
7. Many hands help get the job done.
8. We don't like to admit our small faults.
9. No doubt about the way the wind blows.
10. Dig deep in the earth for pirate's gold.

A-12

**List 69**
1. The steady drip is worse than a drenching rain.
2. A flat pack takes less luggage space.
3. Green ice frosted the punch bowl.
4. A stuffed chair slipped from the moving van.
5. The stitch will serve but needs to be shortened.
6. A thin book fits in the side pocket.
7. The gloss on top made it unfit to read.
8. The hail pattered on the burnt brown grass.
9. Seven seals were stamped on great sheets.
10. Our troops are set to strike heavy blows.

**List 70**
1. The store was jammed before the sale could start.
2. It was a bad error on the part of the new judge.
3. One step more and the board will collapse.
4. Take the match and strike it against your shoe.
5. The pot boiled, but the contents failed to jell.
6. The baby puts his right foot in his mouth.
7. The bombs left most of the town in ruins.
8. Stop and stare at the hard working man.
9. The streets are narrow and full of sharp turns.
10. The pup jerked the leash as he saw a feline shape.

**List 71**
1. Open your book to the first page.
2. Fish evade the net, and swim off.
3. Dip the pail once and let it settle.
4. Will you please answer that phone.
5. The big red apple fell to the ground.
6. The curtain rose and the show was on.
7. The young prince became heir to the throne.
8. He sent the boy on a short errand.
9. Leave now and you will arrive on time.
10. The corner store was robbed last night.

**List 72**
1. A gold ring will please most any girl.
2. The long journey home took a year.
3. She saw a cat in the neighbor's house.
4. A pink shell was found on the sandy beach.
5. Small children came to see him.
6. The grass and bushes were wet with dew.
7. The blind man counted his old coins.
8. A severe storm tore down the barn.
9. She called his name many times.
10. When you hear the bell, come quickly.

## A.2.5. Haskins Syntactic Sentences

*Table 41: Haskins Syntactic Sentences*

| Series 1 | | Series 2 | |
|---|---|---|---|
| 1. | The wrong shot led the farm. | 51. | The new wife left the heart. |
| 2. | The black top ran the spring. | 52. | The mean shade broke the week. |
| 3. | The great car met the milk. | 53. | The hard blow built the truth. |
| 4. | The old corn cost the blood. | 54. | The next game paid the fire. |
| 5. | The short arm sent the cow. | 55. | The first car stood the ice. |
| 6. | The low walk read the hat. | 56. | The hot box paid the tree. |
| 7. | The rich paint said the land. | 57. | The live farm got the book. |
| 8. | The big bank felt the bag. | 58. | The white peace spoke the share. |
| 9. | The sick seat grew the chain. | 59. | The black shout caught the group. |
| 10. | The salt dog caused the shoe. | 60. | The end field sent the point. |
| 11. | The last fire tried the nose. | 61. | The sick word had the door. |
| 12. | The young voice saw the rose. | 62. | The last dance armed the leg. |
| 13. | The gold rain led the wing. | 63. | The fast earth lost the prince. |
| 14. | The chance sun laid the year. | 64. | The gray boat bit the sun. |
| 15. | The white bow had the bed. | 65. | The strong ring shot the nest. |
| 16. | The near stone thought the ear. | 66. | The rich branch heard the post. |
| 17. | The end home held the press. | 67. | The gold glass tried the meat. |
| 18. | The deep head cut the cent. | 68. | The dark cow laid the sea. |
| 19. | The next wind sold the room. | 69. | The deep shoe burned the face. |
| 20. | The full leg shut the shore. | 70. | The north drive hurt the dog. |
| 21. | The safe meat caught the shade. | 71. | The chance wood led the stone. |
| 22. | The fine lip tired the earth. | 72. | The young shore caused the bill. |
| 23. | The plain can lost the men. | 73. | The least lake sat the boy. |
| 24. | The dead hand armed the bird. | 74. | The big hair reached the head. |
| 25. | The fast point laid the word. | 75. | The short page let the knee. |
| 26. | The mean wave made the game. | 76. | The bad bed said the horse. |
| 27. | The clean book reached the ship. | 77. | The bright cent caught the king. |
| 28. | The red shop said the yard. | 78. | The fine bag ran the car. |
| 29. | The late girl aged the boat. | 79. | The old fish called the feet. |
| 30. | The large group passed the judge. | 80. | The late milk made the cold. |
| 31. | The past knee got the shout. | 81. | The clear well asked the air. |
| 32. | The least boy caught the dance. | 82. | The dear hill tried the work. |
| 33. | The green week did the page. | 83. | The full plant cut the voice. |
| 34. | The live cold stood the plant. | 84. | The game boy thought the back. |
| 35. | The third air heard the field. | 85. | The east floor brought the home. |
| 36. | The far man tried the wood. | 86. | The brown chair paid the girl. |
| 37. | The high sea burned the box. | 87. | The plain drink cost the wind. |
| 38. | The blue bill broke the branch. | 88. | The dark road net the hold. |
| 39. | The game feet asked the egg. | 89. | The new truth sat the blow. |
| 40. | The ill horse brought the hill. | 90. | The gray prince called the hall. |
| 41. | The strong rock built the ball. | 91. | The march face spoke the peace. |
| 42. | The dear neck ran the wife. | 92. | The hard heart let the bay. |
| 43. | The dry door paid the race. | 93. | The north king paid the drive. |
| 44. | The child share spread the school. | 94. | The first oil put the drink. |
| 45. | The brown post bit the ring. | 95. | The light eye hurt the lake. |
| 46. | The clear back hurt the fish. | 96. | The bad ice beat the floor. |
| 47. | The round work came the well. | 97. | The best house left the floor. |
| 48. | The good tree set the hair. | 98. | The east show found the cloud. |
| 49. | The bright guide knew the glass. | 99. | The cool lord paid the grass. |
| 50. | The hot nest gave the street. | 100. | The coarse friend shot the chair. |

*Table 41: Haskins Syntactic Sentences*

| Series 3 | | Series 4 | |
|---|---|---|---|
| 101. | The march hall aged the neck. | 151. | The brown bank tired the floor. |
| 102. | The great cloud read the road. | 152. | The deep shop sold the dance. |
| 103. | The past egg passed the shot. | 153. | The gold truth cost the ball. |
| 104. | The round blood grew the wind. | 154. | The big work burned the bird. |
| 105. | The cool rose spread the eye. | 155. | The last arm hurt the shade. |
| 106. | The light ball held the bow. | 156. | The low walk lost the nose. |
| 107. | The salt wing tired the oil. | 157. | The blue eye broke the plant. |
| 108. | The low net set the show. | 158. | The fast face grew the shoe. |
| 109. | The large year ran the bank. | 159. | The large home caused the ear. |
| 110. | The red school hurt the house. | 160. | The rich wave beat the net. |
| 111. | The near bird did the can. | 161. | The light post held the field. |
| 112. | The third press met the arm. | 162. | The dark bill left the branch. |
| 113. | The blue race shut the rock. | 163. | The best man felt the gate. |
| 114. | The ill land put the friend. | 164. | The dear work met the ship. |
| 115. | The green chain knew the man. | 165. | The ill seat read the cent. |
| 116. | The coarse judge saw the walk. | 166. | The live home caught the spring. |
| 117. | The safe hat felt the lord. | 167. | The round shot laid the shout. |
| 118. | The child yard laid the hand. | 168. | The hot door heard the bed. |
| 119. | The dry gate found the wave. | 169. | The brown lord tried the cow. |
| 120. | The best nose gave the corn. | 170. | The mean arm spoke the land. |
| 121. | The good grass held the paint. | 171. | The large hand burned the game. |
| 122. | The high street said the top. | 172. | The blue nest aged the bay. |
| 123. | The wrong room sold the rain. | 173. | The past horse made the shade. |
| 124. | The far ship beat the guide. | 174. | The hard girl caused the blood. |
| 125. | The right spring led the seat. | 175. | The game road found the page. |
| 126. | The wrong head thought the farm. | 176. | The third stone said the net. |
| 127. | The black corn sent the word. | 177. | The young air had the rose. |
| 128. | The strong prince came the grass. | 178. | The dry wind laid the floor. |
| 129. | The short boy paid the school. | 179. | The bright dog saw the glass. |
| 130. | The dark share hurt the earth. | 180. | The bad house hurt the hair. |
| 131. | The north friend gave the drink. | 181. | The gray car knew the wood. |
| 132. | The dead book grew the plant. | 182. | The fast lip ran the field. |
| 133. | The clean show left the men. | 183. | The first wave built the yard. |
| 134. | The safe knee paid the rose. | 184. | The gold walk let the box. |
| 135. | The far voice called the ring. | 185. | The clear shop cost the ball. |
| 136. | The march oil asked the peace. | 186. | The low king bit the wing. |
| 137. | The last tree did the egg. | 187. | The cool sea led the bag. |
| 138. | The next eye shot the ball. | 188. | The old guide beat the well. |
| 139. | The salt bill broke the dance. | 189. | The child top put the shore. |
| 140. | The fine truth tired the ear. | 190. | The rich group stood the press. |
| 141. | The white sun got the boat. | 191. | The high five set the chain. |
| 142. | The coarse paint shut the bird. | 192. | The east face paid the judge. |
| 143. | The red back said the hold. | 193. | The plain post tried the cloud. |
| 144. | The least can sold the chair. | 194. | The chance bank caught the blow. |
| 145. | The end rock lost the shoe. | 195. | The full week reached the race. |
| 146. | The sick neck led the hat. | 196. | The deep heart cut the year. |
| 147. | The green ice passed the hill. | 197. | The good cold held the wife. |
| 148. | The big bow spread the lake. | 198. | The near rain sang the drive. |
| 149. | The late point sat the branch. | 199. | The new feet brought the street. |
| 150. | The great leg armed the milk. | 200. | The light meat ran the fish. |

## A.2.6. MOS-X Test Form

1   Listening Effort: Please rate the degree of effort you had to make to understand the message.
    **Impossible even with much effort    1    2    3    4    5    6    7    No effort required**

2   Comprehension Problems: Were single words hard to understand?
    **All words hard to understand    1    2    3    4    5    6    7    All words easy to understand**

3   Speech Sound Articulation: Were the speech sounds clearly distinguishable?
    **Not at all clear    1    2    3    4    5    6    7    Very clear**

4   Precision: Was the articulation of speech sounds precise?
    **Slurred or imprecise    1    2    3    4    5    6    7    Precise**

5   Voice Pleasantness: Was the voice you heard pleasant to listen to?
    **Very unpleasant    1    2    3    4    5    6    7    Very pleasant**

6   Voice Naturalness: Did the voice sound natural?
    **Very unnatural    1    2    3    4    5    6    7    Very natural**

7   Humanlike Voice: To what extent did this voice sound like a human?
    **Nothing like a human    1    2    3    4    5    6    7    Just like a human**

8   Voice Quality: Did the voice sound harsh, raspy, or strained?
    **Significantly harsh/raspy    1    2    3    4    5    6    7    Normal quality**

9   Emphasis: Did emphasis of important words occur?
    **Incorrect emphasis    1    2    3    4    5    6    7    Excellent use of emphasis**

10  Rhythm: Did the rhythm of the speech sound natural?
    **Unnatural or mechanical    1    2    3    4    5    6    7    Natural rhythm**

11  Intonation: Did the intonation pattern of sentences sound smooth and natural?
    **Abrupt or abnormal    1    2    3    4    5    6    7    Smooth or natural**

12  Trust: Did the voice appear to be trustworthy?
    **Not at all trustworthy    1    2    3    4    5    6    7    Very trustworthy**

13  Confidence: Did the voice suggest a confident speaker?
    **Not at all confident    1    2    3    4    5    6    7    Very confident**

14  Enthusiasm: Did the voice seem to be enthusiastic?
    **Not at all enthusiastic    1    2    3    4    5    6    7    Very enthusiastic**

15  Persuasiveness: Was the voice persuasive?
    **Not at all persuasive    1    2    3    4    5    6    7    Very persuasive**

# A.3. Appendix 4: Conversions to Custom Tagset
## A.3.1. Conversion from CLAWS7 to Custom Tagset

*Table 42: Conversion from CLAWS7 to Custom Tagset*

| CLAWS7 | New | CLAWS7 Definition |
|---|---|---|
| APPGE | D | possessive pronoun, pre-nominal (e.g. my, your, our) |
| AT | D | article (e.g. the, no) |
| AT1 | D | singular article (e.g. a, an, every) |
| BCL | ? | before-clause marker (e.g. in order (that),in order (to)) |
| CC | C | coordinating conjunction (e.g. and, or) |
| CCB | C | adversative coordinating conjunction ( but) |
| CS | C | subordinating conjunction (e.g. if, because, unless, so, for) |
| CSA | C | as (as conjunction) |
| CSN | C | than (as conjunction) |
| CST | C | that (as conjunction) |
| CSW | C | whether (as conjunction) |
| DA | D | after-determiner or post-determiner capable of pronominal function (e.g. such, former, same) |
| DA1 | D | singular after-determiner (e.g. little, much) |
| DA2 | D | plural after-determiner (e.g. few, several, many) |
| DAR | D | comparative after-determiner (e.g. more, less, fewer) |
| DAT | D | superlative after-determiner (e.g. most, least, fewest) |
| DB | D | before determiner or pre-determiner capable of pronominal function (all, half) |
| DB2 | D | plural before-determiner ( both) |
| DD | D | determiner (capable of pronominal function) (e.g any, some) |
| DD1 | D | singular determiner (e.g. this, that, another) |
| DD2 | D | plural determiner ( these,those) |
| DDQ | D | wh-determiner (which, what) |
| DDQGE | D | wh-determiner, genitive (whose) |
| DDQV | D | wh-ever determiner, (whichever, whatever) |
| EX | N | existential there |
| FO | ? | formula |
| FU | ? | unclassified word |
| FW | ? | foreign word |
| GE | ? | germanic genitive marker - (' or's) |
| IF | P | for (as preposition) |
| II | P | general preposition |
| IO | P | of (as preposition) |
| IW | P | with, without (as prepositions) |
| JJ | A | general adjective |
| JJR | A | general comparative adjective (e.g. older, better, stronger) |
| JJT | A | general superlative adjective (e.g. oldest, best, strongest) |
| JK | A | catenative adjective (able in be able to, willing in be willing to) |
| MC | N | cardinal number,neutral for number (two, three..) |
| MC1 | N | singular cardinal number (one) |
| MC2 | p | plural cardinal number (e.g. sixes, sevens) |
| MCGE | N | genitive cardinal number, neutral for number (two's, 100's) |
| MCMC | N | hyphenated number (40-50, 1770-1827) |
| MD | A | ordinal number (e.g. first, second, next, last) |
| MF | A | fraction,neutral for number (e.g. quarters, two-thirds) |
| ND1 | A | singular noun of direction (e.g. north, southeast) |
| NN | N | common noun, neutral for number (e.g. sheep, cod, headquarters) |
| NN1 | N | singular common noun (e.g. book, girl) |
| NN2 | p | plural common noun (e.g. books, girls) |
| NNA | A | following noun of title (e.g. M.A.) |
| NNB | A | preceding noun of title (e.g. Mr., Prof.) |
| NNL1 | N | singular locative noun (e.g. Island, Street) |
| NNL2 | p | plural locative noun (e.g. Islands, Streets) |
| NNO | N | numeral noun, neutral for number (e.g. dozen, hundred) |
| NNO2 | p | numeral noun, plural (e.g. hundreds, thousands) |
| NNT1 | N | temporal noun, singular (e.g. day, week, year) |
| NNT2 | p | temporal noun, plural (e.g. days, weeks, years) |
| NNU | A | unit of measurement, neutral for number (e.g. in, cc) |
| NNU1 | N | singular unit of measurement (e.g. inch, centimetre) |
| NNU2 | p | plural unit of measurement (e.g. ins., feet) |
| NP | N | proper noun, neutral for number (e.g. IBM, Andes) |
| NP1 | N | singular proper noun (e.g. London, Jane, Frederick) |
| NP2 | p | plural proper noun (e.g. Browns, Reagans, Koreas) |
| NPD1 | N | singular weekday noun (e.g. Sunday) |
| NPD2 | p | plural weekday noun (e.g. Sundays) |
| NPM1 | N | singular month noun (e.g. October) |
| NPM2 | p | plural month noun (e.g. Octobers) |
| PN | r | indefinite pronoun, neutral for number (none) |
| PN1 | r | indefinite pronoun, singular (e.g. anyone, everything, nobody, one) |
| PNQO | r | objective wh-pronoun (whom) |
| PNQS | r | subjective wh-pronoun (who) |
| PNQV | r | wh-ever pronoun (whoever) |
| PNX1 | r | reflexive indefinite pronoun (oneself) |
| PPGE | r | nominal possessive personal pronoun (e.g. mine, yours) |
| PPH1 | r | 3rd person sing. neuter personal pronoun (it) |
| PPHO1 | r | 3rd person sing. objective personal pronoun (him, her) |
| PPHO2 | r | 3rd person plural objective personal pronoun (them) |
| PPHS1 | r | 3rd person sing. subjective personal pronoun (he, she) |
| PPHS2 | r | 3rd person plural subjective personal pronoun (they) |
| PPIO1 | r | 1st person sing. objective personal pronoun (me) |
| PPIO2 | r | 1st person plural objective personal pronoun (us) |
| PPIS1 | r | 1st person sing. subjective personal pronoun (I) |
| PPIS2 | r | 1st person plural subjective personal pronoun (we) |
| PPX1 | r | singular reflexive personal pronoun (e.g. yourself, itself) |
| PPX2 | r | plural reflexive personal pronoun (e.g. yourselves, themselves) |
| PPY | r | 2nd person personal pronoun (you) |
| RA | v | adverb, after nominal head (e.g. else, galore) |
| REX | v | adverb introducing appositional constructions (namely, e.g.) |
| RG | v | degree adverb (very, so, too) |
| RGQ | v | wh- degree adverb (how) |
| RGQV | v | wh-ever degree adverb (however) |
| RGR | v | comparative degree adverb (more, less) |
| RGT | v | superlative degree adverb (most, least) |
| RL | v | locative adverb (e.g. alongside, forward) |
| RP | v | prep. adverb, particle (e.g about, in) |
| RPK | v | prep. adv., catenative (about in be about to) |
| RR | v | general adverb |
| RRQ | v | wh- general adverb (where, when, why, how) |
| RRQV | v | wh-ever general adverb (wherever, whenever) |
| RRR | v | comparative general adverb (e.g. better, longer) |
| RRT | v | superlative general adverb (e.g. best, longest) |
| RT | v | quasi-nominal adverb of time (e.g. now, tomorrow) |
| TO | v | infinitive marker (to) |
| UH | ! | interjection (e.g. oh, yes, um) |
| VB0 | V | be, base form (finite i.e. imperative, subjunctive) |
| VBDR | V | were |
| VBDZ | V | was |
| VBG | V | being |
| VBI | V | be, infinitive (To be or not... It will be ..) |
| VBM | V | am |
| VBN | V | been |
| VBR | V | are |
| VBZ | V | is |
| VD0 | V | do, base form (finite) |
| VDD | V | did |
| VDG | V | doing |

| VDI | V | do, infinitive (I may do... To do...) |
|---|---|---|
| VDN | V | done |
| VDZ | V | does |
| VH0 | V | have, base form (finite) |
| VHD | V | had (past tense) |
| VHG | V | having |
| VHI | V | have, infinitive |
| VHN | V | had (past participle) |
| VHZ | V | has |
| VM | V | modal auxiliary (can, will, would, etc.) |
| VMK | V | modal catenative (ought, used) |
| VV0 | V | base form of lexical verb (e.g. give, work) |
| VVD | V | past tense of lexical verb (e.g. gave, worked) |
| VVG | V | -ing participle of lexical verb (e.g. giving, working) |
| VVGK | V | -ing participle catenative (going in be going to) |
| VVI | V | infinitive (e.g. to give... It will work...) |
| VVN | V | past participle of lexical verb (e.g. given, worked) |
| VVNK | V | past participle catenative (e.g. bound in be bound to) |
| VVZ | V | -s form of lexical verb (e.g. gives, works) |
| XX | v | not, n't |
| ZZ1 | N | singular letter of the alphabet (e.g. A,b) |
| ZZ2 | p | plural letter of the alphabet (e.g. A's, b's) |

## A.3.2. Conversion from Brown Corpus Tagset to Custom Tagset

*Table 43: Conversion from Brown Corpus Tagset to Custom Tagset*

| ABL | N | pre-qualifier (quite, rather) |
|---|---|---|
| ABN | A | pre-quantifier (half, all) |
| ABX | A | pre-quantifier (both) |
| AP | A | post-determiner (many, several, next) |
| AT | D | article (a, the, no) |
| BE | C | be |
| BED | C | were |
| BEDZ | C | was |
| BEG | C | being |
| BEM | C | am |
| BEN | C | been |
| BER | C | are, art |
| BEZ | C | is |
| CC | C | coordinating conjunction (and, or) |
| CD | N | cardinal numeral (one, two, 2, etc.) |
| CS | C | subordinating conjunction (if, although) |
| DO | V | do |
| DOD | V | did |
| DOZ | V | does |
| DT | D | singular determiner/quantifier (this, that) |
| DTI | D | singular or plural determiner/quantifier (some, any) |
| DTS | D | plural determiner (these, those) |
| DTX | C | determiner/double conjunction (either) |
| EX | N | existential there |
| FW | N | foreign word (hyphenated before regular tag) |
| HV | A | have |
| HVD | A | had (past tense) |
| HVG | A | having |
| HVN | A | had (past participle) |
| IN | P | preposition |
| JJ | A | adjective |
| JJR | A | comparative adjective |
| JJS | A | semantically superlative adjective (chief, top) |
| JJT | A | morphologically superlative adjective (biggest) |

| MD | A | modal auxiliary (can, should, will) |
|---|---|---|
| NC | N | cited word (hyphenated after regular tag) |
| NN | N | singular or mass noun |
| NN$ | N | possessive singular noun |
| NNS | N | plural noun |
| NNS$ | N | possessive plural noun |
| NP | N | proper noun or part of name phrase |
| NP$ | N | possessive proper noun |
| NPS | N | plural proper noun |
| NPS$ | N | possessive plural proper noun |
| NR | N | adverbial noun (home, today, west) |
| OD | N | ordinal numeral (first, 2nd) |
| PN | r | nominal pronoun (everybody, nothing) |
| PN$ | r | possessive nominal pronoun |
| PP$ | r | possessive personal pronoun (my, our) |
| PP$$ | r | second (nominal) possessive pronoun (mine, ours) |
| PPL | r | singular reflexive/intensive personal pronoun (myself) |
| PPLS | r | plural reflexive/intensive personal pronoun (ourselves) |
| PPO | r | objective personal pronoun (me, him, it, them) |
| PPS | r | 3rd. singular nominative pronoun (he, she, it, one) |
| PPSS | r | other nominative personal pronoun (I, we, they, you) |
| PRP | r | Personal pronoun |
| PRP$ | r | Possessive pronoun |
| QL | A | qualifier (very, fairly) |
| QLP | A | post-qualifier (enough, indeed) |
| RB | v | adverb |
| RBR | v | comparative adverb |
| RBT | v | superlative adverb |
| RN | v | nominal adverb (here, then, indoors) |
| RP | v | adverb/particle (about, off, up) |
| TO | v | infinitive marker to |
| UH | ! | interjection, exclamation |
| VB | V | verb, base form |
| VBD | V | verb, past tense |
| VBG | V | verb, present participle/gerund |
| VBN | V | verb, past participle |
| VBP | V | verb, non 3rd person, singular, present |
| VBZ | V | verb, 3rd. singular present |
| WDT | D | wh- determiner (what, which) |
| WP$ | r | possessive wh- pronoun (whose) |
| WPO | r | objective wh- pronoun (whom, which, that) |
| WPS | r | nominative wh- pronoun (who, which, that) |
| WQL | v | wh- qualifier (how) |
| WRB | v | wh- adverb (how, where, when) |

## A.4. Appendix 4: Test Results
## A.4.1. Harvard Sentences Transcription Scores

*Table 44: Harvard Sentences Transcription Scores*

| BADSPEECH - David | Maximum | BenT | BenP | Kat | KC | Dex | Average | Percent |
|---|---|---|---|---|---|---|---|---|
| 1. The birch canoe slid on the smooth planks. | 5 | 2 | 2 | 0 | 2 | 2 | 1.6 | 32% |
| 2. Glue the sheet to the dark blue background. | 5 | 1 | 3 | 3 | 2 | 1 | 2 | 40% |
| 3. It's easy to tell the depth of a well. | 4 | 4 | 4 | 4 | 4 | 3 | 3.8 | 95% |
| 4. These days a chicken leg is a rare dish. | 6 | 5 | 5 | 6 | 2 | 5 | 4.6 | 77% |
| 5. Rice is often served in round bowls. | 5 | 5 | 5 | 2 | 3 | 4 | 3.8 | 76% |
| 6. The juice of lemons makes fine punch. | 5 | 5 | 5 | 5 | 5 | 5 | 5 | 100% |
| 7. The box was thrown beside the parked truck. | 5 | 5 | 4 | 3 | 3 | 3 | 3.6 | 72% |
| 8. The hogs were fed chopped corn and garbage. | 5 | 1 | 0 | 2 | 1 | 3 | 1.4 | 28% |
| 9. Four hours of steady work faced us. | 6 | 1 | 5 | 0 | 1 | 2 | 1.8 | 30% |
| 10. Large size in stockings is hard to sell. | 5 | 4 | 4 | 4 | 4 | 4 | 4 | 80% |
| Total | 51 | 33 | 37 | 29 | 27 | 32 | 31.6 | **62%** |
| BADSPEECH - Alice | Maximum | BenT | BenP | Kat | KC | Dex | Average | Percent |
| 1. The boy was there when the sun rose. | 5 | 0 | 4 | 2 | 1 | 1 | 1.6 | 32% |
| 2. A rod is used to catch pink salmon. | 5 | 0 | 5 | 3 | 2 | 2 | 2.4 | 48% |
| 3. The source of the huge river is the clear spring. | 5 | 2 | 5 | 3 | 2 | 5 | 3.4 | 68% |
| 4. Kick the ball straight and follow through. | 5 | 0 | 4 | 0 | 0 | 2 | 1.2 | 24% |
| 5. Help the woman get back to her feet. | 6 | 5 | 6 | 5 | 6 | 6 | 5.6 | 93% |
| 6. A pot of tea helps to pass the evening. | 5 | 0 | 5 | 1 | 0 | 2 | 1.6 | 32% |
| 7. Smoky fires lack flame and heat. | 5 | 5 | 4 | 2 | 3 | 2 | 3.2 | 64% |
| 8. The soft cushion broke the man's fall. | 5 | 5 | 5 | 5 | 5 | 5 | 5 | 100% |
| 9. The salt breeze came across from the sea. | 6 | 4 | 4 | 3 | 4 | 5 | 4 | 67% |
| 10. The girl at the booth sold fifty bonds. | 5 | 1 | 3 | 1 | 3 | 1 | 1.8 | 36% |
| Total | 52 | 22 | 45 | 25 | 26 | 31 | 29.8 | **57%** |
| BADSPEECH - Josh | Maximum | BenT | BenP | Kat | KC | Dex | Average | Percent |
| 1. The small pup gnawed a hole in the sock. | 5 | 0 | 3 | 2 | 2 | 0 | 1.4 | 28% |
| 2. The fish twisted and turned on the bent hook. | 5 | 4 | 4 | 1 | 3 | 4 | 3.2 | 64% |
| 3. Press the pants and sew a button on the vest. | 5 | 3 | 4 | 1 | 3 | 2 | 2.6 | 52% |
| 4. The swan dive was far short of perfect. | 5 | 4 | 5 | 3 | 4 | 1 | 3.4 | 68% |
| 5. The beauty of the view stunned the young boy. | 5 | 1 | 5 | 1 | 2 | 1 | 2 | 40% |
| 6. Two blue fish swam in the tank. | 5 | 5 | 5 | 5 | 5 | 5 | 5 | 100% |
| 7. Her purse was full of useless trash. | 6 | 5 | 5 | 5 | 5 | 5 | 5 | 83% |
| 8. The colt reared and threw the tall rider. | 5 | 1 | 3 | 2 | 3 | 3 | 2.4 | 48% |
| 9. It snowed, rained, and hailed the same morning. | 5 | 2 | 5 | 0 | 3 | 2 | 2.4 | 48% |
| 10. Read verse out loud for pleasure. | 5 | 1 | 2 | 3 | 0 | 2 | 1.6 | 32% |
| Total | 51 | 26 | 41 | 23 | 30 | 25 | 29 | **57%** |

| **BADSPEECH - Megan** | Maximum | BenT | BenP | Kat | KC | Dex | Average | Percent |
|---|---|---|---|---|---|---|---|---|
| **1. Hoist the load to your left shoulder.** | 5 | 2 | 2 | 2 | 4 | 0 | 2 | 40% |
| **2. Take the winding path to reach the lake.** | 5 | 4 | 5 | 2 | 4 | 5 | 4 | 80% |
| **3. Note closely the size of the gas tank.** | 5 | 3 | 3 | 3 | 2 | 2 | 2.6 | 52% |
| **4. Wipe the grease off his dirty face.** | 6 | 2 | 6 | 6 | 6 | 6 | 5.2 | 87% |
| **5. Mend the coat before you go out.** | 6 | 0 | 6 | 5 | 3 | 2 | 3.2 | 53% |
| **6. The wrist was badly strained and hung limp.** | 5 | 4 | 5 | 4 | 5 | 4 | 4.4 | 88% |
| **7. The stray cat gave birth to kittens.** | 5 | 4 | 4 | 3 | 2 | 1 | 2.8 | 56% |
| **8. The young girl gave no clear response.** | 6 | 2 | 2 | 2 | 2 | 2 | 2 | 33% |
| **9. The meal was cooked before the bell rang.** | 5 | 2 | 2 | 3 | 2 | 2 | 2.2 | 44% |
| **10. What joy there is in living.** | 5 | 1 | 4 | 2 | 5 | 3 | 3 | 60% |
| Total | 53 | 24 | 39 | 32 | 35 | 27 | 31.4 | **59%** |
| **ODDSPEECH - David** | Maximum | BenT | BenP | Kat | KC | Dex | Average | Percent |
| **1. A king ruled the state in the early days.** | 5 | 3 | 3 | 1 | 2 | 2 | 2.2 | 44% |
| **2. The ship was torn apart on the sharp reef.** | 5 | 5 | 5 | 4 | 5 | 3 | 4.4 | 88% |
| **3. Sickness kept him home the third week.** | 5 | 1 | 3 | 3 | 1 | 2 | 2 | 40% |
| **4. The wide road shimmered in the hot sun.** | 5 | 2 | 2 | 1 | 2 | 1 | 1.6 | 32% |
| **5. The lazy cow lay in the cool grass.** | 5 | 1 | 5 | 3 | 5 | 3 | 3.4 | 68% |
| **6. Lift the square stone over the fence.** | 5 | 2 | 3 | 3 | 1 | 1 | 2 | 40% |
| **7. The rope will bind the seven books at once.** | 6 | 0 | 6 | 2 | 3 | 3 | 2.8 | 47% |
| **8. Hop over the fence and plunge in.** | 5 | 0 | 3 | 2 | 3 | 1 | 1.8 | 36% |
| **9. The friendly gang left the drug store.** | 5 | 2 | 2 | 1 | 0 | 2 | 1.4 | 28% |
| **10. Mesh mire keeps chicks inside.** | 5 | 2 | 2 | 2 | 1 | 2 | 1.8 | 36% |
| Total | 51 | 18 | 34 | 22 | 23 | 20 | 23.4 | **46%** |
| **ODDSPEECH - Alice** | Maximum | BenT | BenP | Kat | KC | Dex | Average | Percent |
| **1. The frosty air passed through the coat.** | 5 | 1 | 4 | 1 | 3 | 1 | 2 | 40% |
| **2. The crooked maze failed to fool the mouse.** | 5 | 0 | 2 | 0 | 1 | 0 | 0.6 | 12% |
| **3. Adding fast leads to wrong sums.** | 5 | 0 | 4 | 2 | 1 | 1 | 1.6 | 32% |
| **4. The show was a flop from the very start.** | 6 | 4 | 6 | 6 | 4 | 4 | 4.8 | 80% |
| **5. A saw is a tool used for making boards.** | 5 | 0 | 4 | 1 | 0 | 2 | 1.4 | 28% |
| **6. The wagon moved on well oiled wheels.** | 5 | 3 | 4 | 1 | 2 | 1 | 2.2 | 44% |
| **7. March the soldiers past the next hill.** | 5 | 1 | 2 | 2 | 1 | 0 | 1.2 | 24% |
| **8. A cup of sugar makes sweet fudge.** | 5 | 3 | 4 | 4 | 2 | 2 | 3 | 60% |
| **9. Place a rosebush near the porch steps.** | 5 | 0 | 0 | 0 | 0 | 0 | 0 | 0% |
| **10. Both lost their lives in the raging storm.** | 5 | 5 | 5 | 4 | 3 | 0 | 3.4 | 68% |
| Total | 51 | 17 | 35 | 21 | 17 | 11 | 20.2 | **40%** |

| ODDSPEECH - Josh | Maximum | BenT | BenP | Kat | KC | Dex | Average | Percent |
|---|---|---|---|---|---|---|---|---|
| **1. We talked of the slide show in the circus.** | 5 | 4 | 3 | 2 | 3 | 1 | 2.6 | 52% |
| **2. Use a pencil to write the first draft.** | 6 | 6 | 6 | 6 | 6 | 6 | 6 | 100% |
| **3. He ran half way to the hardware store.** | 6 | 1 | 6 | 5 | 2 | 6 | 4 | 67% |
| **4. The clock struck to mark the third period.** | 5 | 5 | 5 | 5 | 5 | 5 | 5 | 100% |
| **5. A small creek cut across the field.** | 5 | 5 | 5 | 4 | 5 | 2 | 4.2 | 84% |
| **6. Cars and busses stalled in snow drifts.** | 5 | 2 | 2 | 4 | 2 | 1 | 2.2 | 44% |
| **7. The set of china hit, the floor with a crash.** | 6 | 6 | 6 | 6 | 6 | 6 | 6 | 100% |
| **8. This is a grand season for hikes on the road.** | 6 | 2 | 4 | 3 | 5 | 4 | 3.6 | 60% |
| **9. The dune rose from the edge of the water.** | 5 | 4 | 5 | 3 | 4 | 2 | 3.6 | 72% |
| **10. Those words were the cue for the actor to leave.** | 6 | 2 | 6 | 3 | 4 | 0 | 3 | 50% |
| Total | 55 | 37 | 48 | 41 | 42 | 33 | 40.2 | **73%** |
| **ODDSPEECH - Megan** | Maximum | BenT | BenP | Kat | KC | Dex | Average | Percent |
| **1. A yacht slid around the point into the bay.** | 6 | 0 | 0 | 0 | 1 | 0 | 0.2 | 3% |
| **2. The two met while playing on the sand.** | 5 | 3 | 4 | 2 | 4 | 1 | 2.8 | 56% |
| **3. The ink stain dried on the finished page.** | 5 | 1 | 0 | 0 | 1 | 0 | 0.4 | 8% |
| **4. The walled town was seized without a fight.** | 5 | 3 | 3 | 2 | 3 | 3 | 2.8 | 56% |
| **5. The lease ran out in sixteen weeks.** | 5 | 2 | 5 | 0 | 1 | 0 | 1.6 | 32% |
| **6. A tame squirrel makes a nice pet.** | 5 | 3 | 5 | 3 | 3 | 1 | 3 | 60% |
| **7. The horn of the car woke the sleeping cop.** | 5 | 0 | 0 | 0 | 0 | 0 | 0 | 0% |
| **8. The heart beat strongly and with firm strokes.** | 6 | 0 | 1 | 0 | 0 | 0 | 0.2 | 3% |
| **9. The pearl was worn in a thin silver ring.** | 6 | 2 | 6 | 0 | 2 | 0 | 2 | 33% |
| **10. The fruit peel was cut in thick slices.** | 6 | 4 | 1 | 0 | 1 | 1 | 1.4 | 23% |
| Total | 54 | 18 | 25 | 7 | 16 | 6 | 14.4 | **27%** |

## A.4.2. MOS-X Test Results

*Table 45: MOS-X Test Results*

| **BADSPEECH - David** | **BenT** | **BenP** | **Kat** | **KC** | **Dex** | **Average** | | |
|---|---|---|---|---|---|---|---|---|
| Required effort to understand message: | 2 | 3 | 3 | 3 | 2 | 2.6 | **Intelligibility** | **2.9** |
| Were single words hard to understand? | 2 | 3 | 3 | 3 | 3 | 2.8 | | |
| Were speech sounds clearly distinguishable | 3 | 3 | 3 | 2 | 3 | 2.8 | | |
| Was articulation of speech precise? | 3 | 4 | 3 | 3 | 4 | 3.4 | | |
| Was the voice pleasant to listen to? | 1 | 4 | 1 | 5 | 4 | 3 | **Naturalness** | **2.55** |
| Did the voice sound natural? | 1 | 1 | 2 | 2 | 2 | 1.6 | | |
| Did the voice sound like a human? | 2 | 3 | 2 | 1 | 2 | 2 | | |
| Did the voice sound harsh, raspy, or strained? | 2 | 4 | 6 | 3 | 3 | 3.6 | | |
| Did the rhythm of the speech sound natural? | 3 | 1 | 4 | 2 | 1 | 2.2 | **Prosody** | **2.1** |
| Did the intonation sound smooth? | 3 | 2 | 2 | 1 | 2 | 2 | | |
| **BADSPEECH - Alice** | **BenT** | **BenP** | **Kat** | **KC** | **Dex** | **Average** | | |
| Required effort to understand message: | 2 | 4 | 3 | 2 | 3 | 2.8 | **Intelligibility** | **2.8** |
| Were single words hard to understand? | 1 | 4 | 2 | 3 | 3 | 2.6 | | |
| Were speech sounds clearly distinguishable | 2 | 5 | 2 | 2 | 5 | 3.2 | | |
| Was articulation of speech precise? | 2 | 4 | 1 | 1 | 5 | 2.6 | | |
| Was the voice pleasant to listen to? | 3 | 4 | 1 | 3 | 4 | 3 | **Naturalness** | **3.2** |
| Did the voice sound natural? | 2 | 2 | 4 | 2 | 3 | 2.6 | | |
| Did the voice sound like a human? | 3 | 3 | 3 | 3 | 3 | 3 | | |
| Did the voice sound harsh, raspy, or strained? | 2 | 4 | 6 | 2 | 7 | 4.2 | | |
| Did the rhythm of the speech sound natural? | 4 | 1 | 1 | 1 | 3 | 2 | **Prosody** | **2.1** |
| Did the intonation sound smooth? | 4 | 2 | 1 | 2 | 2 | 2.2 | | |
| **BADSPEECH - Josh** | **BenT** | **BenP** | **Kat** | **KC** | **Dex** | **Average** | | |
| Required effort to understand message: | 2 | 2 | 3 | 2 | 2 | 2.2 | **Intelligibility** | **1.8** |
| Were single words hard to understand? | 2 | 2 | 1 | 2 | 2 | 1.8 | | |
| Were speech sounds clearly distinguishable | 2 | 2 | 1 | 2 | 2 | 1.8 | | |
| Was articulation of speech precise? | 1 | 2 | 1 | 1 | 2 | 1.4 | | |
| Was the voice pleasant to listen to? | 1 | 2 | 1 | 2 | 1 | 1.4 | **Naturalness** | **1.75** |
| Did the voice sound natural? | 1 | 1 | 2 | 1 | 1 | 1.2 | | |
| Did the voice sound like a human? | 2 | 1 | 4 | 2 | 1 | 2 | | |
| Did the voice sound harsh, raspy, or strained? | 2 | 4 | 3 | 1 | 2 | 2.4 | | |
| Did the rhythm of the speech sound natural? | 3 | 1 | 1 | 1 | 1 | 1.4 | **Prosody** | **1.3** |
| Did the intonation sound smooth? | 2 | 1 | 1 | 1 | 1 | 1.2 | | |

| **BADSPEECH - Megan** | **BenT** | **BenP** | **Kat** | **KC** | **Dex** | **Average** | | |
|---|---|---|---|---|---|---|---|---|
| Required effort to understand message: | 2 | 3 | 1 | 4 | 5 | 3 | **Intelligibility** | **2.6** |
| Were single words hard to understand? | 1 | 3 | 1 | 4 | 4 | 2.6 | | |
| Were speech sounds clearly distinguishable | 2 | 3 | 1 | 3 | 3 | 2.4 | | |
| Was articulation of speech precise? | 1 | 3 | 3 | 2 | 3 | 2.4 | | |
| Was the voice pleasant to listen to? | 4 | 3 | 2 | 4 | 4 | 3.4 | **Naturalness** | **3.1** |
| Did the voice sound natural? | 3 | 2 | 2 | 3 | 2 | 2.4 | | |
| Did the voice sound like a human? | 4 | 2 | 2 | 4 | 2 | 2.8 | | |
| Did the voice sound harsh, raspy, or strained? | 4 | 4 | 4 | 3 | 4 | 3.8 | | |
| Did the rhythm of the speech sound natural? | 4 | 2 | 2 | 3 | 3 | 2.8 | **Prosody** | **2.7** |
| Did the intonation sound smooth? | 3 | 2 | 3 | 3 | 2 | 2.6 | | |
| **ODDSPEECH - David** | **BenT** | **BenP** | **Kat** | **KC** | **Dex** | **Average** | | |
| Required effort to understand message: | 1 | 3 | 1 | 2 | 2 | 1.8 | **Intelligibility** | **1.5** |
| Were single words hard to understand? | 1 | 2 | 1 | 1 | 1 | 1.2 | | |
| Were speech sounds clearly distinguishable | 1 | 2 | 1 | 2 | 1 | 1.4 | | |
| Was articulation of speech precise? | 1 | 3 | 1 | 1 | 2 | 1.6 | | |
| Was the voice pleasant to listen to? | 1 | 3 | 3 | 3 | 2 | 2.4 | **Naturalness** | **2.15** |
| Did the voice sound natural? | 1 | 3 | 2 | 2 | 1 | 1.8 | | |
| Did the voice sound like a human? | 1 | 2 | 1 | 2 | 1 | 1.4 | | |
| Did the voice sound harsh, raspy, or strained? | 1 | 4 | 3 | 3 | 4 | 3 | | |
| Did the rhythm of the speech sound natural? | 1 | 2 | 1 | 1 | 1 | 1.2 | **Prosody** | **1.4** |
| Did the intonation sound smooth? | 1 | 3 | 1 | 2 | 1 | 1.6 | | |
| **ODDSPEECH - Alice** | **BenT** | **BenP** | **Kat** | **KC** | **Dex** | **Average** | | |
| Required effort to understand message: | 1 | 3 | 2 | 1 | 1 | 1.6 | **Intelligibility** | **1.6** |
| Were single words hard to understand? | 1 | 3 | 1 | 2 | 1 | 1.6 | | |
| Were speech sounds clearly distinguishable | 1 | 4 | 2 | 1 | 1 | 1.8 | | |
| Was articulation of speech precise? | 1 | 2 | 2 | 1 | 1 | 1.4 | | |
| Was the voice pleasant to listen to? | 1 | 2 | 1 | 3 | 3 | 2 | **Naturalness** | **1.95** |
| Did the voice sound natural? | 1 | 2 | 1 | 2 | 1 | 1.4 | | |
| Did the voice sound like a human? | 1 | 2 | 3 | 1 | 1 | 1.6 | | |
| Did the voice sound harsh, raspy, or strained? | 1 | 3 | 4 | 2 | 4 | 2.8 | | |
| Did the rhythm of the speech sound natural? | 1 | 2 | 1 | 2 | 1 | 1.4 | **Prosody** | **1.4** |
| Did the intonation sound smooth? | 1 | 3 | 1 | 1 | 1 | 1.4 | | |

| ODDSPEECH - Josh | BenT | BenP | Kat | KC | Dex | Average | | |
|---|---|---|---|---|---|---|---|---|
| Required effort to understand message: | 2 | 6 | 3 | 5 | 4 | 4 | Intelligibility | 3.6 |
| Were single words hard to understand? | 2 | 6 | 2 | 5 | 4 | 3.8 | | |
| Were speech sounds clearly distinguishable | 2 | 6 | 2 | 5 | 2 | 3.4 | | |
| Was articulation of speech precise? | 2 | 5 | 3 | 4 | 2 | 3.2 | | |
| Was the voice pleasant to listen to? | 2 | 4 | 3 | 3 | 3 | 3 | Naturalness | 3.2 |
| Did the voice sound natural? | 2 | 5 | 3 | 4 | 2 | 3.2 | | |
| Did the voice sound like a human? | 2 | 3 | 3 | 4 | 1 | 2.6 | | |
| Did the voice sound harsh, raspy, or strained? | 2 | 5 | 5 | 4 | 4 | 4 | | |
| Did the rhythm of the speech sound natural? | 2 | 4 | 3 | 3 | 2 | 2.8 | Prosody | 2.8 |
| Did the intonation sound smooth? | 2 | 5 | 3 | 3 | 1 | 2.8 | | |
| **ODDSPEECH - Megan** | **BenT** | **BenP** | **Kat** | **KC** | **Dex** | **Average** | | |
| Required effort to understand message: | 1 | 3 | 1 | 1 | 1 | 1.4 | Intelligibility | 1.4 |
| Were single words hard to understand? | 2 | 3 | 1 | 1 | 1 | 1.6 | | |
| Were speech sounds clearly distinguishable | 1 | 3 | 1 | 1 | 1 | 1.4 | | |
| Was articulation of speech precise? | 2 | 1 | 1 | 1 | 1 | 1.2 | | |
| Was the voice pleasant to listen to? | 2 | 1 | 3 | 3 | 1 | 2 | Naturalness | 1.65 |
| Did the voice sound natural? | 1 | 1 | 2 | 2 | 1 | 1.4 | | |
| Did the voice sound like a human? | 1 | 1 | 2 | 1 | 1 | 1.2 | | |
| Did the voice sound harsh, raspy, or strained? | 1 | 2 | 3 | 3 | 1 | 2 | | |
| Did the rhythm of the speech sound natural? | 1 | 1 | 1 | 2 | 1 | 1.2 | Prosody | 1.4 |
| Did the intonation sound smooth? | 1 | 4 | 1 | 1 | 1 | 1.6 | | |